\ifx\pdfminorversion\undefined
  \ifx\pdfoptionpdfminorversion\undefined
  \else\pdfoptionpdfminorversion=6\fi
\else\pdfminorversion=6\fi
\documentclass[12pt]{article}
\usepackage{color}

\usepackage{slashed}

\usepackage{graphicx}
\usepackage{amsmath}
\usepackage{amssymb}
\usepackage{array}
\usepackage{wrapfig}
\usepackage{units}

\newcommand\slurp[1]{#1}
\newcommand\sslurp[2]{#1{#2}}
{\catcode`/=\active \expandafter}%
\slurp{\newcommand/}{/\penalty1000\hskip0pt\relax}
{\catcode`:=\active \expandafter}%
\slurp{\newcommand:}{:\penalty1000\hskip0pt\relax}

\newcommand\addspace{\ifcat\nextchar a\spacefactor999. \else.\fi}
{\catcode`\.=\active \expandafter}%
\slurp{\newcommand.}{\futurelet\nextchar\addspace}

\usepackage[unicode]{hyperref} 
\ifx\href\undefined\def\href#1{}\fi
\ifx\texorpdfstring\undefined\def\texorpdfstring#1#2{#1}\fi

\newcommand\myslash{/} \newcommand\mycolon{:}
\newcommand\doi{{\catcode`/=\active \catcode`:=\active \expandafter}\sslurp\realdoi}
{\catcode`/=\active \catcode`:=\active \expandafter}%
\slurp{\newcommand\realdoi[1]{{\let/=\myslash \let:=\mycolon
                               \edef\raw{{http://dx.doi.org/#1}}\expandafter}%
                               \expandafter\href\raw{doi:#1}}}
\newcommand\eprint[2]{{\escapechar-1%
                       \edef\a{\expandafter\string\csname arXiv\endcsname}%
                       \edef\b{\expandafter\string\csname #1\endcsname}%
                       \edef\c{\expandafter\string\csname #2\endcsname}%
                       \edef\d{\noexpand\href{http://arXiv.org/abs/\c}}%
                       \ifx\a\b\expandafter\d\fi{\tt #1:#2}}}

\newcommand{\fleq}[2][0]{%
  \hbox to 0pt{\hss
    \hbox to\columnwidth{\hspace{#1em}$\displaystyle#2$\hfil}%
   \hss}%
}
\newcommand{\ket}[1]{\:|\,{#1}\rangle\:}                      
\newcommand{\br}[1]{\:\langle{#1}|\:}

\newcommand{\beq}{\begin{equation}}
\newcommand{\eeq}{\end{equation}}
\newcommand{\bea}{\begin{eqnarray}}

\newcommand{\eea}{\end{eqnarray}}

\newcommand{\bal}{\begin{align}}
\newcommand{\eal}{\end{align}}

\newcommand{\ba}{\begin{array}}
\newcommand{\ea}{\end{array}}

\newcommand{\bset}{\begin{subequations}\begin{eqaligntwo}}
\newcommand{\eset}{\end{eqaligntwo}\end{subequations}}

\newcommand{\bseu}{\begin{subequations}\begin{eqalignno}}
\newcommand{\eseu}{\end{eqalignno}\end{subequations}}

\newcommand{\OMIT}[1]{{}}

\newcommand\spur{\raise.15ex\hbox{/}\kern-.57em } 

\newcommand{\ignore}[1]{}

\newcommand\inputwithmichaelmacro[1]{
  \let\ba=\bea\let\ea=\eea\let\l=\ell\let\sb=\tsb
  \input{#1}%
}

\newcommand\eq{\end{equation}}

\newcommand\centeron[2]{{\setbox0=\hbox{#1}\setbox1=\hbox{#2}\ifdim
\wd1>\wd0\kern.5\wd1\kern-.5\wd0\fi
\copy0\kern-.5\wd0\kern-.5\wd1\copy1\ifdim\wd0>\wd1
\kern.5\wd0\kern-.5\wd1\fi}}
\newcommand\ltap{\;\centeron{\raise.35ex\hbox{$<$}}{\lower.65ex\hbox{$\sim$}}\;}
\newcommand\gtap{\;\centeron{\raise.35ex\hbox{$>$}}{\lower.65ex\hbox{$\sim$}}\;}
\newcommand\gsim{\mathrel{\gtap}}
\newcommand\lsim{\mathrel{\ltap}}

\newcommand{\nc}{\newcommand}
\nc{\barrayn}{\begin{eqnarray*}}
\nc{\earrayn}{\end{eqnarray*}}
\nc{\er}[1]{(\ref{eq:#1})}
\nc{\ttbar}{t\bar  t}
\nc{\spacer}{\phantom{spacer}}
\nc{\ev}{\;\mathrm{eV}}
\nc{\mev}{\;\mathrm{MeV}}
\nc{\gev}{\;\mathrm{GeV}}
\nc{\tev}{\;\mathrm{TeV}}
\nc{\mc}{\mathcal}
\nc{\mcP}{\mc{P}}
\nc{\mcL}{\mathcal{L}}
\nc{\mcM}{\mathcal{M}}
\nc{\Z}{{\mathbb Z }}

\nc{\barray}{\begin{eqnarray}}
\nc{\earray}{\end{eqnarray}}
\nc{\bcenter}{\begin{center}}
\nc{\ecenter}{\end{center}}

\nc{\0}{\ket{0}}
\nc{\onehalf}{\frac{1}{2}}
\nc{\partialbar}{\bar{\partial}}
\nc{\psit}{\widetilde{\psi}}
\nc{\Tr}{\mbox{Tr}}

\newcommand\tsb{\tilde{b}}

\ifx\DeclareTextCommand\undefined\else
  \DeclareTextCommand{\textrightarrow}{PU}{\9041\222}
  \DeclareTextCommand{\textsupplus}{PU}{\9040\172}
\fi

\begin{document}
\newcommand\mystyle{\small\sl}
\newcommand\authorspace{\\[0.2cm]}
\newcommand\separatorspace{\\[1cm]}
\newcommand\addressspace{\\[0.25cm]}
\newcommand\titlefont{\bf}

\title{
\begin{flushright}
\vspace* {-0.2in}
\tiny LA-UR-11-11460 \\
\tiny NT@UW-11-16 \\
\tiny IFIC/11-57 \\
\tiny FTUV/11-1007 \\
\tiny NPAC-11-14 
\end{flushright}
\Large \bf Probing Novel Scalar and Tensor Interactions \\ 
from (Ultra)Cold  Neutrons to the LHC}
\author{Tanmoy Bhattacharya$^1$,
        Vincenzo Cirigliano$^1$,
        Saul D. Cohen$^{2,5}$,\authorspace
        Alberto Filipuzzi$^3$,
        Mart\'in Gonz\'alez-Alonso$^4$,\authorspace
        Michael L. Graesser$^1$,
        Rajan Gupta$^1$,
        Huey-Wen Lin$^5$ \separatorspace 
        \mystyle$^1$ Theoretical Division, Los Alamos National Laboratory, Los Alamos, NM 87545, USA\addressspace 
        \mystyle$^2$Center for Computational Science, Boston University, Boston, MA 02123, USA\addressspace 
        \mystyle$^3$Departament de F\'{\i}sica Te\`orica, IFIC, 
                    Universitat de Val\`encia -- CSIC  \\ 
        \mystyle    Apt.  Correus  22085, E-46071 
                    Val\`encia, Spain\addressspace 
        \mystyle$^4$Department of Physics, University of Wisconsin-Madison, \\
        \mystyle    1150 University Ave., Madison, WI, 53706, USA\addressspace 
        \mystyle$^5$Department of Physics, University of Washington, Seattle, WA 98195, USA}
\maketitle

    \begin{abstract}
      Scalar and tensor interactions were once competitors to the now
well-established $V-A$ structure of the Standard Model weak interactions.
We revisit these interactions and survey constraints 
from  low-energy probes (neutron, nuclear, and pion decays) 
as well as collider searches.
Currently, the most stringent limit on scalar  and tensor interactions arise from 
$0^+ \to 0^+$ nuclear decays and the radiative pion decay $\pi \to e \nu \gamma$, respectively. 
For the future, 
we find that  upcoming neutron beta decay and LHC measurements
will compete in setting the most stringent bounds. 
For neutron beta decay, we demonstrate the importance of lattice computations of the
neutron-to-proton matrix elements to setting limits on these interactions,
and provide the first  lattice estimate of the scalar charge  and a new average 
of existing results for the tensor charge.  
Data taken at the LHC is currently probing these interactions at the
$10^{-2}$ level (relative to the standard weak interactions), with the potential to reach the 
$\lsim 10^{-3}$ level.  
We show that,  with  some theoretical assumptions, the discovery of a charged spin-0 resonance 
decaying to an electron and missing energy implies a lower limit on the strength of  scalar interactions 
probed at low energy.

    \end{abstract}

\setcounter{page}{0}    
\thispagestyle{empty}
\newpage

    \section{Introduction\label{sec:intro}}
       Nuclear and neutron beta decays  have historically played a central role 
in determining the $V-A$ structure of weak interactions and in shaping 
what we now call the Standard Model (SM)~\cite{Weinberg:2009zz,Severijns2006dr}. 
Nowadays, precision measurements of  low-energy processes such as neutron decay 
can be used to probe the existence of non-SM interactions, such as novel scalar and tensor  structures.   
Considerable experimental efforts using both cold and ultracold neutrons  are underway worldwide, 
with the aim to improve the precision of  various neutron decay observables~\cite{Abele:2008zz,Dubbers:2011ns}:  
lifetime~\cite{Dewey2009189,Arzumanov2009186,Walstrom200982,Materne2009176,Leung2009181},  
beta asymmetry $A$~\cite{PERKEOIII:2009,Plaster:2008si,abBA,Dubbers:2007st}
neutrino asymmetry $B$~\cite{WilburnUCNB,abBA}, 
electron-neutrino correlation $a$~\cite{Pocanic:2008pu,aSPECT:2008,Wietfeldt:2005wz}, 
and Fierz interference term $b$~\cite{Pocanic:2008pu,UCNb}.
In some of the asymmetry measurements there are prospects to reach 
experimental sensitivities between  $10^{-3}$ and $10^{-4}$;  this makes  these 
observables  very interesting 
probes of new physics effects originating at  the TeV scale  that have expected size 
$(v/\Lambda_{\rm BSM})^2$, where $v = (2 \sqrt{2} G_F)^{-1/2} \approx 174$ GeV 
and $\Lambda_{\rm BSM}$ denotes the mass scale where physics beyond the Standard Model (BSM) 
appears. 

The  overall goal of this work is to assess the  discovery potential  and discriminating power 
of planned precision beta-decay  measurements  with cold and ultracold neutrons.  In particular 
we wish to study  the sensitivity of neutron decay to new physics in the
context of and in competition with:    
(i) other low-energy precision measurements in nuclear beta decays and pion decays; and 
(ii) high-energy collider searches (Tevatron, LHC).
In order to achieve our goal,  we work  within an effective field theory (EFT) setup, in which the dynamical 
effects of new heavy BSM  degrees of freedom are  parameterized by local operators  of dimension higher than four 
built with SM fields. 
In the absence of a clear new-physics signal from collider searches, we find this way of proceeding 
the most attractive and general: all specific model analyses of beta decays (see Ref.~\cite{Profumo:2006yu} for 
a discussion within supersymmetry) can be cast in the 
EFT language and the constraints on effective operators that we will derive can be readily 
converted into constraints on the parameters of any SM extension. 

Among various BSM contributions we identify new  scalar and tensor operators 
involving left-handed neutrinos  
as the most promising to probe with neutron decay, 
because they interfere  with the SM amplitude and thus  
contribute at  {\it linear} order to decay parameters. 
Motivated by this,  in the unified EFT framework 
we present a  comprehensive analysis of constraints on such scalar and tensor BSM interactions 
from  a broad range of  low-energy probes (neutron decay, nuclear decays, pion decays) as well as collider 
searches.\footnote{The EFT analysis of collider searches is valid as long as 
the particles that mediate the new interactions are above threshold for production at colliders.}
To our knowledge such an analysis is missing in the literature, 
despite being essential to judging the relative merits of various low-energy experiments. 

Extracting bounds on short-distance scalar and tensor couplings from neutron and nuclear beta decays 
requires knowledge of the nucleon scalar and tensor form factors 
at zero momentum transfer, denoted here by   $g_{S,T}$. 
In  previous beta-decay studies,   $g_{S}$ and $g_{T}$ have been assumed to be  $O(1)$ based on quark-model estimates 
(see,  for example,  Ref.~\cite{Herczeg2001vk}). 
The importance of the hadronic form factors can be appreciated by considering the extreme case in which 
 $g_{S,T} \ll 1$,  which  would dilute the  sensitivity  of beta decays to new physics. 
Concerning the hadronic form factors, the main results of this work are:
\begin{itemize}
\item We provide the first lattice-QCD estimate of $g_{S}$  and a new average of existing $g_T$ results. 
Current lattice uncertainties are at the level of $50\%$ for $g_S$ and $35\%$  for $g_T$. 
This already enables  much improved phenomenology (see for example Fig.~\ref{fig:constraints3}).  
\item We show that a precision of  20\% in $g_S$  will be 
needed to take full advantage of  $10^{-3}$-level neutron-decay measurements.  
We identify and discuss the key systematic effects that need to be brought under control 
in order to achieve $\delta g_{S}/g_{S} \sim 20\%$.  
\end{itemize}

Besides the new estimates of  $g_{S}$ and $g_T$ with lattice QCD (LQCD),  
the main new  findings of our analysis  can be summarized as follows:

\begin{itemize}

\item  Currently,  the most stringent constraints on the scalar and 
tensor effective couplings  (denoted by $\epsilon_S$ and $\epsilon_T$) arise from low-energy probes.
$\epsilon_S$ is constrained by $0^+ \to 0^+$ nuclear beta decays, while 
$\epsilon_T$ is constrained by the  Dalitz-plot analysis of the radiative pion decay $\pi \to e \nu \gamma$. 
There are also potentially very strong constraints on $\epsilon_{S,T}$ from 
the ratio of $\pi \to e \nu$ to $\pi \to \mu \nu$ decay rates. This constraint arises from operator mixing: 
once a scalar or tensor interaction is generated by new physics,  SM radiative corrections 
will generate an effective pseudoscalar operator that mediates the helicity-suppressed mode $\pi \to e \nu$. 
If the flavor structure of the SM extension is known, this constraint could be the strongest. 

\item Future neutron-decay measurements of the Fierz interference term $b$ and 
the analogue term $b_\nu$ in the neutrino asymmetry $B$ can greatly improve  
existing constraints on tensor interactions:  precision levels  
$\delta b, \delta b_\nu  \sim 10^{-3}$ would 
provide a four-fold or higher improvement in the bound (depending on the sign of $\epsilon_T$),
as shown in Figs.~\ref{fig:constraints1} and \ref{fig:constraints3}. 
On the other hand,   $\delta b, \delta b_\nu  \sim 10^{-4}$-level measurements would improve 
current bounds on $\epsilon_T$  by one order of magnitude and 
current bounds on $\epsilon_S$ by a factor of two  (see Fig.~\ref{fig:constraints1v2}).

\item Current collider bounds from the LHC are  not yet  competitive with low-energy constraints 
(see Fig.~\ref{LHCbounda}). 
Folding in the current uncertainty on $g_S$,  the LHC  bounds  on $\epsilon_S$  and $\epsilon_T$ 
are weaker by a factor of  about $4$ and $3$, respectively,   than those obtained  from nuclear decays 
and $\pi \to e \nu \gamma$.

\item Future LHC results, based on higher center-of-mass energy and higher integrated luminosity, 
 would definitely improve on current low-energy  bounds on $\epsilon_{S,T}$,  
 and would compete with improved low-energy constraints 
based on $\delta b, \delta b_\nu  \sim 10^{-3}$  in future neutron-decay measurements 
(see Fig.~\ref{LHCboundb}). 

\item Finally, we have explored the possibility that a mediator of new scalar 
interactions can be produced at the LHC. 
In this case, the EFT approach breaks down at collider energies and we have derived a general correlation between production cross-section for a scalar resonance at colliders and new-physics signal in neutron decay. 
This correlation links the discovery of a scalar resonance in $p p \to e \nu + X$ at the LHC with a 
lower bound (i.e. guaranteed signal) on $\epsilon_S$.  This is illustrated in Figs.~\ref{plot:eps-tau}, \ref{plot:eps-bound1}, 
and \ref{plot:eps-bound2}. 

\end{itemize}

The paper is organized as follows. 
In Section~\ref{sec:1} we present the effective theory description of
low-energy charged-current processes and briefly  discuss how the
coefficients may be constrained. In Section~\ref{sec:1a} we explain
our notation for the matrix elements required to describe the neutron
beta decay  and discuss how this decay constrains the parameters in
the effective field theory. In Section~\ref{sec:2}, we discuss the
low-energy phenomenological constraints on chirality-violating scalar and 
tensor operators  in the effective Lagrangian.  Section~\ref{sec:3} discusses current
and planned lattice analyses for the matrix elements of the 
quark bilinear structures $\bar{u} \Gamma d$  between  neutron and proton states, 
with special emphasis on the scalar and tensor structures. 
We provide the first estimate of $g_{S}$ from lattice QCD and a new average 
of existing calculations of $g_T$. 
In Section~\ref{sect:latticepheno} we summarize the impact of 
lattice estimates of $g_{S,T}$ on the phenomenology of scalar and tensor BSM interactions. 
In Section~\ref{sec:4}, we present the constraints on the
short-distance couplings obtained  from an analysis of high-energy scattering 
experiments  and discuss the improvement expected in the next few
years. We present our concluding remarks in Section~\ref{sec:discuss}. Two
appendices provide details of the operators contributing to charged-current processes and of the 
neutron-decay  differential decay distribution.

\section{\hspace*{-3pt}Effective theory  description of low-energy charged-current processes     
\label{sec:1}}
       
Following Ref.~\cite{Cirigliano:2009wk},  
we describe  new physics  contributions to low-energy charged-current (CC) processes 
in a model-independent  effective-theory setup, 
paying special attention to neutron-decay observables and their interplay with 
other low-energy and collider measurements.

We parameterize the effect of new  degrees of freedom and interactions beyond the SM 
via a series of  higher-dimensional operators constructed with low-energy SM
fields, assuming the existence of a mass gap between the SM
and its ultraviolet completion.    If the SM extension is weakly coupled, the resulting
TeV-scale effective Lagrangian linearly realizes the electroweak (EW)
symmetry $SU(2)_L \times U(1)_Y$ and contains a SM-like Higgs
doublet~\cite{Buchmuller:1985jz}.
We also assume that potential right-handed neutrino fields (sterile with respect to the 
SM gauge group) are heavy compared to the weak scale and therefore have been integrated out 
of the low-energy effective theory. 
This method  is quite general and allows us to study the implications
of precision measurements  on a large class of  models.

In our analysis we truncate the expansion of the effective Lagrangian to the lowest non-trivial order, 
given by dimension-six operators.  The contribution from the dimension-six operators to physical amplitudes  involves terms proportional to
$v^2/\Lambda_{\rm BSM}^2$ and $E^2/\Lambda_{\rm BSM}^2$, where $v= \langle \varphi^0 \rangle \approx
 174 \,  \rm{GeV}$
is the vacuum expectation value (VEV) of the Higgs field and $E$ is the characteristic energy
scale of a given process. We will work to linear order in these ratios of scales.

 \subsection{Effective Lagrangian}

In Ref.~\cite{Cirigliano:2009wk} a minimal basis of $SU(2) \times U(1)$ invariant  
dimension-six operators  contributing to low-energy charged-current processes was identified 
(see Appendix~\ref{sect:operators} for details). 
Denoting with  $\Lambda_i$  the effective dimensionful coupling associated with the operator $O_i$, 
we can write the effective Lagrangian as 
\bea
{\cal L}^{(\rm{eff})}
= {\cal L}_{\rm{SM}} + \sum_{i} \frac{1}{\Lambda_i^2}~ O_i \ \longrightarrow \ 
{\cal L}_{\rm{SM}} +  \frac{1}{v^2}  \, \sum_{i}  \, \hat{\alpha}_i   ~ O_i \, ,
\qquad 
{\rm with}  \  \  \hat{\alpha}_i = \frac{v^2}{\Lambda_i^2}~, 
\eea
where in the last step we have set the correct dimensions by the 
Higgs VEV $v$ and defined the dimensionless  new-physics couplings $\hat{\alpha}_i$, 
which are $O(10^{-3})$  for $\Lambda_i \sim {\rm TeV}$.  

In this framework one can derive the low-scale  $O(1 \ {\rm GeV})$  effective Lagrangian 
for semi-leptonic transitions.  It  receives contributions from both $W$-exchange diagrams (with modified $W$-fermion couplings)
and the four-fermion operators  
$O_{l q}^{(3)}$, $O_{qde}$, $O_{l q}$,  $O_{l q}^t$ defined in Appendix~\ref{sect:operators}. 
This matching procedure leads to~\cite{Cirigliano:2009wk}
\bea
{\cal L}_{\rm CC}
&=&  \frac{-g^2}{2 M_W^2} \, V_{ij} \, \Bigg[
 \Big(1 + [v_L]_{\ell \ell ij} \Big) \   \bar{\ell}_L \gamma_\mu  \nu_{\ell L}    \ \bar{u}_L^i \gamma^\mu d_L^j
 \ + \  [v_R]_{\ell \ell ij}  \   \bar{\ell}_L \gamma_\mu  \nu_{\ell L}    \ \bar{u}_R^i \gamma^\mu d_R^j
\nonumber\\
&+&  [s_L]_{\ell \ell ij}  \   \bar{\ell}_R   \nu_{\ell L}    \ \bar{u}_R^i  d_L^j
\ + \  [s_R]_{\ell \ell ij}  \   \bar{\ell}_R   \nu_{\ell L}    \ \bar{u}_L^i  d_R^j
\nonumber \\
& + &   [t_L]_{\ell \ell ij}  \   \bar{\ell}_R   \sigma_{\mu \nu} \nu_{\ell L}    \ \bar{u}_R^i   \sigma^{\mu \nu} d_L^j
\Bigg]~+~{\rm h.c.}~.
\label{eq:leffq}
\eea
where we use  $\sigma^{\mu \nu} = i [\gamma^\mu, \gamma^\nu]/2$. 
The SM effective Lagrangian corresponds to $v_L = v_R = s_L = s_R = t_L = 0$. 
The effective couplings $v_L , v_R, s_L,  s_R, t_L  \sim  v^2/\Lambda_i^2$ are functions of the 
coupling $\hat{\alpha}_i$  of $SU(2)\times U(1)$ invariant weak-scale operators. 
While their explicit expressions can be found in 
Appendix~\ref{sect:operators}, 
here we simply point out  two important features: 
\begin{itemize}
\item   $v_L$  involves a linear combination of  three weak-scale effective couplings: 
 a quark-gauge boson vertex correction, a lepton-gauge boson vertex
 correction, and a four-fermion operator coupling left-handed quarks
 and leptons (same chirality structure as the SM).  An important
 consequence is that by $SU(2) \times U(1)$ gauge invariance, $v_L$ is
 related to $Z^0$ fermion-antifermion vertex corrections and
 neutral-current four-fermion vertices.

\item  $v_R$ and  $s_L,s_R,t_L$ are in one-to-one correspondence with  weak-scale 
effective couplings. $v_R$ describes a right-handed charged-current quark coupling, while 
$s_L,s_R,t_L$ correspond to scalar and tensor four-quark operators.  Again, $SU(2)$ gauge invariance 
implies that these couplings mediate not only charged-currrent processes but also  
processes such as $\bar{e} e \leftrightarrow \bar{u} u, \bar{d} d$, with scalar or tensor Dirac  structure. 

\end{itemize}

In what follows, we will work in the limit in which the effective non-standard couplings $v_{L,R}$,  $s_{L,R}$, and $t_L$ are real
and we will focus only on CP-even observables (for a discussion of CP-odd observables 
refer to Ref.~\cite{Herczeg2001vk}). 
To simplify the notation, we will omit flavor indices,  e.g.  $[v_L]_{eeud}  \to v_L$. 
In addition, we will use the tree-level definition of the Fermi constant $g^2/(8 M_W^2) \equiv G_F^{(0)}/\sqrt{2}$.
Working to linear order in the non-standard couplings, and focusing on the $ij = ud$ component,  
the semi-leptonic effective Lagrangian 
can be written in the following useful form:
\bea
{\cal L}_{\rm CC} 
&=&
- \frac{G_F^{(0)} V_{ud}}{\sqrt{2}} \,
\Big(1 + \epsilon_L + \epsilon_R  \Big)
\Bigg[
\bar{\ell}  \gamma_\mu  (1 - \gamma_5)   \nu_{\ell} 
\cdot \bar{u}   \Big[ \gamma^\mu \ - \  \big(1 -2  \epsilon_R  \big)  \gamma^\mu \gamma_5 \Big] d \nonumber\\
&+& \bar{\ell}  (1 - \gamma_5) \nu_{\ell}
\cdot \bar{u}  \Big[  \epsilon_S  -   \epsilon_P \gamma_5 \Big]  d
+ \epsilon_T     \,   \bar{\ell}   \sigma_{\mu \nu} (1 - \gamma_5) \nu_{\ell}    \cdot  \bar{u}   \sigma^{\mu \nu} (1 - \gamma_5) d
\Bigg]+{\rm h.c.}, \ \ \ \  \
\label{eq:leffq2} 
\eea
where we have defined the effective scalar, pseudoscalar, and tensor couplings as follows: 
\beq
\epsilon_{L,R} \equiv v_{L,R}
\qquad
\epsilon_S \equiv s_L+s_R 
\qquad 
\epsilon_P \equiv  s_L-s_R
\qquad
\epsilon_T  \equiv  t_L~.
\eeq
While the physical amplitudes  are  renormalization scale  and scheme independent, the individual 
effective couplings   $\epsilon_{i}$   and hadronic matrix elements 
can display a strong scale dependence. 
Throughout the paper, we will quote estimates and bounds for the $\epsilon_i$ at the renormalization 
scale  $\mu=2$~GeV in the 
$\overline{\rm MS}$ scheme, unless otherwise specified.

The Lagrangian (\ref{eq:leffq2}) mediates all low-energy charged-current weak
processes involving up and down quarks. 
For a recent analysis of flavor-dependent constraints,  see Ref.~\cite{Carpentier:2010ue}.
In some of the charged-current  processes involving first-generation quarks 
 the theoretical and experimental precision has reached or will reach
in the near future a level that allows stringent bounds on
the new-physics effective couplings.  In this work we are interested
in assessing the sensitivity of neutron decay to new physics in the
context of (i) other low-energy constraints from nuclear beta decays
and pion decays; and (ii) constraints from high-energy colliders (LEP,
Tevatron, LHC).  To set the stage for the discussion, we summarize the
observables that give us access to the couplings appearing in
Eq.~(\ref{eq:leffq2}) (we will come back in detail to these in
following sections):

\begin{itemize}
\item  
The combination $(\epsilon_L+\epsilon_R)$ affects the overall normalization of the
effective Fermi constant.  This is phenomenologically accessible
through quark-lepton universality tests (precise determination of
$V_{ud}$ from $0^+ \to 0^+$ nuclear decays under the assumption that
$G_F = G_\mu$, where $G_{\mu}$ is the Fermi constant extracted from
muon decay).  An extensive analysis of the constraints on $(\epsilon_L+\epsilon_R)$
from universality tests and precision electroweak observables from  the
$Z$-pole was performed in Ref.~\cite{Cirigliano:2009wk}, within BSM 
scenarios with  minimal flavor violation.
In this context it was shown that constraints from low-energy are at the same level or
stronger (depending on the operator) than from $Z$-pole observables and
$e^+ e^- \to q \bar{q}$ cross-section measurements at LEP.

\item 
The right-handed coupling $\epsilon_R$ affects the relative normalization of
the axial and vector currents.  In neutron decay $\epsilon_R$ can be
reabsorbed in a redefinition of the axial coupling and experiments are
only sensitive to the combination $(1 - 2 \epsilon_R)g_A/g_V$ ($g_V$ and
$g_A$ are the vector and axial form factors at zero momentum transfer,
to be precisely defined below).  Disentangling $\epsilon_R$ requires
precision measurements of $(1 - 2 \epsilon_R)g_A/g_V$ and precision
calculations of $g_A/g_V$ in LQCD.

\item 
The effective pseudoscalar combination $\epsilon_P \equiv s_L - s_R$
contributes to leptonic decays of the pion.  It is strongly
constrained by the helicity-suppressed ratio $R_\pi \equiv \Gamma(\pi \to e \nu
[\gamma])/\Gamma(\pi \to \mu \nu [\gamma])$.  Moreover, as discussed
in Refs.~\cite{Voloshin:1992sn,Herczeg:1994ur, Campbell:2003ir}, the
low-energy coupling $\epsilon_P$ receives contributions proportional
to $\epsilon_{S,T}$ through electroweak radiative corrections.  We
will discuss the resulting constraints on $\epsilon_{S,P,T}$
in  Section~\ref{sect:pienu}.

\item  
Both the scalar combination $\epsilon_S \equiv s_L + s_R$ and the
tensor coupling $\epsilon_T \equiv t_L$ contribute at  linear
order  to the Fierz interference
terms in beta decays of neutrons and nuclei, and the neutrino-asymmetry
correlation coefficient $B$ in polarized neutron and nuclear decay
(see Appendix~\ref{sect:ndecaydetails} for notation).  Because of the
peculiar way in which the Fierz interference term appears in many
asymmetry measurements, bounds on $\epsilon_S$ and $\epsilon_T$ can
also be obtained by observation of the beta-asymmetry correlation
coefficient $A$,  electron-neutrino correlation $a$,   and positron
polarization measurements in various nuclear beta decays.  Finally,
the tensor coupling $\epsilon_T$ can also be constrained through
Dalitz-plot studies of the radiative pion decay $\pi \to
e \nu \gamma$.

\item  
All of the above operators can provide signatures at colliders. 
Currently there are no competitive
collider bounds on the chirality-flipping scalar and tensor couplings 
$\epsilon_{S,P,T}$, because their interference with the SM amplitude
carries  factors of $m_f/E_f$ (where $m_f$ is a light fermion mass,
$f \in \{e,u,d\}$), which at collider energies strongly suppresses the whole
effect.  So we immediately see that low-energy physics provides a
unique opportunity to probe these couplings, to which collider 
searches are sensitive only quadratically (i.e. via non-interference terms). 
We will derive  in Section~\ref{sec:4}  the current bounds on $\epsilon_{S,T}$ 
from searches at the LHC,  and we will show that 
with higher center-of-mass energy and integrated luminosity 
they will become  competitive with low-energy searches.

\end{itemize} 

Next, we review the analysis of neutron decay in the SM and beyond
within the  EFT  framework described above.

\section{Neutron \texorpdfstring{$\beta$}{\textbeta} decay\label{sec:1a}}

The amplitude for neutron decay $n (p_n) \to p (p_p) e^-
(p_e) \bar{\nu}_e (p_\nu) $ mediated by the effective Lagrangian
(\ref{eq:leffq2}) involves in principle  the matrix elements between the neutron and proton 
of all possible quark bilinears. These 
can be parameterized in terms of Lorentz-invariant form factors as
follows~\cite{Weinberg:1958ut}: 
\begin{subequations}
\label{eq:nucleonmatching}
\bea
\br{p (p_p) } \bar{u} \gamma_\mu d \ket{n (p_n)} &=&
\bar{u}_p (p_p)  \left[
g_V(q^2)  \,  \gamma_\mu    
+ \frac{\tilde{g}_{T(V)} (q^2)}{2 M_N}   \, \sigma_{\mu \nu}   q^\nu  
+ \frac{\tilde{g}_{S} (q^2)}{2 M_N}   \,  q_\mu  
\right]  
\hspace{-0.1cm}
 u_n (p_n)  \nonumber\\&&
\\
\br{p (p_p) } \bar{u} \gamma_\mu \gamma_5  d \ket{n (p_n)} &=&
\bar{u}_p (p_p)  \left[
g_A(q^2)    \gamma_\mu    
\hspace{-0.1cm}
+ \frac{\tilde{g}_{T(A)} (q^2)}{2 M_N}   \sigma_{\mu \nu}   q^\nu  
+ 
\hspace{-0.1cm}
\frac{\tilde{g}_{P} (q^2)}{2 M_N}   q_\mu  
\right]  \hspace{-0.15cm}  \gamma_5  u_n (p_n)  \nonumber\\&&
\label{eq:inducedgP}
\\
\br{p (p_p) } \bar{u} \,   d \ket{n (p_n)} &=&
g_S(q^2)  \ \bar{u}_p (p_p)  \, u_n (p_n)  
\\
\br{p (p_p) } \bar{u} \,  \gamma_5 \,  d \ket{n (p_n)} &=&
g_P(q^2)  \ \bar{u}_p (p_p)  \, \gamma_5 \, u_n (p_n)  
\label{eq:defgP}
\\
\br{p (p_p) } \bar{u} \, \sigma_{\mu \nu}  \,  d \ket{n (p_n)} &=&
 \bar{u}_p (p_p)   
 \left[
g_T(q^2) \, \sigma_{\mu \nu}   +  g_{T}^{(1)} (q^2)  \left(q_\mu \gamma_\nu - q_\nu \gamma_\mu \right) 
\nonumber  \right. \\
&+& 
\left.  g_{T}^{(2)} (q^2)  \left( q_\mu P_\nu - q_\nu P_\mu  \right) 
+
g_{T}^{(3)} (q^2) 
 \left(
\gamma_\mu  \slashed{q}  \gamma_\nu - 
\gamma_\nu  \slashed{q} \gamma_\mu
 \right) 
\right] 
u_n (p_n) 
\nonumber \\ & &  
\eea
\end{subequations}
where $u_{p,n}$ are the proton and neutron spinor amplitudes,   $P = p_n + p_p$,  
$q = p_n - p_p$ is the momentum transfer,  and $M_N= M_n = M_p$ denotes a common nucleon 
mass.\footnote{In the case of vector and axial bilinears, the induced tensor term proportional to 
$\sigma_{\mu \nu} q^\nu$ can be traded for an independent ``scalar" form factor proportional to $P_\mu$. 
Here we choose to follow the parameterization of Ref.~\cite{Weinberg:1958ut}.}
Note that all the above spinor contractions are $O(1)$, except for $\bar{u}_p \gamma_5 u_n $ which is $O(q/M_N)$. 
Moreover, as discussed below,  second-class current contributions   $\tilde{g}_S$ and $\tilde{g}_{T(A)}$  affect the amplitude  
at levels below the expected 
experimental sensitivities.

Our goal here is  to identify  TeV-induced new physics contaminations to the amplitude  of typical size  
$\epsilon_{P,S,T} \sim (v/\Lambda_{\rm BSM})^2  \sim 10^{-3}$. 
The effect we are after is of the same size as recoil corrections $q/M_N \sim 10^{-3}$ as well as radiative corrections $\alpha/\pi$. 
So in our analysis  we perform a simultaneous expansion in new physics contributions,  recoil, and radiative corrections 
keeping terms up to first order and neglecting higher-order terms, as they are 
smaller than the current and planned experimental sensitivity.
In light of this simultaneous expansion in $\epsilon_{P,S,T}$, $q/M_N$, and $\alpha/\pi$, 
we now discuss contributions from all quark-bilinear operators: 

\begin{itemize}
\item  {\bf Vector current}:  
The form factor $g_V(0)$ contributes at $O(1)$ to the amplitude and
$\tilde{g}_{T(V)} (0)$ contributes at first order in $q/M_N$. Also, up
to isospin-breaking corrections of order $(M_n - M_p)/M_N \sim q/M_N$,
the weak magnetism form factor $\tilde{g}_{T(V)} (0)$ can be related
to the difference of proton and neutron magnetic moments, that are
well known. On the other hand, the induced-scalar form factor
$\tilde{g}_S (q^2)$ vanishes in the isospin limit
\cite{Weinberg:1958ut}, so it is of order $(M_n - M_p)/M_N \sim q/M_N$. 
Since it multiplies one power of $q_\mu/M_N$,  its contribution to the amplitude is  effectively
second order in the recoil expansion, so we drop it. 

\item  {\bf Axial current}:  
From the axial current only $g_A (0)$ contributes up to first order. 
The induced-tensor  form factor  $\tilde{g}_{T(A)} (q^2)$  vanishes in the isospin limit 
\cite{Weinberg:1958ut}, and since it multiplies one power of $q_\mu/M_N$  its contribution to the amplitude 
is of second order in $q/M_N$, so we drop it.  Similarly, the
contribution associated with the induced-pseudoscalar form factor
$\tilde{g}_P$ is quadratic in our counting, because the pseudoscalar
bilinear is itself of order $q/M_N$, and it comes with an explicit
$q/M_N$ suppression, so we neglect it.\footnote{This effect is, however, 
enhanced. Using partially conserved axial current one can show that
the form factor $\tilde{g}_P$ is of order $M_N/m_q \sim 100$, making
the contribution to the amplitude of order $10^{-4}$.  In
Section~\ref{ssec:inducedps} we review the status of experimental data
and LQCD calculations showing this enhancement.  The effect of
$\tilde{g}_P$ on the neutron beta-decay rate has been worked out in
Ref.~\cite{Holstein:1974zf}, and it should be included when the
experiments reach that level of precision.}

\item  {\bf Pseudoscalar bilinear}:
The pseudoscalar bilinear $\bar{u}_p \gamma_5 u_n$ is itself of order $q/M_N$.
Since it necessarily multiplies a new-physics effective coupling $\epsilon_P$
(there is no pseudoscalar coupling in the SM), this term is also of
second order in our expansion,  and we drop it. 

\item {\bf Scalar and tensor bilinears}: 
These  bilinears enter into the analysis multiplied by new-physics effective couplings $\epsilon_{S,T}$. 
So we need the matrix elements  to zeroth order in the recoil expansion,  which leaves us 
with $g_{S} (0)$ and $g_T(0)$. $g_T^{(1,2,3)}(q^2)$ are all multiplied by one power of $q$ 
and $g_T^{(3)}$ vanishes in the isospin limit~\cite{Weinberg:1958ut}. 

\end{itemize}

In summary,  to the order we are working, the  amplitudes depend only  
on $g_i \equiv g_i(0)$ ($i \in \{ V,A,S,T \}$) and $\tilde{g}_{T(V)} (0)$.
Up to second-order corrections in isospin breaking, one has $g_V = 1$~\cite{Ademollo:1964sr,Donoghue:1990ti}.
For notational convenience, it is also useful to define the ratio of the 
axial to vector form factors as $\lambda \equiv g_A/g_V$.  As
noted earlier, in presence of non-standard right-handed interactions
the axial form factor is always multiplied by the correction factor
$(1 - 2 \epsilon_R)$, so that the neutron-decay amplitude is actually a
function of $\tilde{\lambda} \equiv \lambda (1 - 2 \epsilon_R)$.

Finally, in order to make contact with the existing standard
references on neutron and nuclear beta-decay
phenomenology~\cite{Lee:1956qn,Jackson1957zz,Severijns2006dr}, let us
note here that Eq.~(\ref{eq:nucleonmatching}) can be viewed as the
matching conditions from our quark-level effective
theory Eq.~(\ref{eq:leffq2}) to a nucleon-level effective theory, such as
the one originally written down by Lee and Yang~\cite{Lee:1956qn}.
The Lee-Yang effective couplings $C_i$, $C_i'$ ($i \in \{V,A,S,T\}$) can be
expressed in terms of our parameters as 
\begin{subequations}
\bea
C_{i} &=& \frac{G_F}{\sqrt{2}} \,  V_{ud} \,  \bar{C}_{i}    \\
\bar{C}_V &=& g_V  \left(1 + \epsilon_L + \epsilon_R \right)  \\
\bar{C}_A &=& - g_A  \left(1 + \epsilon_L - \epsilon_R \right)  \\
\bar{C}_S &=&   g_S  \, \epsilon_S \\
\bar{C}_T &=&    4 \, g_T \, \epsilon_T  ~, 
\eea
\end{subequations}
with $C_i ' = C_i$, since we only have left-handed neutrinos in our
low-energy effective theory. 
Operators involving right-handed neutrinos do not interfere with the SM amplitude 
and therefore contribute at second order to all observables. 
An analysis involving such operators will be presented elsewhere~\cite{VCMGMGA}.
Finally, notice that Ref.~\cite{Herczeg2001vk}
defines the couplings $C_A, C'_{V,S,T}$ with an  
overall  minus sign compared to ours. 

\subsection{Differential decay distribution}
 
Including the effect of recoil corrections, radiative corrections, and BSM couplings, the differential decay rate for polarized neutrons reads~\cite{Wilkinson:1982hu,Gluck:1995hs,Ando:2004rk,Gudkov:2005bu}
\bea
\label{eqn:jtw2}  
\frac{d\Gamma}{dE_e d \Omega_e d \Omega_\nu} &=& \frac{(G_F^{(0)})^2  \,  |V_{ud}|^2 }{(2\pi )^5}  
\left( 1 + 2 \epsilon_L + 2 \epsilon_R \right) \, 
 \left( 1 + 3 \, \tilde{\lambda}^2 \right) 
\cdot  w (E_e) \cdot D  (E_e,  \mathbf{p} _e, \mathbf{p} _\nu, \boldsymbol{\sigma} _n)   \, ,   
\nonumber\\
\eea
where 
$ \mathbf{p} _e$ and  $\mathbf{p} _\nu$ denote the electron and neutrino three-momenta, 
while $\boldsymbol{\sigma} _n$ denotes the neutron polarization. 
The bulk of the electron spectrum is described by 
\bea
w (E_e)  & = &    p_e E_e (E_0 - E_e)^2  \ F (Z=1, E_e)  \
 \left( 1 + \frac{\alpha}{2 \pi} e_V^R   +
\frac{\alpha}{2 \pi} \delta_\alpha^{(1)} (E_e)  
\right)  
\eea
where $E_0 = \Delta - (\Delta^2 - m_e^2)/(2 M_n)$ (with $\Delta = M_n
- M_p$) is the electron endpoint energy, $m_e$ is the electron mass,
and $F(Z,E_e)$ is the Fermi function that captures the Coulomb
radiative corrections ($Z$ denotes the charge of the daughter nucleus,
which coincides with the proton in this case).  The function
$\delta_\alpha^{(1)} (E_e)$~\cite{Ando:2004rk,Gudkov:2005bu} captures
model-independent (``outer") radiative corrections, while the coupling
$e_V^R$ is sensitive to the short-distance (``inner") radiative
correction~\cite{Czarnecki:2004cw,Ando:2004rk}.  The differential
decay distribution function $D(E_e, \mathbf{p} _e, \mathbf{p}
_\nu, \boldsymbol{\sigma} _n)$ is given
by~\cite{Ando:2004rk,Gudkov:2005bu}
\bea
D (E_e, \mathbf{p} _e, \mathbf{p} _\nu, \boldsymbol{\sigma} _n)  &  =  & 
1 +  c_0  + c_1 \, \frac{E_e}{M_N}   + \frac{m_e}{E_e} \bar{b}  +\bar{a} (E_e)   \frac{\mathbf{p }_e \cdot    \mathbf{p} _\nu}{E_e E_\nu}  +  \bar{A} (E_e)    \frac{\boldsymbol{\sigma}_n \cdot  \mathbf{p} _e}{E_e} 
\nonumber \\
&+& 
   \bar{B} (E_e)   \frac{\boldsymbol{\sigma}_n \cdot \mathbf{p }_\nu}{E_\nu}
+ \bar{C}_{(aa)} (E_e)   \left(\frac{\mathbf{p} _e \cdot    \mathbf{p} _\nu}{E_e E_\nu}\right)^2
+ \bar{C}_{(aA)} (E_e)   \frac{\mathbf{p} _e \cdot    \mathbf{p} _\nu}{E_e E_\nu}\frac{\boldsymbol{\sigma} _n \cdot 
 \mathbf{p} _e}{E_e}
\nonumber \\
&+& 
  \bar{C}_{(aB)} (E_e)  \frac{\mathbf{p} _e \cdot    \mathbf{p}_\nu}{E_e E_\nu} \frac{\boldsymbol{\sigma}_n \cdot \mathbf{p}_\nu}{E_\nu}    ~, 
\label{eq:diffn}
\eea
where  
$\bar{b}$ 
is an effective Fierz interference term 
and $\bar{a} (E_e)$, $\bar{A} (E_e)$, $\bar{B} (E_e)$ and $\bar{C}_{aa,aA,aB}(E_e)$ are  effective energy-dependent correlation coefficients, whose full  expressions~\cite{Ando:2004rk,Gudkov:2005bu,Gardner:2000nk} we report in Appendix~\ref{sect:ndecaydetails}, where one can also find the coefficients $c_{0,1}$  generated by recoil corrections.\footnote{
See also Ref.~\cite{Sjue:2005ks} for a discussion of recoil corrections to the proton asymmetry.}
In absence of radiative corrections, recoil corrections and BSM contributions, the effective correlation coefficients $\bar{a} (E_e)$, $\bar{A} (E_e)$ and $\bar{B} (E_e)$ reduce to the following well-known leading-order expressions 
\bea
\bar{a}(E_e) \to  \frac{1 - \lambda^2}{1 + 3 \lambda^2} ~,
~~~~~~ \bar{A}(E_e) \to    \frac{2 \lambda (1 - \lambda)}{1 + 3 \lambda^2} ~,
~~~~~~ \bar{B}(E_e) \to     \frac{2 \lambda ( 1 + \lambda)}{1 + 3 \lambda^2}~,
\eea
with the rest of coefficients ($c_{0,1},\bar{b},\bar{C}_{(aa,aA,aB)} (E_e)$) vanishing in this limit. 

The  impact of new-physics contributions can be summarized as follows: 
\begin{itemize}
\item  The effect of $\epsilon_{L/R}$ was already evident from the effective Lagrangian of Eq.~\ref{eq:leffq2}: they induce 
(i) an overall correction proportional to $(1 + 2  \epsilon_L + 2 \epsilon_R)$, and
(ii) the shift $\lambda \to \tilde{\lambda} = \lambda (1 - 2 \epsilon_R)$. As
a consequence of this second effect, working to linear order in new-physics 
contributions, the measurements of different correlation
coefficients by themselves cannot disentangle $\lambda$ and $\epsilon_R$; 
they simply provide independent measures of $\tilde{\lambda}$. 
In order to probe $\epsilon_R$ from correlation measurements, one needs to independently know 
$g_A/g_V$ from LQCD calculations. 

\item The scalar and tensor interactions $\epsilon_{S,T}$ 
appear to linear order only  through the Fierz interference term $\bar{b}$ and the analogue term
$b_\nu$ in the neutrino-asymmetry parameter ($b_\nu$ is the part of
$\bar{B} (E_e)$ proportional to $m_e/E_e$, see Appendix~\ref{sect:ndecaydetails} for a precise
definition)
\begin{subequations}
\label{eq:bbsm}
\bea
b^{\rm BSM}
&=&  \frac{ 2 }{1 + 3 \lambda^2}  \Bigg[ g_S \,  \epsilon_S  -   12 \lambda  \, g_T \, \epsilon_T  \Bigg]  
\approx  0.34\, g_S \,  \epsilon_S  -   5.22 \, g_T \, \epsilon_T~,   \\
b_\nu^{\rm BSM}
&=&  \frac{2}{1 + 3 \lambda^2}  \Bigg[ g_S \,  \epsilon_S \,  \lambda   - 4 g_T \, \epsilon_T  \, (1 + 2 \lambda)  \Bigg] 
\approx  0.44\, g_S \,  \epsilon_S   - 4.85\, g_T \, \epsilon_T~.
\eea
\end{subequations}
To the order we are working,  in the above expressions we can use either $\lambda$ or $\tilde{\lambda}$. 
\end{itemize}

Experimentally, one can probe the new-physics contributions in $\tilde{\lambda}$,  $b^{\rm BSM}$, and $b_\nu^{\rm BSM}$ 
through (i) measurements of the electron spectrum, aimed to isolate the term $\bar{b}$ in Eq.~(\ref{eq:diffn}); 
or (ii)  correlation measurements, aimed to isolate  $\bar{a}  (E_e)$, 
$\bar{A}  (E_e)$, and  $\bar{B}  (E_e)$ in Eq.~(\ref{eq:diffn}). 
Correlation measurements involve the construction of asymmetry ratios~\cite{Gluck:1995hs}. 
For example, in order to isolate $\bar{A} (E_e)$ one constructs the ratio 
$A_{\rm exp} (E_e) = (N_+ (E_e) - N_- (E_e))/(N_+ (E_e) + N_- (E_e))$ 
where $N_{\pm} (E_e)$ are the spectra corresponding to events with 
$\boldsymbol{\sigma}_n \cdot  {\bf p} _e >0$ and 
$\boldsymbol{\sigma}_n \cdot  {\bf p} _e <0$.
Similarly,  in order to isolate $\bar{B} (E_e)$ one can use the simple ratio
$B_{\rm exp} (E_e) = (Q_{++} (E_e) - Q_{--} (E_e))/(Q_{++} (E_e) + Q_{--} (E_e))$, 
where $Q_{++} (E_e)$  and $Q_{--} (E_e)$ are the spectra of events with 
$\boldsymbol{\sigma}_n \cdot  {\bf p} _e >0$, $\boldsymbol{\sigma}_n \cdot  {\bf p} _p >0$ 
and 
$\boldsymbol{\sigma}_n \cdot  {\bf p} _e < 0$,  $\boldsymbol{\sigma}_n \cdot  {\bf p} _p <0$, respectively. 
One can immediately see that through the total spectra in the denominator, 
both $A_{\rm exp} (E_e)$ and $B_{\rm exp} (E_e)$ are sensitive to the 
Fierz interference term $\bar{b}$,  so that asymmetry measurements involving simple 
ratios as described above really measure 
\beq
\tilde{Y} (E_e)  = \frac{\bar{Y} (E_e)}{ 1 +  \bar{b}  \, m_e/E_e}~,
\label{eq:obstilde}
\eeq
where  $Y \in \{ A,B, a, ... \}$.  Moreover, each individual experiment 
applies optimization cuts in $E_e$, thus mesuring a 
specific weighted average of Eq.~(\ref{eq:obstilde}). 

The above observation has important consequences for the phenomenology of
neutron decay: 
(i) The $m_e/E_e$ component of  $B_{\rm exp} (E_e)$ is sensitive not to $b_\nu^{\rm BSM}$ but 
rather to the combination 
 $ (1 + 3 \lambda^2)/(2 \lambda ( 1 + \lambda )) \, b_\nu^{\rm BSM}
- b^{\rm BSM} \approx b_\nu^{\rm BSM}  - b^{\rm BSM}$. 
Besides $B_{\rm exp} (E_e)$, it might be possible to construct  
a set of observables that disentangle the contribution of $b^{\rm BSM}$ and 
$b_{\nu}^{\rm BSM}$~\cite{ayoung}. In this case the BSM sensitivity of $b_\nu^{\rm BSM}$ alone 
is of interest.  In our phenomenological analysis we will study both cases (constraints from $b_\nu^{\rm BSM}  - b^{\rm BSM}$ 
and $b_\nu^{\rm BSM}$). 
(ii) More generally, correlation coefficients  measurements 
traditionally used to determine
$\lambda = g_A/g_V$ within the SM ($\epsilon_{L/R} =0, b=b_\nu=0$), provide
information on three independent parameters in our EFT setup:
$\tilde{\lambda} = \lambda (1 - 2 \epsilon_R)$, $b^{\rm BSM}$, and
$b_\nu^{\rm BSM}$.\footnote{
In other words, if $\epsilon_{S,T}$ are larger than the experimental errors, one has to observe an unexpected energy dependence of the form $m/E$ in the measurements of the correlation coefficients (in addition to the various expected energy dependences due to sub-leading standard effects that are detailed in Appendix~\ref{sect:ndecaydetails}). Thus, for a certain energy, a determination of $\lambda$ from $a(A)$ would be actually extracting the quantity $\tilde{\lambda}\left( 1 +  n_{a(A)} b^{\rm BSM} m/E \right)$, whereas in a $B$-based determination of $\lambda$, we would have
$\tilde{\lambda}\left( 1 +  n_B (b_\nu^{\rm BSM} - b^{\rm BSM}) m/ E \right)$, where
$n_a = \frac{(1-\lambda^2)(1+3\lambda^2)}{8\lambda^2} \approx -0.28$,
$n_A = -\frac{(1-\lambda)(1+3\lambda^2)}{(1+\lambda)(1-3\lambda)} \approx -0.25$ and
$n_B = \frac{(1+\lambda)(1+3\lambda^2)}{(1-\lambda)(1+3\lambda)} \approx -10.2$~.}
A fit to the current
data~\cite{Abele:2002wc,Liu:2010ms,Schumann:2007qe,Byrne:2002tx} (with
precision $\delta A/A \sim 0.005$, $\delta a/a \sim 0.05$, $\delta
B/B \sim 0.005$) yields $ - 0.3 < b^{\rm BSM}, b_\nu^{\rm BSM}  < 0.5$ at the 95\%
C.L.~\cite{Dubbers:2011ns}, which,  as we will see, is not competitive
with other bounds. It will be interesting, however, to explore the
implications of future experimental improvements in the
combined extraction of $\tilde{\lambda}$, $b^{\rm BSM}$ and $b_\nu^{\rm BSM}$ from $a$, $A$, and $B$ 
measurements, along the lines described in 
Refs.~\cite{Dubbers:2011ns,Konrad:2010wz}.

The main conclusion from the above discussion is that measurements of the 
differential neutron-decay  distribution  
are mostly sensitive to new physics through $b^{\rm BSM}$ and $b_\nu^{\rm BSM}$, 
which depend on the scalar and tensor couplings, $\epsilon_S$ and $\epsilon_T$,
to  linear order.  
Therefore,  apart from the next section, which we include for completeness, 
in the rest of this paper we restrict our 
discussion on these exotic scalar and tensor interactions, 
comparing the physics reach of neutron decay to other low-energy and collider probes.

\subsection{Total decay rate and determination of \texorpdfstring{$V_{ud}$}{V\string_ud}}

 For completeness, we discuss here the BSM corrections to the neutron decay rate and the extraction of $V_{ud}$ 
 from neutron decay. 
Expressing $G_F^{(0)}$ in terms of the Fermi constant determined in muon decay $G_{\mu}$
(this involves    non-standard contributions to the purely leptonic charged-current interaction encoded in the 
coefficient  $\tilde{v}_L$~\cite{Cirigliano:2009wk})
and performing the phase-space integrations, the total decay rate reads 
\begin{equation}
\Gamma   =  \frac{G_\mu^2  |V_{ud}|^2  m_e^5}{2 \pi^3}  \left(1 + 3 \tilde{\lambda}^2\right) \cdot f \cdot  \left(1 + \Delta_{\rm RC} \right) 
\,  \Bigg[  1  +  2 \epsilon_L - 2 \tilde{v}_L  + 2 \epsilon_R + b^{\rm BSM} \frac{I_1(x_0)  }{I_0(x_0)} \Bigg]~.
\label{eq:taun}
\end{equation}
In the above expression, the corrections from BSM physics are encoded in $\tilde{\lambda}$ and 
the terms in square brackets. 
$\Delta_{\rm RC}= 3.90 (8)  \times 10^{-2}$ is the SM electroweak radiative correction~\cite{Czarnecki:2004cw}, 
and the phase-space integrals are defined by
\beq
I_k  (x_0)= \int_{1}^{x_0}  x^{1 - k} \, (x_0 - x)^2 \, \sqrt{x^2 - 1} \ dx  \qquad \qquad 
f  = I_0 (x_0)  (1 + \Delta_f) ~, 
\eeq
where $x_0 = E_0/m_e$ and  
$\Delta_f$ encodes  Coulomb and recoil corrections that are numerically quite important, 
$I_0 ( x_0 ) =  1.629, \ 
f  = 1.6887, \ 
I_1 ( x_0)/I_0 ( x_0) =  0.652$ (See Ref.~\cite{Czarnecki:2004cw} for details).
In order to extract $V_{ud}$ from neutron decays one needs (see Eq.~\ref{eq:taun}) 
experimental input on the neutron lifetime $1/\Gamma$~\cite{Serebrov:2004zf,Pichlmaier:2010zz}  and  
$\tilde{\lambda}$, which is usually extracted
 from beta-asymmetry $A_{\rm exp} (E_e)$ measurements~\cite{Abele:2002wc,Liu:2010ms} 
(after accounting for recoil and radiative corrections). 
 Taking into account Eq.~(\ref{eq:obstilde}), the usual method for extracting $\tilde{\lambda}$  actually 
 determines  $\tilde{\lambda}\left( 1+ c \,b^{\rm BSM} \right)$, where $c$ is a certain $O(1)$ number that depends on the specific experimental analysis.  
In summary what we really extract from neutron beta decay is not $V_{ud}$ but the combination 
\bea
|V_{ud}|^2 \Big|_{n \to p e \bar{\nu}}
&=& |V_{ud}|^2  \Bigg[  1  +  2 \epsilon_L - 2 \tilde{v}_L  + 2 \epsilon_R + b^{\rm BSM} \left( \frac{I_1(x_0)}{I_0(x_0)} -  \frac{6 \lambda^2}{1+3\lambda^2}c \right) \Bigg]~
\nonumber\\
&\approx& |V_{ud}|^2  \Bigg[  1  +  2 \epsilon_L - 2 \tilde{v}_L  + 2 \epsilon_R +  b^{\rm BSM} \left( 0.65 -  1.66\,c \right) \Bigg]~. 
\label{eq:Vud1}
\eea

\section{Low-energy phenomenology of scalar and tensor interactions}
\label{sect:pheno}
\label{sec:2}

\subsection{Other probes of scalar and tensor interactions}

In order to assess the discovery potential of experiments planning to measure 
$\bar{b}$ and $\tilde{B}$ at the level of  $10^{-3}$ and $10^{-4}$, it is crucial to 
identify existing constraints on new scalar and tensor operators. 
As we discuss below in some detail,  the most stringent constraint 
on the scalar coupling $\epsilon_S$  arises from $0^+ \to 0^+$ 
nuclear beta decays. On the other hand, the most stringent bound 
on the tensor effective coupling $\epsilon_T$ arises from the Dalitz-plot 
study of the radiative pion decay $\pi \to e \nu \gamma$. 
For completeness, we will also briefly review 
(i) constraints on $\epsilon_{S,T}$  from other nuclear beta-decay observables, 
showing that they  are not competitive at the moment; 
and (ii)  constraints on $\epsilon_{S,P,T}$  arising from the 
helicity-suppressed $\pi \to e \nu$ decay. 
As we will show, the latter provides  potentially the strongest constraints on $\epsilon_{S,T}$, 
once the flavor structure of the underlying theory is known. 
This provides very stringent constraints on model building. 

\subsubsection{\texorpdfstring{$0^+ \to 0^+$}{0\textsupplus \textrightarrow 0\textsupplus} transitions and scalar interactions}

At leading order within the SM and new physics, the differential decay rate for an  unpolarized nucleus 
is~\cite{Jackson1957zz}
\bea
\label{eqn:jtw2a}  
\frac{d\Gamma_{0^+ \to 0^+}}{dE_e d \Omega_e d \Omega_\nu} &=& 
2 \frac{(G_F^{(0)})^2  \,  |V_{ud}|^2 }{(2\pi )^5}  
\left( 1 + 2 \epsilon_L + 2 \epsilon_R \right) \, 
p_e E_e (\tilde{E}_0 - E_e)^2 F(- Z, E_e) \nonumber \\
& \times &  
\left\{ 1 + a_{0^+}   \frac{\bf{p}_e \cdot
\bf{p}_\nu}{E_e E_\nu} +
b_{0^+}  \frac{m_e}{E_e}  
\right\}
\eea
where $\tilde{E}_0= M_P - M_D$ is the electron endpoint energy expressed in terms of the masses of parent and daughter nuclei,  
$F(- Z, E_e)$ is the Fermi function,  $Z$ is the atomic number of the daughter nucleus (the minus sign  
applies to $\beta^+$ emitters for which the most  precise measurements exist). 
For $0^+ \to 0^+$ transitions the coefficients  
$a, b$ are 
\begin{subequations}
\bea
a_{0^+}  &=& 1\\
b_{0^+}  & = &  -2 \gamma \, g_S \, \epsilon_S  \qquad \qquad \gamma = \sqrt{ 1- \alpha^2  Z^2}~, 
\eea
\end{subequations}
and the total  rate  is given by 
\beq
\Gamma_{0^+ \to 0^+}  =
\frac{G_\mu^2   |V_{ud}|^2  m_e^5}{ \pi^3} \, f_{0^+ \to 0^+} \left( 1 + \Delta_{\rm RC}^{(0^+ \to 0^+)} \right) \, 
\Bigg[ 1 + 2 \epsilon_L - 2 \tilde{v}_L  + 2 \epsilon_R  + b_{0^+} \frac{I_1(\tilde{x}_0)}{I_0 (\tilde{x}_0)} \Bigg]
\label{eq:taunucl}
\eeq
where  $\tilde{x}_0 = \tilde{E}_0/m_e$. In this last expression, the SM sub-effects have been included through $\Delta_{\rm RC}^{(0^+ \to 0^+)}$ and also inside $f_{0^+ \to 0^+}$, that up to Coulomb, nuclear distortion and recoil effects, is $f_{0^+ \to 0^+} = I_0 (\tilde{x}_0)$, similarly to what happens in the neutron-decay case. The various radiative corrections (including  $\Delta_{\rm RC}^{(0^+ \to 0^+)}$)  are  discussed in detail in Refs.~\cite{Marciano:2005ec,Hardy:2008gy}. Comparing the values of  $V_{ud}$ as extracted from neutron and nuclear decays, we find (see Eq.~(\ref{eq:Vud1}) and the preceding discussion)
\beq
\frac{|V^{0^+\to 0^+}_{ud}|^2}{|V^{n\to p e \bar{\nu}}_{ud}|^2} =
1  
+   b^{\rm BSM}_{0^+}  \frac{I_1 (\tilde{x}_0)}{I_0(\tilde{x}_0)}  - b^{\rm BSM}_n \left( \frac{I_1(x_0 )}{I_0(x_0 )} - \frac{6 \lambda^2}{1+3\lambda^2}c \right) ~, 
\eeq
which in principle provides another handle on scalar and tensor interactions. 

Let us now come to the point of greatest interest for this paper's discussion.
From a comparison of precisely known half-lives corrected by phase-space factors $f_{0^+ \to 0^+}$,  
Hardy and Towner~\cite{Hardy:2008gy}
found $b_{0^+} = - 0.0022(26)$, which translates into  the following  bound on the product of 
nucleon scalar form factor and short-distance scalar coupling: 
\beq
 - 1.0 \times 10^{-3}  <   g_S \, \epsilon_S  <  3.2 \times 10^{-3}  
 \qquad \qquad (90 \% \  \rm{C.L.})~. 
\label{eq:bFbound}
\eeq
This is the most stringent bound on scalar interactions from low-energy probes.

\subsubsection{Radiative pion decay and the tensor interaction}
An analysis of the Dalitz plot of  the radiative pion decay 
$\pi^+  \to  e^+ \nu_e \gamma$ is sensitive to the same tensor operator 
that can be probed in beta decays.   The experimental results from the PIBETA  
collaboration~\cite{Bychkov:2008ws}
put constraints on the product $\epsilon_T \times f_T$ of the short-distance coupling $\epsilon_T$ 
and the hadronic form factor $f_T$  defined by~\cite{Mateu:2007tr}
\beq
\langle \gamma (\epsilon, p) | \bar{u} \sigma_{\mu \nu} \gamma_5 d | \pi^+  \rangle = 
- \frac{e}{2}  \, f_T  \, \left(p_\mu \epsilon_\nu - p_\nu \epsilon_\mu  \right)~, 
\eeq
where $p_\mu$ and $\epsilon_\mu$ are the photon four-momentum and polarization vector, respectively. 
The analysis of Ref.~\cite{Mateu:2007tr}, based on a large-$N_c$-inspired resonance-saturation model 
provides  $f_T = 0.24(4)$ at the renormalization scale $\mu=1$~GeV, 
with parametric uncertainty induced by the uncertainty in the quark condensate. 
The $90\%$-C.L. experimental constraint\footnote{Note that there is a factor of 2 difference in the 
normalization of the tensor coupling $\epsilon_T$ compared to what was used in 
Refs.~\cite{Herczeg:1994ur,Bychkov:2008ws}.} 
$ - 2.0 \times 10^{-4}  \ < \ \epsilon_T  \times f_T \  <  \   2.6  \times 10^{-4} $,  when combined with 
the  above estimate for $f_T$ run to 2~GeV implies
\beq
- 1.1  \times 10^{-3}  \ < \ \epsilon_T \  <  \  1.36  \times 10^{-3} \qquad \qquad (90 \%  \ \rm{C.L.})~. 
\label{eq:tradpi}
\eeq
Again, this is the most stringent constraint on the tensor coupling from low-energy experiments. 
The next best constraints, which we report in the next section,  arise from measurements of nuclear beta decays.

\subsubsection{Bounds on scalar and tensor structures from other nuclear beta decays}

Bounds on scalar and tensor interactions can be obtained from a number of observables 
in nuclear beta decays,  other than $0^+ \to 0^+$ transitions. 
Although these bounds are currently not competitive,   we summarize them here for completeness. 

The leading sensitivity to scalar and tensor operators appears through the Fierz interference term $b$, which in the limit 
of pure Gamow-Teller transitions is proportional to the tensor coupling ($b_{\rm GT}= - (8 \gamma  g_T \epsilon_T)/\lambda$), while in   pure Fermi transitions is proportional to the scalar coupling ($b_{\rm F}  =  2 \gamma  g_S \epsilon_S$).
Significant constraints on  $b$ arise from  electron-polarization observables~\cite{Jackson1957zz} 
as well as in measurements of  $\tilde{A}$  and $\tilde{a}$  
in both Fermi and Gamow-Teller transitions. 
Here is a summary of  current  bounds on $\epsilon_{S,T}$:
\begin{itemize}
\item The most stringent constraint from the beta asymmetry in pure Gamow-Teller transitions 
($\tilde{A}_{\rm GT}$)  arises from  $^{60} {\rm Co}$ measurements and implies~\cite{Wauters:2010gh}
\beq
- 2.9 \times 10^{-3}  \ < \ g_T \, \epsilon_T \  <  \   1.5  \times 10^{-2} \qquad \qquad (90 \%  \ \rm{C.L.})~. 
\eeq
Similar bounds can be obtained from  measurements of $\tilde{A}_{\rm GT}$ in $^{114}{\rm In}$ decay~\cite{Wauters:2009jw}:  $- 2.2 \times 10^{-2}  \ < \ g_T \, \epsilon_T \  <  \  1.3 \times 10^{-2}$ (90 \% C.L.). 

\item Measurements of the ratio $P_{\rm F}/P_{\rm GT}$ of longitudinal polarization in the positron emitted in pure Fermi and Gamow-Teller 
transitions~\cite{Carnoy:1991jd,Wichers:1986es} imply 
\beq
- 0.76  \times 10^{-2}  \ < \    g_S \, \epsilon_S   +    \frac{4}{\lambda}   g_T \, \epsilon_T \  <  \   1.0  \times 10^{-2} \qquad \qquad (90 \%  \ \rm{C.L.})~. 
\eeq

\item    Preliminary results have been reported on the measurement of the longitudinal polarization of positrons emitted by 
polarized $^{107} {\rm In}$ nuclei~\cite{Severijns:2000vc}.  The corresponding  90 \% C.L. sensitivity to tensor interactions,  
 $|   g_T \, \epsilon_T |   <    3.1 \times 10^{-3}$, is quite promising although not yet competitive with the radiative pion decay. 

\item Finally,  the beta-neutrino correlation $a$ has been measured in a number of nuclear  
transitions~\cite{Vetter:2008zz,Gorelov:2004hv,Adelberger:1999ud,Johnson:1963zza}.  
The resulting constraints on scalar and tensor interactions are nicely summarized in 
Fig.~7 of Ref.~\cite{Vetter:2008zz}. 
In terms of the coupling constants used here,  the 90 \% C.L. combined  bound on the tensor interaction  
reads  $| g_T \, \epsilon_T | <  5  \times 10^{-3}$, again not competitive with the radiative pion decay. 

\end{itemize}

We observe that in order to improve on the existing bound on $\epsilon_T$ from $\pi \to e \nu \gamma$, 
future measurements sensitive to $b_{\rm GT}$  should aim at sensitivities of 
$\delta b_{\rm GT} \lsim   6.3 \times g_T \times 10^{-3}$.    
For example, a  $10^{-3}$ measurement of $b_{\rm GT}$ would probe $g_T \epsilon_T$ at the 
$2 \times 10^{-4}$-level, providing a very competitive bound.

\subsubsection{Constraints on \texorpdfstring{$\epsilon_{S,P,T}$}{\textepsilon\string_S,P,T} from 
               \texorpdfstring{$\pi \to e \nu$}{\textpi \textrightarrow e\textnu}}
\label{sect:pienu}

The ratio $R_\pi \equiv \Gamma ( \pi \to e \nu [\gamma]) /
\Gamma ( \pi \to \mu \nu [\gamma])$ 
probes more than just the effective low-energy pseudoscalar coupling
$\epsilon_P$ defined earlier as the coefficient of the operator
$\bar{e}(1 - \gamma_5) \nu_e \cdot \bar{u} \gamma_5 d$.  In fact,
since (i) $R_\pi$ is defined as the ratio of electron-to-muon decay
and (ii) the neutrino flavor in both the decays is not observed, this
observable is sensitive to the whole set of parameters
$\epsilon_P^{\alpha \beta}$ defined by
\beq
{\cal L}_{\rm eff} \quad  \supset    \quad
\frac{G_F}{\sqrt{2}} V_{ud} \ \epsilon_P^{\alpha \beta} \  \ 
\bar{e}_\alpha (1 - \gamma_5) \nu_\beta \cdot \bar{u} \gamma_5 d~, 
\eeq
where $\alpha \in \{e, \mu \}$ refers to the flavor of the charged lepton and  $\beta \in \{ e, \mu, \tau \}$ 
refers to the neutrino flavor. 
One generically expects SM extensions to generate non-diagonal components in $\epsilon_{P,S,T}^{\alpha \beta}$, 
In the new notation the previously defined pseudoscalar, scalar, and tensor  couplings 
reads  $\epsilon_{P,S,T}  \equiv \epsilon_{P,S,T}^{ee}$. 
It is important to note here that only $\epsilon_P^{ee}$ and 
$\epsilon_P^{\mu \mu}$ can interfere with the SM amplitudes, while the remaining 
$\epsilon_P^{\alpha \beta}$ contribute incoherently to both the numerator and denominator in $R_\pi$.\footnote{
While in our setup the incoherent contribution arises from ``wrong-flavor" neutrinos,  
in general it could have a different nature.  For example, 
the incoherent  contribution to $R_\pi$  discussed in Refs.~\cite{Herczeg:1994ur,Herczeg2001vk}
is due to a right-handed light neutrino.}
In summary, allowing for non-standard interactions and factoring out the SM prediction for $R_\pi$, 
one can write\footnote{Here we are neglecting the overall effect of  $v_{L/R}$,  not enhanced by helicity arguments.}: 
\beq
\label{eq:Rpiconstraint1}
\frac{R_\pi}{R_\pi^{\rm SM}} = 
\frac{\left[ \left(1  - \frac{B_0}{m_e} \epsilon_P^{ee} \right)^2   
+  \left(\frac{B_0}{m_e} {\epsilon}_P^{e\mu} \right)^2 
+  \left(\frac{B_0}{m_e}{\epsilon}_P^{e\tau} \right)^2 
 \right]}{
\left[ \left(1  - \frac{B_0}{m_\mu} \epsilon_P^{\mu \mu} \right)^2   
+  \left(\frac{B_0}{m_\mu}{\epsilon}_P^{\mu e} \right)^2  
+  \left(\frac{B_0}{m_\mu}{\epsilon}_P^{\mu \tau} \right)^2  
\right]}~.
\eeq
In the above equation the factors of $B_0/m_{e, \mu} \epsilon_P$ represent the ratio 
of new-physics amplitude over SM amplitude.  The latter is proportional to 
the charged-lepton mass due to angular-momentum conservation arguments, while the former 
is proportional to  $\langle 0| \bar{u} \gamma_5 d| \pi\rangle$, characterized by
the  scale- and scheme-dependent parameter\footnote{Note that the scale and scheme dependence  
of $B_0(\mu)$ is compensated in physical quantities  by the scale  and scheme dependence of the Wilson coefficients 
$\epsilon_P^{\alpha \beta}$.} 
\beq
B_0 (\mu)   \equiv  \frac{M_\pi^2}{m_u (\mu) + m_d(\mu)}~.
\eeq
Since $B_0^{\overline{\rm MS}} (\mu = 1 \  {\rm GeV}) = 1.85 \, {\rm GeV}$ and consequently 
$B_0/m_e =  3.6 \times 10^3$,  $R_\pi$ has enhanced sensitivity to $\epsilon_P^{\alpha \beta}$, and 
one needs to keep quadratic terms in these new physics coefficients.\footnote{
This feature is specific to purely leptonic decays of pseudoscalar mesons. 
In beta decays one never encounters relative enhancement factors such as $B/m_e$,  because  
$\epsilon_P$ is always multiplied by nucleon velocity factors and the SM amplitude 
does not suffer anomalous suppression (as the helicity argument 
implies in the case of $\pi \to e \nu$).}

\begin{figure}[t!]
\centering
\includegraphics[width=0.60\textwidth]{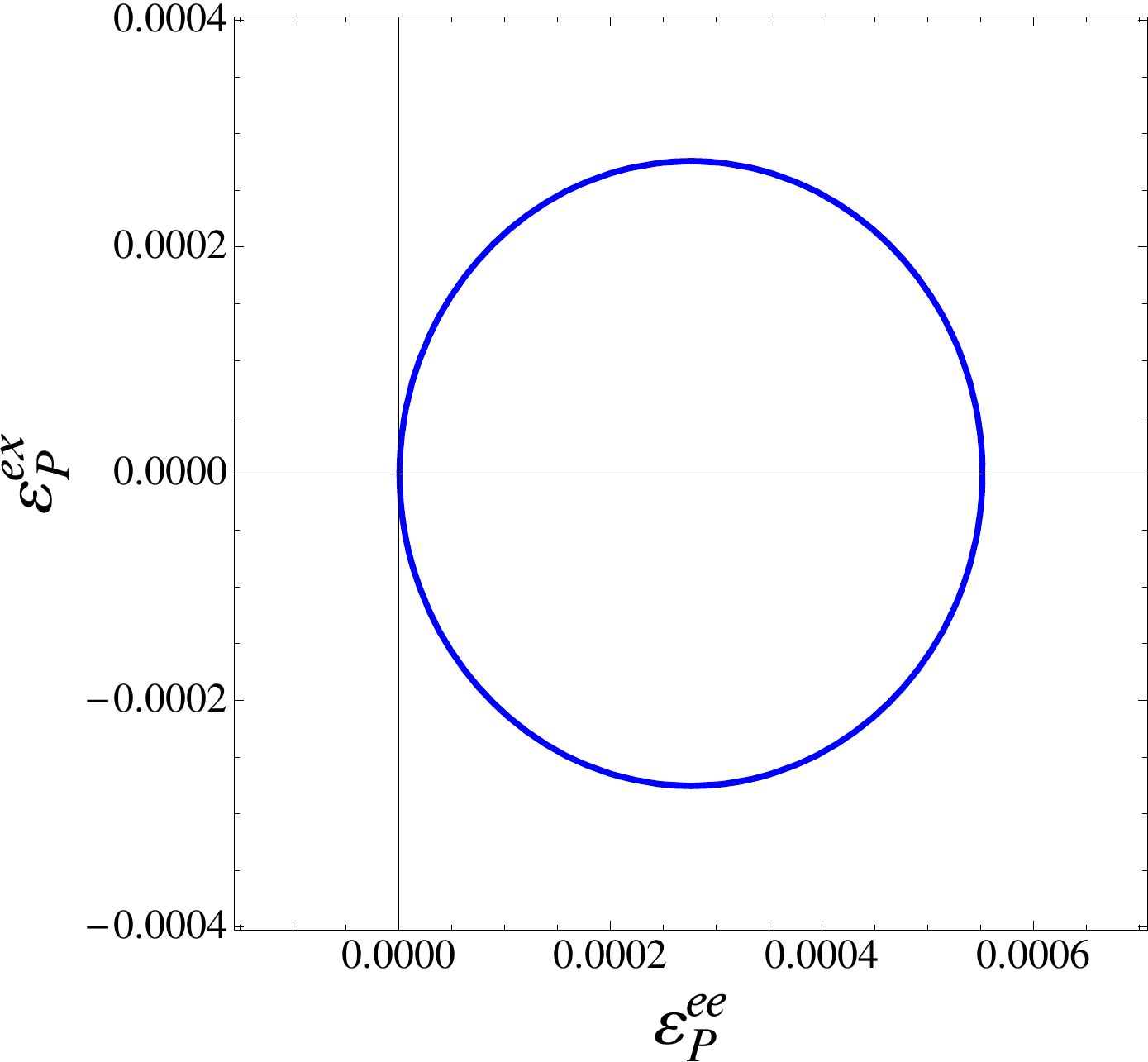}
\caption{
The allowed region in the two-dimensional plane  $\epsilon_P^{ee}$-$\epsilon_P^{ex}$ 
determined by 
$R_\pi$ is given by  an annulus  of   thickness $1.38 \times 10^{-6}$.   
 In the  absence of information on $\epsilon_P^{ex}$, 
the 90 \% C.L. bound on $\epsilon_P^{ee}$ is 
$- 1.4  \times 10^{-7}  <  \epsilon_P^{ee}  <  5.5 \times 10^{-4}$.}
\label{fig:Remu}
\end{figure}

Inspection of  Eq.~(\ref{eq:Rpiconstraint1}) reveals that
 if the new-physics couplings respect 
 $\epsilon_P^{e \alpha}/m_e =   \epsilon_P^{\mu  \alpha}/m_\mu$, 
 then $R_\pi/R_\pi^{\rm SM} =1$, and there are no constraints on these couplings. 
On the other hand, if the effective  couplings $\epsilon_P^{\alpha \beta}$ are all of similar size, 
one can neglect the entire denominator in Eq.~(\ref{eq:Rpiconstraint1}), 
a it is suppressed with respect to the numerator by powers of $m_e/m_\mu$. 
We will assume to be in this second scenario. 
In this case the constraint in 
Eq.~(\ref{eq:Rpiconstraint1}) forces the couplings 
$\epsilon_P^{ee},\epsilon_P^{e\mu},\epsilon_P^{e \tau}$ to live in a spherical shell 
of radius   $m_e/B_0 \sqrt{R_\pi^{\rm exp}/R_\pi^{\rm SM}} \approx  2.75 \times 10^{-4}$  centered at 
$\epsilon_P^{ee} = m_e/B_0 \approx 2.75 \times 10^{-4}$, $\epsilon_P^{e\mu}= \epsilon_P^{e \tau}=0$.
The thickness of the shell is numerically $1.38 \times 10^{-6}$ 
and  is determined by the current combined uncertainty 
in  $R_\pi^{\rm exp}$ ~\cite{Britton:1992pg,Czapek:1993kc}
and $R_\pi^{\rm SM}$~\cite{Cirigliano:2007ga,Cirigliano:2007xi}: 
$R_\pi^{\rm exp}/R_\pi^{\rm SM} = 0.996(5)$ (90\% C.L.).
This is illustrated in Fig.~\ref{fig:Remu}, where we plot the allowed region 
in the two-dimensional plane given by $\epsilon_P^{ee}$ and a generic 
``wrong-flavor" coupling denoted by $\epsilon_P^{ex}$.
Note that the allowed region is given by   the 
thickness of the curve in the figure, 
thus enforcing a strong correlation between 
 $\epsilon_P^{ee}$ and $\epsilon_P^{ex}$. 
Since $\epsilon_P^{\alpha \neq \beta}$ are essentially unconstrained by other measurements  
and can be of order $10^{-3}$,  we can marginalize over either one of the couplings to obtain 
a bound on the other.  The resulting  90\%-C.L. bounds  are
\beq
\label{eq:epsPconstraint}
- 1.4  \times 10^{-7}  <  \epsilon_P^{ee}  <  5.5 \times 10^{-4},  \quad  {\rm or} \quad 
- 2.75  \times 10^{-4}  <  \epsilon_P^{e \alpha}  <  2.75 \times 10^{-4} \ \ (\alpha \neq e)~, 
\eeq
in qualitative agreement with the findings of  Refs.~\cite{Herczeg:1994ur,Herczeg2001vk}.

As originally discussed in Refs.~\cite{Voloshin:1992sn,Herczeg:1994ur, Campbell:2003ir},  
the pseudoscalar coupling $\epsilon_P^{ee}$ can be radiatively generated 
starting from nonzero $\epsilon_{S,T}$. Hence, the stringent constraint in Eq.~(\ref{eq:epsPconstraint}) 
puts constraints on the same $\epsilon_{S,T}$ that can be probed in beta decays. 
The physics of this effect is very simple: once the scalar, pseudoscalar, and tensor operators are generated 
by some non-standard physics at the matching scale $\Lambda$,  electroweak radiative corrections 
induce mixing among these three operators. So even if one engineers a small pseudoscalar 
contribution $\epsilon_P (\Lambda)$ at the matching scale,  known SM physics 
generates a nonzero $\epsilon_P (\mu)$ at some lower energy  scale $\mu$  via loop diagrams. 
The general form of the constraint can be worked out by using the 
three-operator mixing results from Ref.~\cite{Campbell:2003ir}\footnote{The authors of Ref.~\cite{Campbell:2003ir} 
focused only on the phenomenology of scalar-to-pseudoscalar mixing.}. 
The leading-order result is
\begin{subequations} 
\bea
\epsilon_P^{\alpha \beta} (\mu) &=& \epsilon_P^{\alpha \beta}  (\Lambda) \left(1 + \gamma_{PP} \,  \log \frac{\Lambda}{\mu} \right) 
+ \epsilon_S^{\alpha \beta}  (\Lambda)  \ \gamma_{SP} \,  \log \frac{\Lambda}{\mu} 
+ \epsilon_T^{\alpha \beta}  (\Lambda)\  \gamma_{TP} \,  \log \frac{\Lambda}{\mu} 
\\
\gamma_{PP} &=& \frac{3}{4} \frac{\alpha_2}{\pi}  + \frac{113}{72} \frac{\alpha_1}{\pi}  \approx 1.3 \times 10^{-2}
\\
\gamma_{SP} &=&  \frac{15}{72} \frac{\alpha_1}{\pi}  \approx 6.7 \times 10^{-4}
\\
\gamma_{TP} &=& -  \frac{9}{2} \frac{\alpha_2}{\pi}  -  \frac{15}{2} \frac{\alpha_1}{\pi} \approx - 7.3 \times 10^{-2}~,
\eea
\end{subequations}
where $\alpha_1 = \alpha/\cos^2\theta_W$ and $\alpha_2 = \alpha/\sin^2\theta_W$ are the $U(1)$ and $SU(2)$ 
weak couplings, expressed in terms of the fine-structure constant and the weak mixing angle.  
Setting $\epsilon_P^{ee} (\Lambda) = 0$  and neglecting the small $O(\alpha/\pi)$ fractional difference between
$\epsilon_{S,T} (\Lambda)$ and the observable 
$\epsilon_{S,T} (\mu)$ at the low scale, 
the 90\% C.L. constraint on the $\epsilon_S$-$\epsilon_T$ plane reads 
\beq
\frac{- 1.4 \times 10^{-7} }{\log (\Lambda/\mu)}
 \ <  \  \gamma_{SP}  \ \epsilon_S  \ + \   \gamma_{TP}  \ \epsilon_T   \ <  \ \frac{5.5 \times 10^{-4}}{\log (\Lambda/\mu)}~.
\eeq
Even assuming $\log (\Lambda/\mu) \sim 10$ (e.g. $\Lambda \sim 10\, {\rm TeV}$ and $\mu \sim  1 \, {\rm GeV}$), 
using the numerical values of $\gamma_{SP,TP}$, 
one can verify that the individual constraints are at the  level of 
$| \epsilon_S |  \lsim  8 \times 10^{-2}$ and $| \epsilon_T |  \lsim 10^{-3}$, 
implying that this constraint on $\epsilon_T$ is roughly equivalent to the one arising from 
$\pi \to e \nu \gamma$. 
Of course, these bounds become logarithmically more stringent as the new-physics scale $\Lambda$ grows.

\begin{figure}[!ht]
\centering
\includegraphics[width=0.45\textwidth]{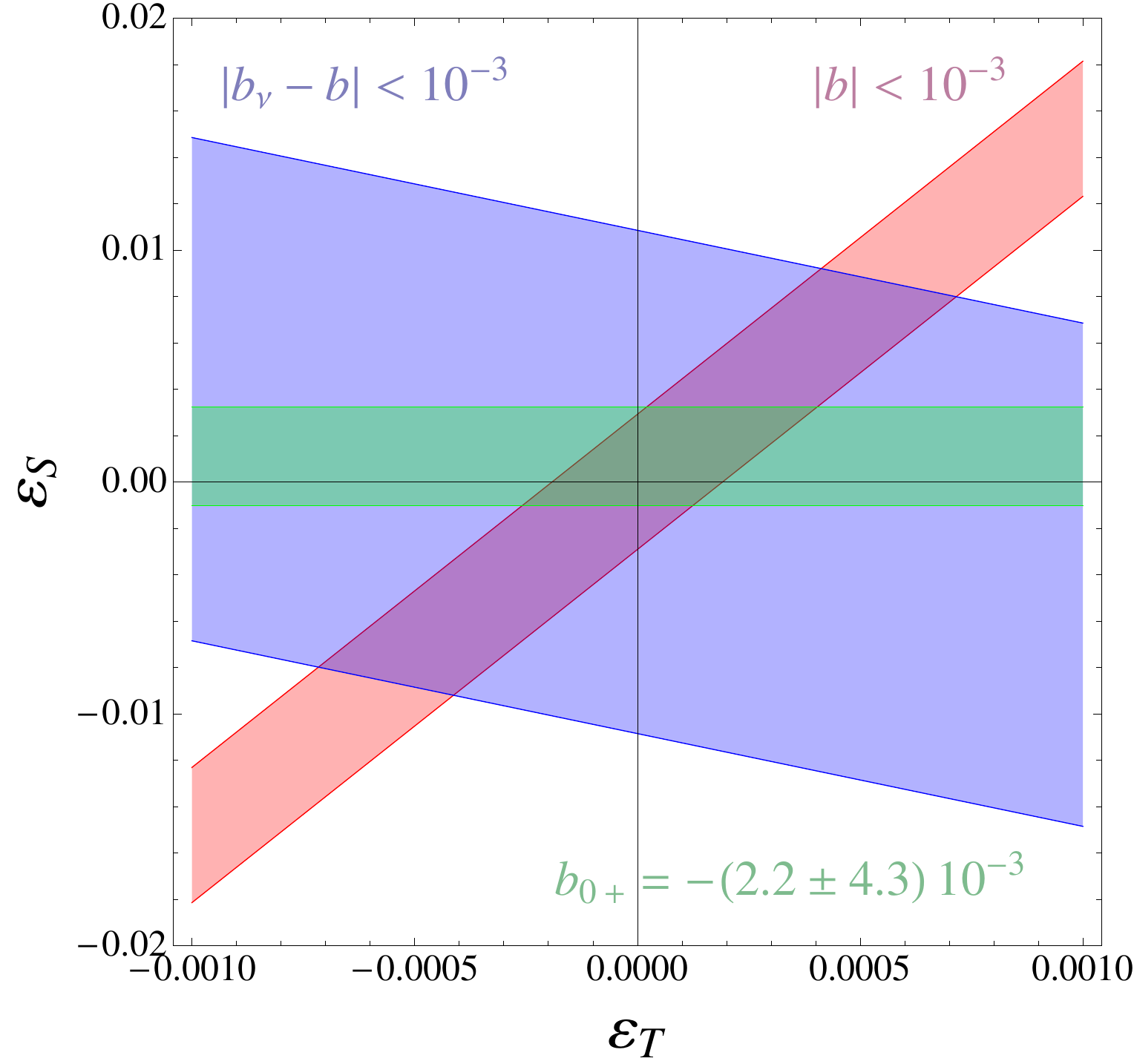}
\hspace{0.2in}
\includegraphics[width=0.45\textwidth]{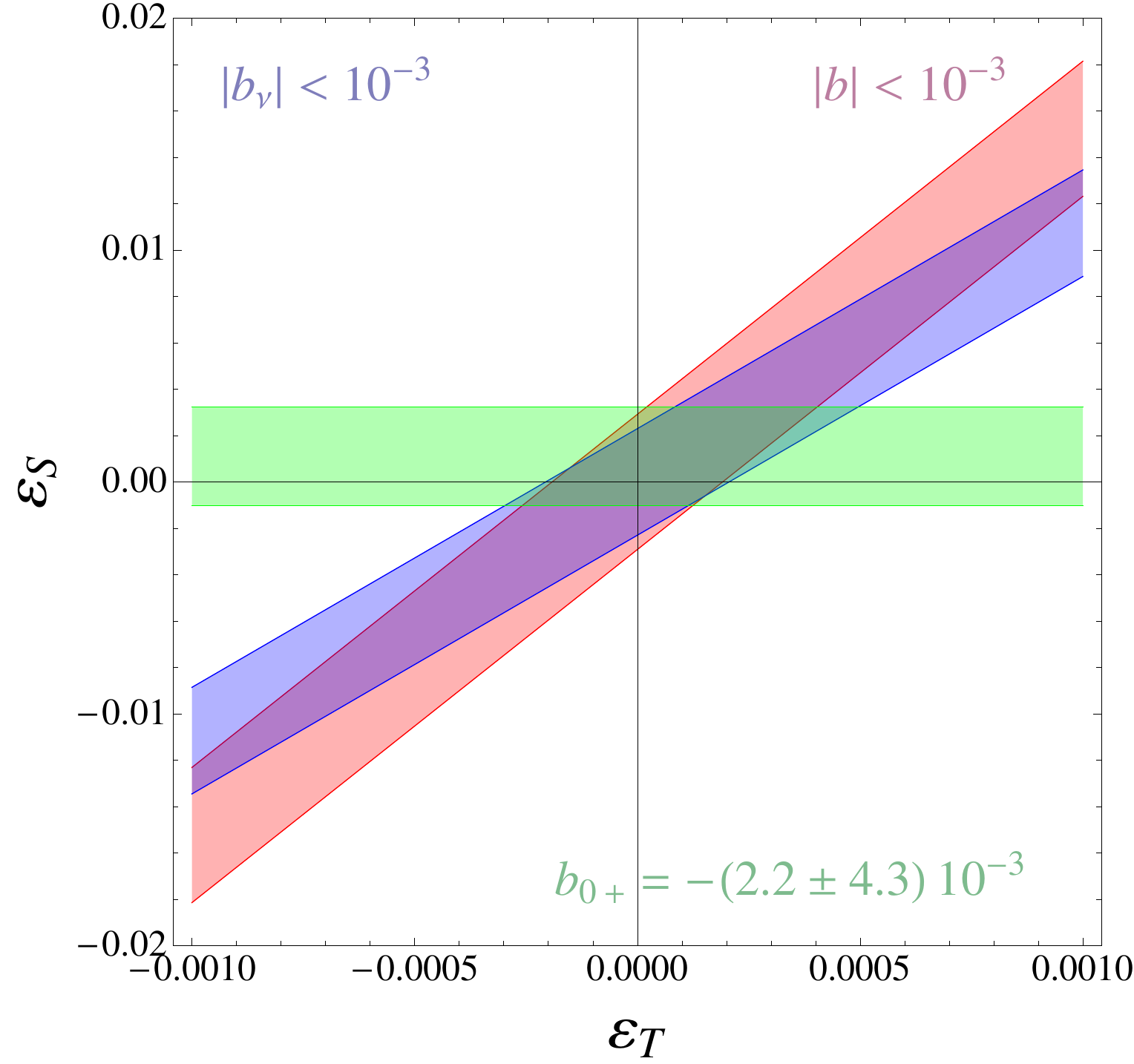}
\caption{
$90\%$ C.L. allowed regions  in 
the  $\epsilon_S$-$\epsilon_T$ plane
implied by (i) the existing bound on $b_{0^+}$  (green horizontal band);  
(ii)  projected $10^{-3}$-level  limits  on  $b$  (red band),  
 $b_\nu-b$ (blue band, left panel), and 
 $b_\nu$ (blue band, right panel). 
The hadronic form factors are taken to be $g_S = g_T = 1$ in the ideal scenario of no uncertainty. 
The impact of hadronic uncertainties is discussed in Section~\ref{sect:latticepheno}. 
} 
\label{fig:constraints1}
\end{figure}

\begin{figure}[!ht]
\centering
\includegraphics[width=0.45\textwidth]{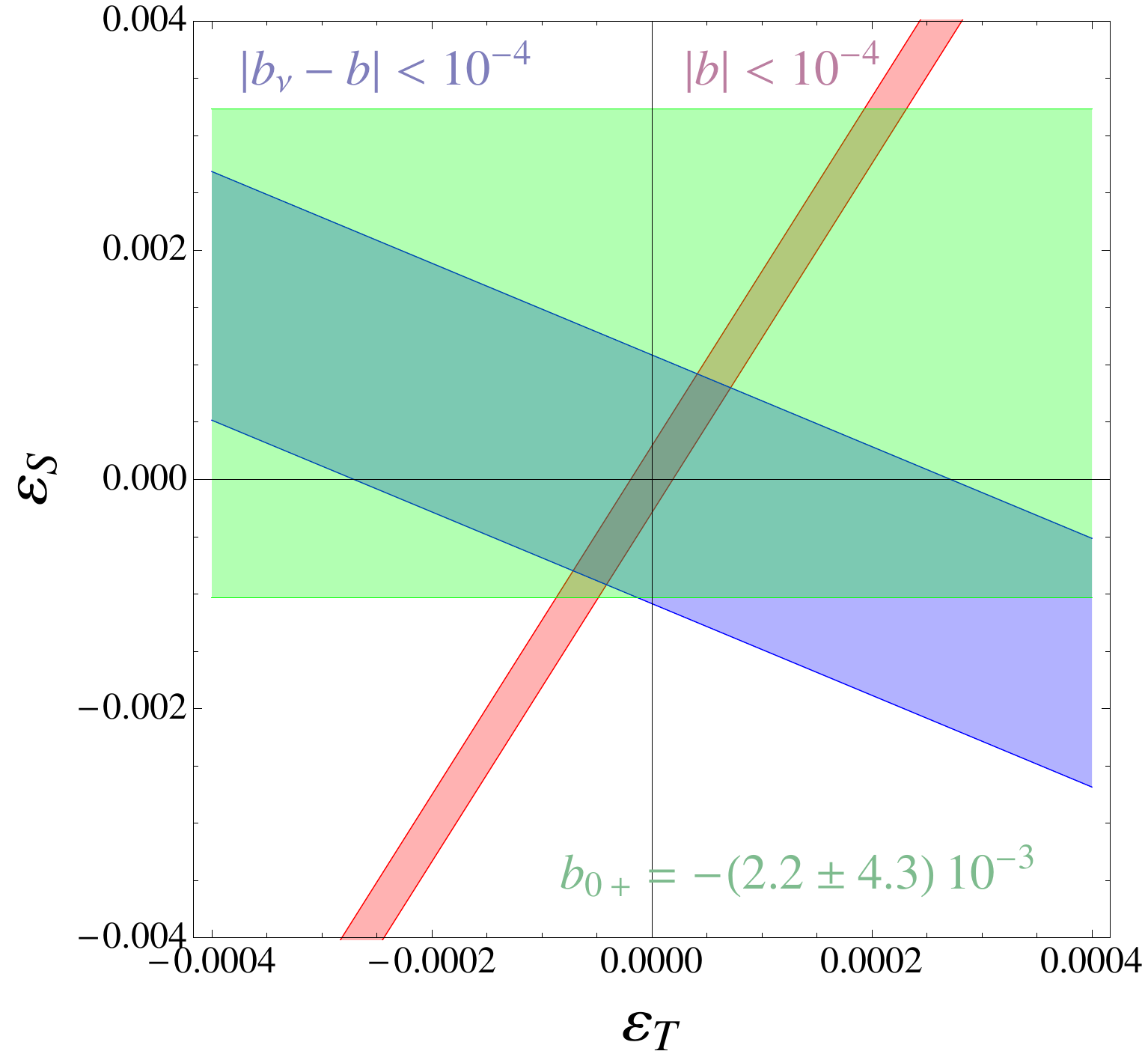}
\hspace{0.2in}
\includegraphics[width=0.45\textwidth]{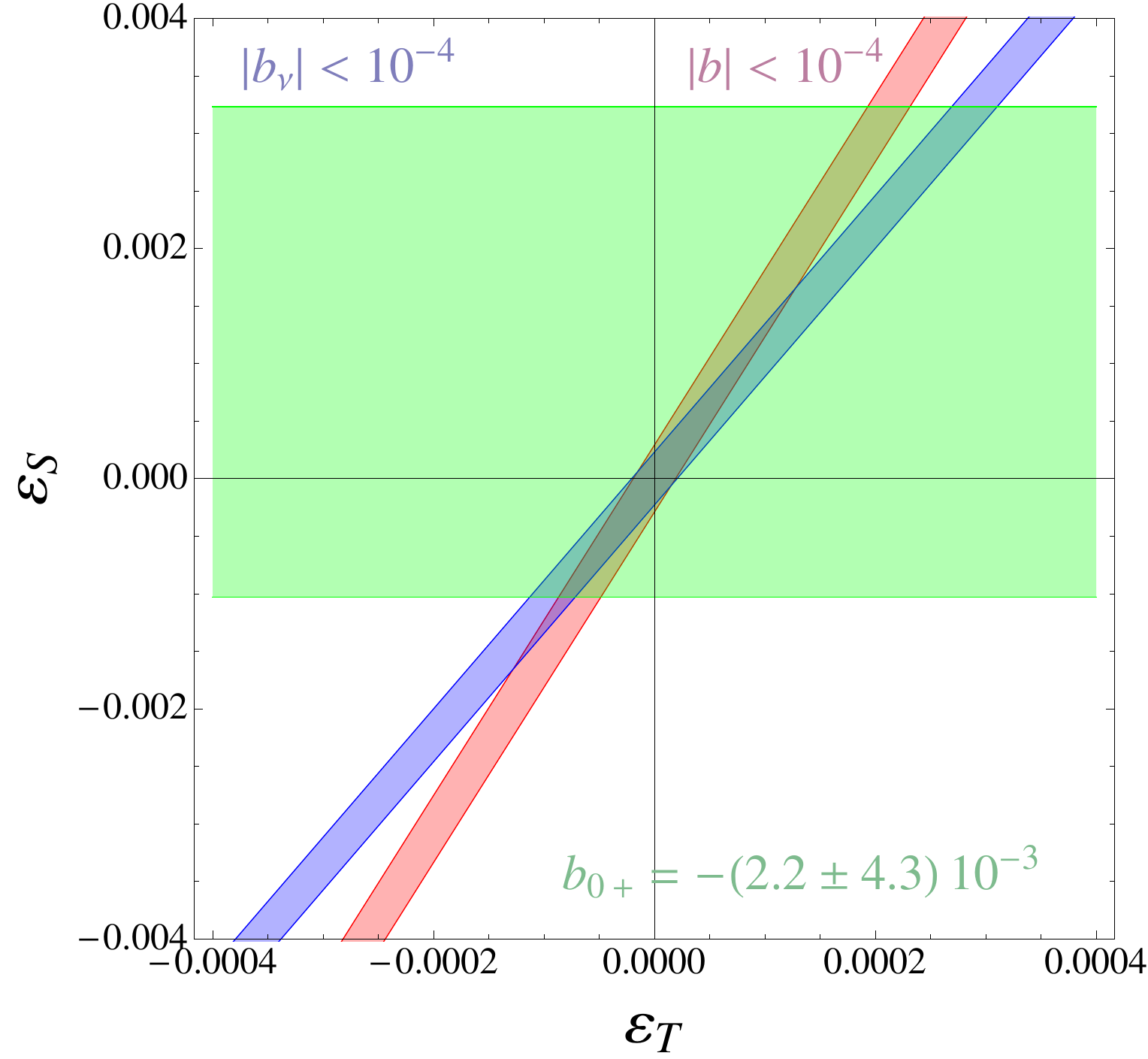}
\caption{
$90\%$ C.L. allowed regions  in 
the  $\epsilon_S$-$\epsilon_T$ plane
implied by (i) the existing bound on $b_{0^+}$  (green horizontal band);  
(ii)  projected $10^{-4}$-level  limits  on  $b$  (red band),  
 $b_\nu-b$ (blue band, left panel), and 
 $b_\nu$ (blue band, right panel). 
The hadronic form factors are taken to be $g_S = g_T = 1$ in the ideal scenario of no uncertainty. 
The impact of hadronic uncertainties is discussed in Section~\ref{sect:latticepheno}. 
} 
\label{fig:constraints1v2}
\end{figure}

\subsection{The impact of future \texorpdfstring{$b$}{b} and \texorpdfstring{$B$}{B}  neutron measurements}
\label{sect:impact1}

The discussion in the preceding subsection  has shown that currently the most stringent low-energy  constraints on novel scalar  
and tensor  interactions arise, respectively, from the Fierz interference term in $0^+ \to 0^+$ nuclear beta decays   
(Eq.~(\ref{eq:bFbound})) and from the radiative pion decay $\pi \to e \nu \gamma$  (Eq.~(\ref{eq:tradpi})).    
It is important to realize that 
the  allowed $\epsilon_S$ interval   
derived from Eq.~(\ref{eq:bFbound}) 
depends on the nucleon form factor $g_S$ 
(as do  all the constraints arising from neutron and nuclear beta decays).
For a given experimental accuracy, 
the constraint  on the short-distance couplings $\epsilon_{S,T}$  becomes stronger as  $\delta g_{S,T}/g_{S,T} \to 0$.   
In this section,  we will first explore the maximal constraining power of nuclear and neutron 
measurements in the ideal scenario of  {\it no uncertainty} on $g_{S,T}$, and for 
illustrative purposes we assume the central values  $g_S = g_T = 1$.  
We will quantify  the implications of finite uncertainties on $g_{S,T}$ on the  $\epsilon_{S,T}$ constraints  in 
Section~\ref{sect:latticepheno}.

With the above assumptions on $g_{S,T}$,  the currently  allowed region (at 90\% C.L.)  on the 
$\epsilon_S$-$\epsilon_T$ plane is given by the green horizontal band 
in Figs.~\ref{fig:constraints1} and \ref{fig:constraints1v2}.  
The vertical ($\epsilon_S$)   boundaries of this region are determined by the constraint from $b_{0^+}$, 
while  essentially the entire horizontal ($\epsilon_T$)  range  on the scale of these plots is allowed 
by the $\pi \to e \nu \gamma$ limit (see Eq.~(\ref{eq:tradpi})). 

In this  ideal scenario of no uncertainty on $g_{S,T}$,  we can quantify the impact 
of future neutron measurements by plotting the $90\%$ C.L. allowed region in 
the  $\epsilon_S$-$\epsilon_T$ plane  implied by 
projected limits on $b$,  $b_\nu-b$, and $b_\nu$. 
The neutron constraints are derived using Eqs.~(\ref{eq:bbsm}) and in generating the plots we use 
the central value  $\lambda = 1.269$.
In Fig.~\ref{fig:constraints1} we focus on the case in which the experimental 
sensitivity on $b$,  $b_\nu-b$, and $b_\nu$   is at  the $10^{-3}$ level. 
In the left panel we show the constraints from the existing $b_{0^+}$ limit 
(green horizontal band) and  $10^{-3}$-level limits on $b$ and $b_\nu - b$, 
(red and blue bands, respectively).
In the  right panel we replace  the $10^{-3}$-level limit on $b_\nu - b$ 
with the $10^{-3}$ limit on $b_\nu$, which in principle can be isolated experimentally~\cite{ayoung}. 
In Fig.~\ref{fig:constraints1v2} we plot the constraints resulting from 
projected limits on $b$,  $b_\nu-b$, and $b_\nu$ at the $10^{-4}$ level. 
The intersection of the various bands in Figs.~\ref{fig:constraints1}  and \ref{fig:constraints1v2} denotes 
the combined allowed region in the $\epsilon_S$-$\epsilon_T$ plane that would result 
after future neutron measurements. 
Two important remarks are in order here:
\begin{itemize}
\item   For a given experimental sensitivity, the combination $b_\nu - b$ gives 
weaker constraints on $\epsilon_{S,T}$ than $b$ or $b_\nu$. 
This is easily understood:  by taking the difference of  Eqs.~(\ref{eq:bbsm}) 
one sees that $b_\nu - b \propto  \lambda -1$, which for $\lambda \approx 1.27$ provides 
a suppression factor. 

\item  There is an almost exact  ``degeneracy"  in the  constraints from $b$ and $b_\nu$, 
again controlled by the form of  Eqs.~(\ref{eq:bbsm})  and the numerical value of $\lambda$. 
For the purposes of constraining $\epsilon_{S,T}$, an upper  limit  on $b$ is essentially equivalent to 
an upper  limit on $b_\nu$. This provides strong motivation to pursue experimental determinations 
of both $b_\nu - b$ and $b_\nu$ via neutrino asymmetry ($B$) measurements.
From the theoretical point of view, we can use either $b$ or $b_\nu$, and in subsequent sections 
we will use $b$ for illustrative purposes. 

\end{itemize}

Fig.~\ref{fig:constraints1} clearly illustrates that 
with experimental sensitivity in neutron decay at the $10^{-3}$ level,
the most stringent constraint arises from a combination of $b_{0^+}$ and $b$  or 
$b_{0^+}$   and  $b_\nu$. 
The complementarity of these  measurements would lead to 
a significant  (four-fold) improvement in the bound on $\epsilon_T$, compared  to Eq.~(\ref{eq:tradpi}).    
The impact of $10^{-4}$ measurements of $b$, $b_\nu$,  and  $b_\nu-b$ in neutron decay 
is even more dramatic (Fig.~\ref{fig:constraints1v2}), as in that case the constraint 
from $b_{0^+}$ would   become irrelevant and the combination of $b$ and 
$b_\nu - b$  or   $b$ and $b_\nu$ would imply an improvement of one order of magnitude in 
the bound on $\epsilon_T$ and a factor of two in $\epsilon_S$. 

In Section~\ref{sect:latticepheno} we will revisit the impact of proposed neutron measurements 
on $\epsilon_{S,T}$ in light of nonzero uncertainties in the hadronic matrix elements $g_{S,T}$.

    \section{Lattice calculation of matrix elements\label{sec:3}}
       

To connect the measurements of $b$ and $b_\nu$ in neutron decays to
new physics at the TeV scale requires precision measurements of the
matrix elements of isovector bilinear quark operators between an
initial neutron and final proton state, in particular of the scalar
and tensor operators. Lattice QCD is a path-integral
formulation of QCD on a discrete, four-dimensional Euclidean
spacetime, and numerical simulations of it provide the best
nonperturbative method for evaluating these matrix elements.  It
has been successfully employed to calculate hadron masses and their
decay properties, such as matrix elements, with control over
statistical and all systematic errors, in many cases at higher
precision than can be measured
experimentally~\cite{Laiho:2009eu,Sachrajda:2011tg}.

To obtain continuum results, estimates from LQCD obtained at a number
of values of lattice spacing $a$ and spacetime volume $L^3 \times T$
are extrapolated to $a \rightarrow 0$ and $L \rightarrow \infty$ to
eliminate the artifacts introduced by formulating QCD in a finite
discretized box. Another source of systematic uncertainty is
introduced when estimates obtained at multiple values of $u$ and $d$
quark masses heavier than in nature are extrapolated to the physical
point. One typically uses chiral perturbation theory to carry out this
extrapolation, with low-energy constants determined by
over-constraining the fits using experimental and lattice
data~\cite{Colangelo:2010et}. Current state-of-the-art simulations are
beginning to provide results at physical light-quark masses obviating
the need for a chiral extrapolation. Recent calculations by the BMW
collaboration~\cite{Durr:2008zz,Durr:2011ap} at multiple lattice
spacings, volumes and pion masses as light as 120~MeV provide an
excellent demonstration of how hadronic properties can be extracted
with fully understood and controlled systematics.

In this section we review
current LQCD calculations of the nucleon isovector matrix elements
in order to highlight what needs to be done to obtain the precision
required to probe new physics at the TeV scale in neutron-decay experiments. We also 
present our current best estimates of $g_S$ and $g_T$, which are used in 
the phenomenological analysis presented in Sec.~\ref{sect:latticepheno}.

\subsection{Lattice methodology}
\label{ssec:LQCDmethod}

A lattice calculation proceeds in two steps: First, a Monte-Carlo
sampling of the QCD vacuum, called an ``ensemble of gauge-field
configurations'', is generated using an appropriate discretization of
the gauge and fermion actions. The particular choices of the actions
have important implications for the computational cost of the
calculation, for the size of the discretization errors and for which
symmetries are violated at finite lattice spacing. We will review the existing
calculations, summarized in Table~\ref{tab:LQCDsummary}, with two
light flavors (2-flavor) and two light and one strange flavor
(2+1-flavor) as these are close approximations to the real world.

The second step is to calculate expectation values on these ensembles
of gauge configurations and from these extract estimates of the
desired observables. For hadronic observables, the fermion action used
at this stage may differ from the one used in producing the gauge
configurations, in which case it is called a ``mixed-action''
calculation. Further details on the domain-wall fermion (DWF)
formulation are given in
Refs.~\cite{Kaplan:1992bt,Kaplan:1992sg,Shamir:1993zy,Furman:1994ky,Neuberger:1997fp};
clover fermions in Ref.~\cite{Sheikholeslami:1985ij}; twisted-mass
fermions in Ref.~\cite{Frezzotti:2000nk}; and improved staggered
fermions in Refs.~\cite{Naik:1986bn,Orginos:1998ue,Follana:2006rc}.

Calculation of the isovector nuclear matrix elements requires two
separate optimizations in addition to the choice of the actions. The
first is to tune the size of smearing applied to the local 
interpolating operator with the correct quantum
numbers of the nucleon
\begin{equation}
\label{eq:lat_B-op}
 \chi^N (x) =  \epsilon_{abc} [{\psi}_1^{a\mathrm{T}}(x)C\gamma_5\psi_2^b(x)]\psi_1^c(x),
\end{equation}
where $a,\ b, \ c$ are color indices, $C$ is the charge-conjugation
matrix, and $\psi_1$ and $\psi_2$ are $u$ or $d$ quarks; for example,
to create a proton, we want $\psi_1=u$ and $\psi_2=d$. This local
operator, unfortunately, couples to the nucleon and all its excited
states with the same quantum numbers. To improve the overlap with the
desired ground state, the quark fields in this operator may be
``smeared'' around the point $x$. The goal of this smearing is to
approximate the ground-state nucleon wavefunction. We adopt the
commonly used application of the three-dimensional gauge-invariant
Laplacian to smear around the source point $x$ and tune the smearing
size to improve the overlap with the ground-state nucleon in the two-
and three-point correlation functions. The two-point function,
projected to a definite momentum at either the source or sink time by
making a three-dimensional Fourier transformation, is given by
\begin{eqnarray}
\Gamma^{(2)}_{AB}(t;\mathbf{p})
  &=& \left\langle \chi^N_{A}(t, \mathbf{p})(\chi^N_B)^\dagger(\mathbf{p})\right\rangle  \nonumber\\
  &=& \sum_n \left\langle 0|\chi^N_A(t, \mathbf{p})| n\right\rangle
             \left\langle n|(\chi^N_B)^\dagger(0)|0\right\rangle
            \frac{1}{2E_n(\mathbf{p})} e^{-E_n(\mathbf{p})t}, 
\label{eq:two-pt}
\end{eqnarray}
where the indices $A$ and $B$ indicate the choice of operator
smearing. The nucleon states are normalized as $\langle 0
|(\chi^N_A)^\dagger|p,s\rangle = X_A u_s(\mathbf{p})$ where $X_A$ is the
overlap of the operator with the state, and the spinors satisfy
$\sum_{s} u_s(\mathbf{p})\overline u_s(\mathbf{p}) = E(\mathbf{p})\gamma^t
-i \boldsymbol\gamma \cdot \mathbf{p} + m$.  In the limit of large time
separation $t$, the correlator is dominated by the ground-state
nucleon, and the above form simplifies to
\begin{equation}\label{eq:naive_2pt}
\Gamma^{(2)}_{AB}(t;\mathbf{p}) = \frac{E(\mathbf{p})+M_n}{2E(\mathbf{p})} X_A(\mathbf{p}) X_B(\mathbf{p}) e^{-E(\mathbf{p})t} .
\end{equation}

To calculate the nucleon matrix elements, we also need to construct
nucleon three-point functions with insertion operators $O_\Gamma(x)
\equiv Z_\Gamma O_\Gamma^b = Z_\Gamma \overline{u}(x)\Gamma d(x)$, 
where $O^b$ is the bare operator, $\Gamma$ represents one of the
sixteen Dirac matrices and $Z_\Gamma$ is the associated
renormalization constant of the operator. The three-point functions
take the form
\begin{equation}
\Gamma^{(3)}_{AB} (t_i,t,t_f;\mathbf{p}_i,\mathbf{p}_f)
  = Z_\Gamma \left\langle \chi^N_B(t_f, \mathbf{p}_f)
  O_\Gamma^b(t) \overline{\chi^N_A}(t_i, \mathbf{p}_i) \right\rangle.
\end{equation}
By inserting a complete set of states $\{n,n^\prime\}$ between the operators, 
this three-point function can be written as 
\begin{eqnarray}
\label{eq:3-point}
\Gamma^{(3),T}_{AB}(t_i,t,t_f,\mathbf{p}_i,\mathbf{p}_f)  &=&
 a^3 Z_\Gamma \sum_n \sum_{n^\prime} \frac{ X_{n^\prime,B}(p_f) X_{n,A}(p_i)}{4 E_n^\prime(\mathbf{p}_f)E_n(\mathbf{p}_i)}
     e^{-(t_f-t)E_n^\prime(\mathbf{p}_f)}e^{-(t-t_i)E_n(\mathbf{p}_i)} \nonumber \\
  &\times & \sum_{s,s^\prime}  T_{\alpha\beta}
  u_{n^\prime}(\mathbf{p}_f,s^\prime)_\beta
  \left<N_{n^\prime}(\mathbf{p}_f,s^\prime)\left|O_\Gamma^b\right|N_n(\mathbf{p}_i,s)\right>
     \overline{u}_n(\mathbf{p}_i,s)_\alpha, \nonumber \\
\end{eqnarray}
where $T$ is an appropriate projection on the baryon spinors. At 
sufficiently large source-sink separation ($t_{\rm sep} = t_f-t_i$), the signal due
to excited states dies out exponentially, and the sum over states
reduces to just the ground states $n=n^\prime=0$. The operator overlap
factors $X_{A,B}$ and the exponential time dependence can be canceled
out by constructing a ratio of three- and two-point functions, which
for the simple case of $\mathbf{p}_i = \mathbf{p}_f = 0$ is
\begin{eqnarray}
\label{eq:Ratio_Gen}
R_O &=& \frac{
      \Gamma^{(3),T}_{AB}(t_i,t,t_f;\mathbf{p}_i = 0,\mathbf{p}_f=0)}{\Gamma^{(2),T}_{AB}(t_i,t_f;\mathbf{p}=0)} \ .
\end{eqnarray}
In practice, choosing a sufficiently large source-sink separation
$t_{\rm sep}$ to make the excited-state contamination negligible is
challenging because the statistical signal in both the two- and
three-point functions involving nucleons degrades exponentially with $t_{\rm
sep}$. Thus, the second optimization required is over $t_{\rm
sep}$. In ongoing LQCD calculations we are exploring multiple values of $t_{\rm sep}$ and
will explicitly include excited states in our analysis to understand and
reduce this systematic error.

\begin{table}
\scriptsize\setlength{\tabcolsep}{1pt}
\hfuzz=3pt
  \begin{center}
	\begin{tabular}{lccccccc}
\hline
Collaboration & Action & $N_f$ & $M_\pi$ (MeV) & $L$ (fm) & $(M_\pi L)_{\rm min}$ & $a$ (fm) & $g_\Gamma$ Calculated  \\
\hline\hline
QCDSF\cite{Khan:2006de}     & clover & 2 & 595--1000 & 1.0--2.0            & $4.6$ & 0.07--0.116 &  $g_A$ \\
QCDSF\cite{Pleiter:2011gw}  & clover & 2 & 170--270 & 2.1--3.0             & $2.6$ & 0.08--0.116  &  $g_A$, $g_T$ \\
CLS\cite{Brandt:2011sj}     & clover & 2 & 290--575 & 1.7--3.4             & $4.2$ & $\{0.05,0.07,0.08\}$ &   $g_A$ \\
ETMC\cite{Alexandrou:2010hf}& twisted Wilson & 2 & 260--470 & \{2.1, 2.8\} & $3.3$ & $\{0.056,0.070,0.089\}$ &   $g_A$ \\
RBC\cite{Lin:2008uz}        & DWF    & 2 & 490--695 & 1.9 & $4.75$ & 0.117  &  $g_A$, $g_P^*$, $g_T$, $g_V$ \\
\hline
RBC/UKQCD\cite{Yamazaki:2008py,Aoki:2010xg} & DWF & 2+1 & 330--670 & $\{1.8,2.7\}$                          & $3.8$   & 0.114 & $g_A$, $g_T$  \\
LHPC\cite{Edwards:2005ym,Edwards:2006qx,Bratt:2010jn}   & DWF on staggered & 2+1 & 290--870 & $\{2.5,2.7\}$ & $3.68$  & 0.1224 & $g_A$, $g_P^*$, $g_T$  \\
QCDSF\cite{Gockeler:2011ze} & clover & 2+1 &350--480 & 1.87                                                 & $3.37$  & 0.078 & $g_A$  \\
HSC\cite{Lin:2011sa} & anisotropic clover & 2+1 &450--840 & 2.0                                             & $4.57$  & 0.125 ($a_t=0.036$) & $g_A$  \\
\hline
\end{tabular}
\caption{A summary of recent LQCD calculations of $g_A$, $g_P^*$ and $g_T$ by different collaborations 
using two and three flavors of dynamical quarks and $O(a)$-improved
actions. For brevity, we use $g_P^*$ for the induced-pseudoscalar
charge discussed in Sec.~\ref{ssec:inducedps} and $(M_\pi L)_{\rm min}$ for the minimum value of 
$M_\pi L$ used in that set of calculations.
\label{tab:LQCDsummary}
}
\end{center}
\end{table}


\subsection{Issues in extracting the matrix elements}
\label{ssec:LQCDissues}

The matrix elements of most interest to us are those of the scalar and tensor
bilinear operators, $\bar{u} d $ and
$\bar{u}\sigma_{\mu \nu}  d$; however, we are calculating
all five Lorentz structures as the additional cost is
negligible. There is independent interest in high-precision
measurements of $g_A$, and it provides a cross-check of the lattice
systematics. The three-point correlation functions of the vector operator
will be used to construct ratios of matrix elements and
renormalization constants to reduce systematic errors.  In this section we
summarize issues relevant to the LQCD calculations of these matrix elements.

The desired matrix elements of isovector bilinear operators 
$O_\Gamma(x) = Z_\Gamma \overline{u}(x)\Gamma d(x)$
have a number of simplifying features and allow us to make certain approximations:
\begin{itemize}
\item
There are no disconnected Feynman diagrams contributing to the three-point
functions.  These typically arise when quark fields in composite
operators can be contracted between themselves.
\item
There are no lower-dimensional operators with which isovector bilinear operators
mix, so there are no power-law divergences. Only multiplicative
renormalization factors $Z_\Gamma$ need to be calculated.
\item
Current lattice simulations are done with degenerate $u,\ d$ quarks,
at zero momentum transfer, and do not include electromagnetic effects.
The momentum transfer in neutron decay, $q^2 = 1.7 $ MeV${}^2$ is
sufficiently small that the matrix elements can be calculated at $q_\mu=0$.  Also,
the isospin-breaking and electromagnetic contributions are expected
to be smaller than the statistical errors.
\item
Protons and neutrons are both stable asymptotic states of
strong interactions, so there are no other hadronic final states 
that complicate the calculations.
\end{itemize}

The issues that need to be addressed to obtain precision
results are the following:
\begin{itemize}
\item
The signal-to-noise ratio in both two- and three-point correlators
decreases rapidly with the time separation $t_{\rm sep}$ between the
source and the sink in Eq.~\ref{eq:Ratio_Gen}. It is, therefore,
necessary to improve the signal by increasing the overlap of the
operators used as sources/sinks with the nucleon ground state.  As discussed in
Sec.~\ref{ssec:LQCDmethod}, our current approach is to (i) smear the quark
fields in the interpolating operator given in
Eq.~\ref{eq:lat_B-op} and tune the smearing size, and (ii) explicitly include 
excited states in the analysis.
\item
A careful optimization of the Eucledian time interval $t_{\rm sep}$
between the source and sink in the three-point functions has to be
carried out for each lattice spacing $a$ and light-quark mass. On the
one hand, this interval should be as large as possible to isolate the
nucleon ground state on either side of the operator insertion, and on
the other hand the statistical noise limits the time separation.
While there is no $a\ priori$ minimum value of $t_{\rm sep}$ as it
depends on how well the source and sink operators are tuned, in
Sec.~\ref{ssec:LQCDgA} we show that current data suggest that
asymptotic estimates are obtained with $t_{\rm sep} \ge 1.2 $ fm for
the operators used.  Our focus will be on improving the operators and
investigating 2--3 values of $t_{\rm sep}$ to reduce and quantify this
systematic error.
\item
One needs to demonstrate that the lattices are large enough
that finite-size effects are under control, especially for
proposed calculations with pions masses below $350$~MeV.  When the
spatial volume used is too small, finite-volume effects arise due to
the coarseness of the available lattice momenta, squeezing of the
wavefunction due to the interaction of a spatially extended particle
with itself and contamination from partons wrapping around 
the lattice. Previous
studies have shown, as a rule of thumb, that finite-size effects are
smaller than statistical errors for $M_\pi L \gtrsim 4$. The detailed
form of the finite-volume corrections is quantity-dependent.
\item
Very high-statistics measurements, typically on a few thousand
gauge configurations, will be needed to improve the signal
in the two- and three-point correlation functions to overcome the
rapid growth in noise with $t_{\rm sep}$.  Our ongoing calculations
show that the statistics needed will be determined by $g_S$ as it has
the smallest signal-to-noise ratio.
\item
The calculations need to be performed at a sufficient number of values
of the light-quark mass to extrapolate results to the
physical value $m_l = 0.037 m_s$, and at sufficient number of values
of the lattice spacing $a$ to extrapolate to the continuum limit.
\item
The renomalization constants $Z_\Gamma$ depend on the choice of both the
gauge and fermion actions and have to be calculated for each ensemble
of gauge configurations. In past calculations, $Z_A$ typically varied
between 0.75--0.9 for the lattice spacings that have been simulated.
The scale-dependent $Z_S$, $Z_P$, and $Z_T$ (given in the
$\overline{\rm MS}$ scheme at $2$ GeV) show larger variations and
dependence on the lattice action. One-loop tadpole improved
perturbation theory can underestimate corrections to $|1-Z_\Gamma|$ by
50$\%$.  Nonperturbative methods, such as calculating $Z_\Gamma$ in the
RI-MOM scheme~\cite{Martinelli:1994ty,Gockeler:2010yr,Aoki:2010yq},
are preferred as they reduce this uncertainty to a few percent, and
we will use them in our calculations.
\end{itemize}
In the next four sub-sections we summarize the extent to which these
issues are under control in current calculations of each of the matrix
elements in order to highlight what needs to be done to achieve the
desired precision of 10--20$\%$. The analyses of $g_A$, $g_P$ and
$g_T$ are reviews of existing calculations, and the new estimate of
$g_S$ we present is preliminary.

\subsection{Nucleon axial charge \texorpdfstring{$g_A$}{g\string_A}}
\label{ssec:LQCDgA}

The axial charge of the nucleon $g_A \equiv g_A(q^2=0)$, defined in
Eq.~\ref{eq:nucleonmatching}, is a fundamental hadronic observable,
well measured in neutron beta-decay experiments: $g_A\! =\!
1.2695(29) \times g_V$~\cite{Nakamura:2010zzi}, where the vector
charge $g_V = 1$ since $V_{ud}$ has been factored out in the
Lagrangian given in Eq.~(\ref{eq:leffq2}).  Since the axial charge is
experimentally well known, it has long served as a benchmark quantity
for LQCD calculations, particularly for estimating systematic errors
in other nucleon matrix elements that are either poorly measured in
experiments or completely unknown.

\begin{figure}
\begin{center}
\includegraphics[height=.24\textheight]{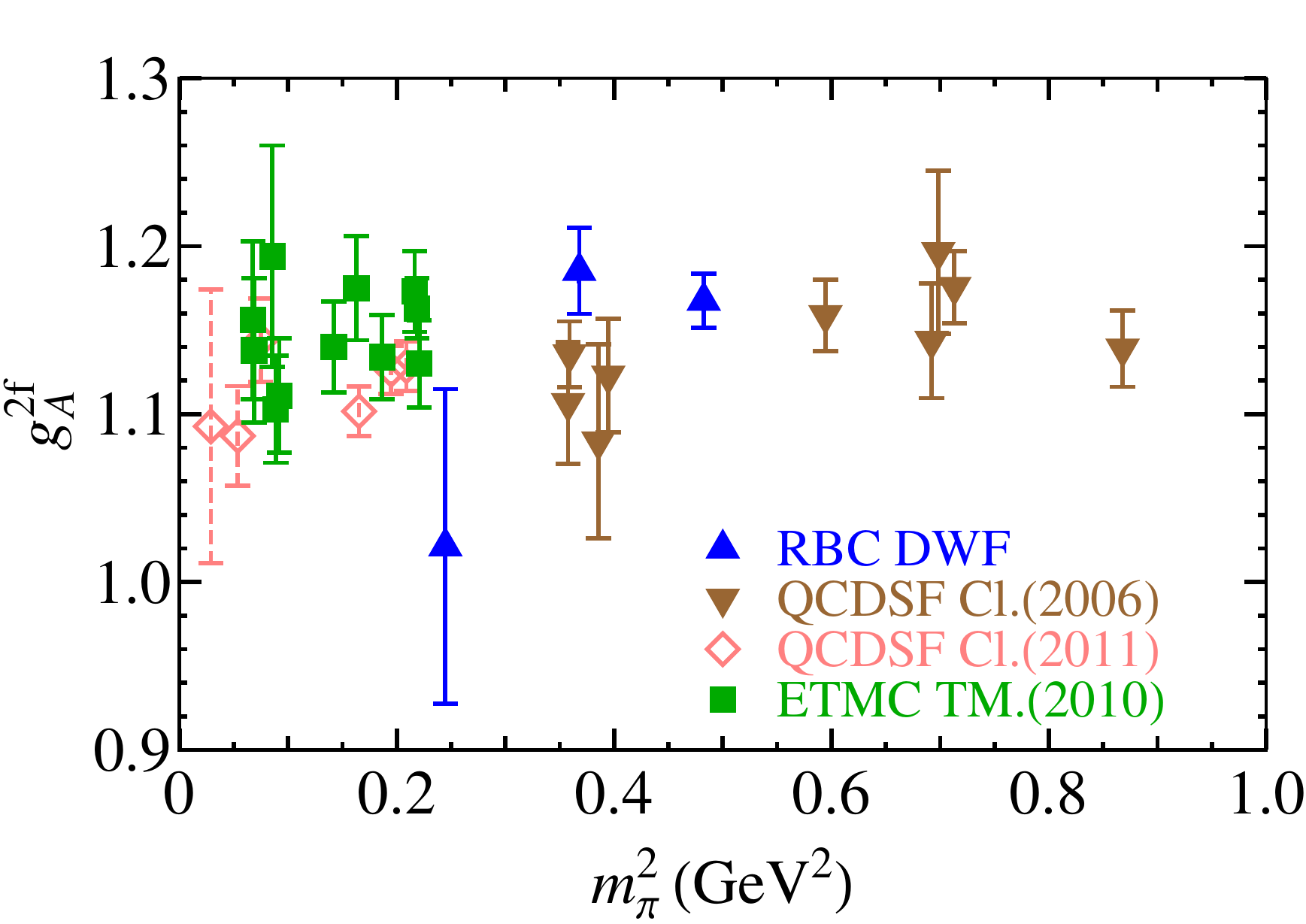}
\includegraphics[height=.24\textheight]{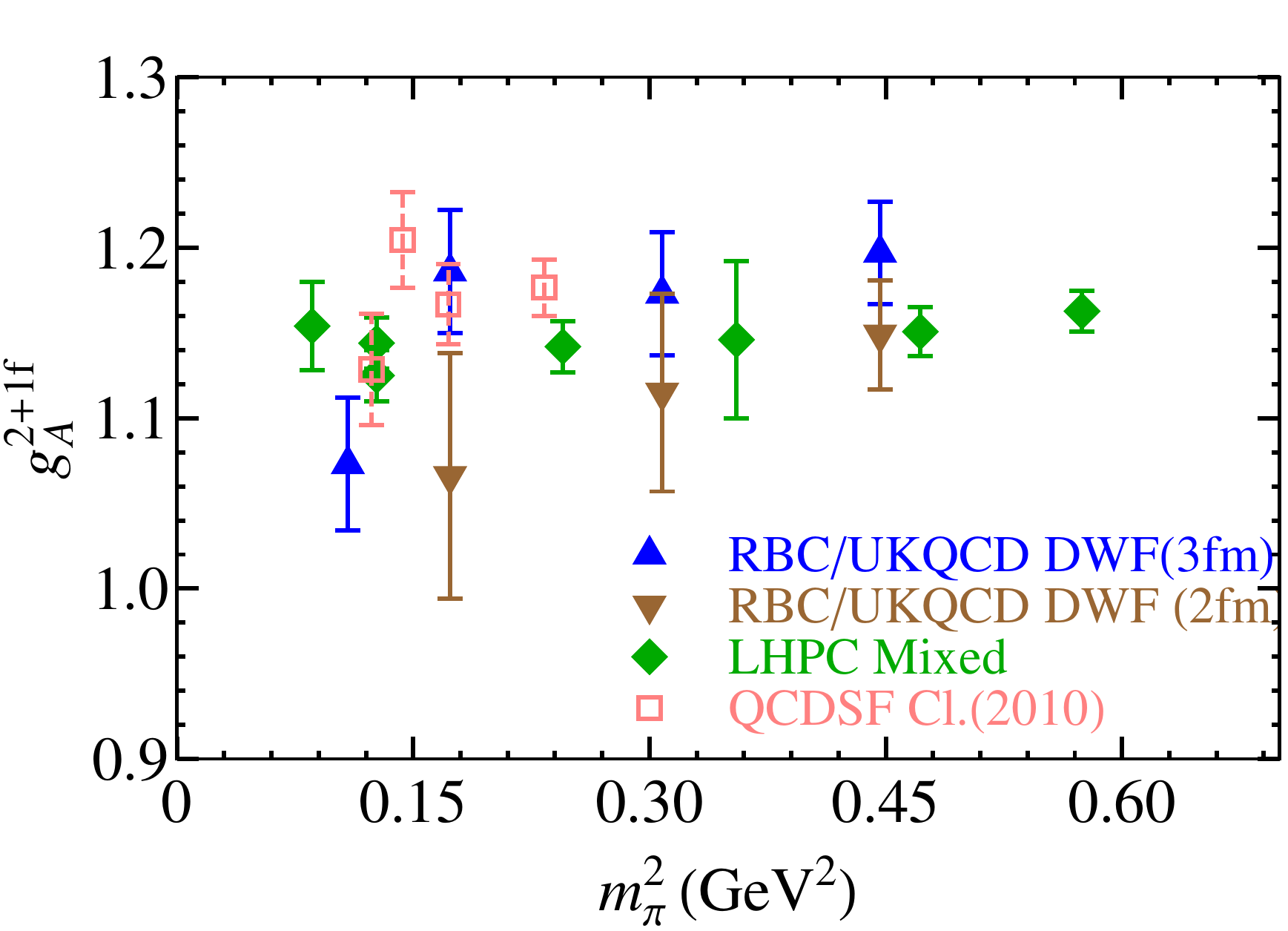}
\includegraphics[height=.27\textheight]{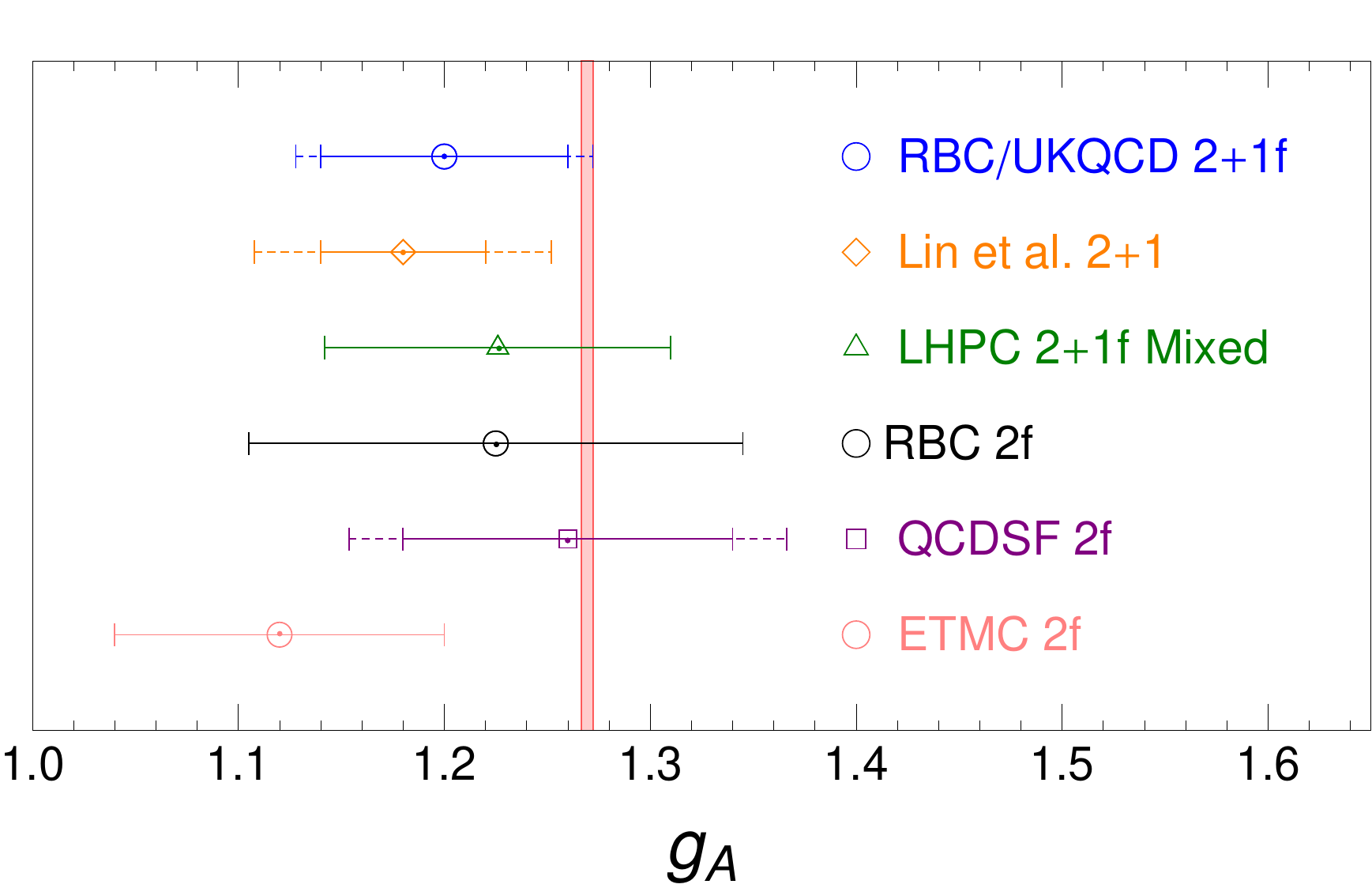}
\end{center}
\vspace{-0.1in}
\caption{(Upper row) The axial charge versus $M_\pi^2$ 
from 
$N_f=2$~\cite{Khan:2006de,Pleiter:2011gw,Brandt:2011sj,Alexandrou:2010hf,Lin:2008uz}
(left) and
$N_f=2+1$~\cite{Yamazaki:2008py,Bratt:2010jn,Edwards:2005ym,Gockeler:2011ze,Lin:2011sa}
(right) calculations with different types of
$O(a)$-improved fermion actions.  The filled symbols and
solid errorbar (open symbols and dashed errorbar) denote results
taken from published papers (the latest lattice proceedings).
(Lower panel) Comparison of the published values of $g_A$ after chiral
and continuum extrapolations with experimental measurements~\cite{Nakamura:2010zzi} (vertical band). The solid
lines indicate statistical error while the dashed lines include
systematic errors.  Lin~et~al.~\cite{Lin:2007ap} find that an
SU(3)-constrained fit to the $g_A$ for octet baryons reduces the
statistical error in the chiral extrapolation. This is illustrated by
the larger errors in the LHPC result~\cite{Bratt:2010jn} compared
to those in Ref.~\cite{Lin:2007ap}, which are obtained using similar lattice
parameters.}
\label{fig:gA_all}
\end{figure}

Many groups worldwide have calculated $g_A$ using various gauge
ensembles and fermion actions as summarized in
Table~\ref{tab:LQCDsummary} and shown in the first two panels of
Fig.~\ref{fig:gA_all} for two- and three-flavor simulations,
respectively. The results from each study, after a chiral
extrapolation to the physical pion mass, are shown in the third panel
of Fig.~\ref{fig:gA_all}.

The overall observations are: (i) The central values vary between
$1.12 < g_A < 1.26$, and the errors are much larger than the
experimental uncertainty.  The deviations from the experimental value
are large, considering that corrections due to strong interactions
determine $g_A - 1$.  (ii) There is no significant difference between
2- and 2+1-flavor estimates or dependence on the light-quark mass at these 
unphysically large $M_\pi^2 \propto m_q$. More high-precision
calculations are needed to determine whether the chiral behavior
changes at smaller quark masses and to gain control over the
extrapolation to the physical $M_\pi$.  (iii) Within errors, the
lattice data are consistent between the different groups (with
different lattice actions), different lattice
spacings and between 2- and $2+1$-flavor theories. Our
understanding of systematic errors, discussed in
Sec.~\ref{ssec:LQCDissues}, are summarized next.

Investigations of finite-volume effects have been carried out by
the RBC/\allowbreak UKQCD collaboration~\cite{Yamazaki:2008py}. They used
domain-wall fermions at a fixed lattice spacing of $1/a=1.73(3)$ GeV
(equivalently, $a = 0.114(2)$ fm) on two lattice sizes $L=1.8$ and
$2.7$ fm.  They found that at fixed $M_\pi^2 \sim 0.1$ GeV${}^2$ there are
significant finite-volume effects for $L < 2.5$ fm, and these lower the value of
$g_A$. They also analyzed $g_A$ as a function of $M_\pi L$ and found
that the data scale in this variable; $i.e.$ data from a given action
and for a given number of flavors collapse onto a single curve. For
small $M_\pi L$, the value of $g_A$ is underestimated and to get within 
$1\%$ of the infinite-volume result requires $M_\pi L \gsim 6$. 

The QCDSF collaboration~\cite{Khan:2006de} analyzed $g_A$ at four
lattice spacings ranging from 0.07 to 0.116~fm, and found no
significant dependence on the lattice spacing.  They, and the ETMC
collaboration~\cite{Alexandrou:2010hf}, have also analyzed their data
using finite-volume corrections suggested by heavy-baryon chiral
perturbation theory (HB$\chi$PT) 
with small-scale expansion. They find that
correcting their data for finite-volume effects at each lattice
spacing improves their extrapolation to the physical pion mass. On the
other hand, the RBC collaboration~\cite{Lin:2008uz} finds that such
corrections do not account for their data either qualitatively or
quantitatively. An understanding of finite-volume effects, therefore,
needs more work.

A source of potentially large systematic error is excited-state
contamination when the source-sink separation ($t_{\rm sep} = t_f
-t_i$) is insufficient. The 2008 RBC 2-flavor study~\cite{Lin:2008uz}
used $t_{\rm } \approx 1.0$ and 1.2~fm to check whether there is
any significant dependence of $g_A$ on $t_{\rm sep}$. In this case,
the values of $g_A$ were consistent within statistical error, and the
central values increase by less than 5\% between $t_{\rm sep}=1.0$ and
$1.2$ fm.  Similarly, LHPC~\cite{Bratt:2010jn} observe a tiny shift of
the central value when changing $t_{\rm sep }$ from 1.1 to 1.2~fm.  The
recent work by the CLS collaboration~\cite{Brandt:2011sj} investigated
smaller source-sink separations, $t_{\rm sep} \approx 0.56$, 0.70 and
1.05~fm, on their $a \approx 0.07$ fm lattices. They find that the
value of $g_A$ increases by about $ 10\%$ with $t_{\rm sep}$, and at
the same time the statistical error increases by a factor of $5$.
They use a linear extrapolation in $t_{\rm sep}$ to reduce the effect
of excited-state contamination and conclude that for the interpolating
operators used $t_{\rm sep} > 1.1$ fm is needed to approximate the
asymptotic value. In any case, one
should include the excited states explicitly in the analysis of the
matrix elements as demonstrated in
Refs.~\cite{Lin:2011sa,Brandt:2011sj}.

The uncertainty in the lattice determinations of $g_A$, which still do
not fully include all systematic errors discussed in
Sec.~\ref{ssec:LQCDissues}, is much larger than the experimental one,
limiting its utility as a probe for physics beyond the Standard Model.
Our conclusion is that a combination of high statistics, use of
multiple $t_{\rm sep}$ and investigation of correlators with different
overlap of source with ground versus excited states will be needed to
extract the matrix elements with high precision.  A promising
direction for reducing the statistical error in $g_A$ is to use a
simultaneous chiral extrapolation of the octet baryons since the axial
charges of the $\Sigma$ and $\Xi$ baryons are calculated with
significantly smaller errors~\cite{Lin:2007ap}.  A major limitation to
testing whether excited-state contamination is a significant factor in
the underestimate of $g_A$ is the computational resources needed to
simulate close to (and eventually at) the physical light-quark masses,
high statistics and extrapolations to the continuum limit.  The
U.S. national report on the future of extreme-scale
computing~\cite{DOEReport} has made the high-precision calculation of
$g_A$ a milestone to achieve, so we anticipate steady improvement in
lattice estimates of all such matrix elements with increasing
computational power.

\subsection{Nucleon induced-pseudoscalar charge \texorpdfstring{$g_P^*$}{g\string_P}}
\label{ssec:inducedps}

There has been renewed interest in the induced-pseudoscalar form
factor $\tilde{g}_P(q^2)$, defined in Eq.~\ref{eq:inducedgP}, due to the recent MuCap
Collaboration~\cite{Andreev:2007wg} high-precision experiment studying
ordinary muon capture (OMC) by protons, $\mu^- p \rightarrow \nu_\mu
n$. We define the induced-pseudoscalar coupling as\looseness-1
\begin{equation}
g_{P}^* = \frac{m_\mu}{2 M_N} \tilde{g}_P(q^2=0.88 m_\mu^2),
\end{equation}
where $m_\mu$ is the muon mass. Improved calculations of electroweak
radiative corrections~\cite{Czarnecki:2007th} allow precise extraction
of the form factor from these experiments.
The new MuCap experiment yields
$g_P^*=7.3\pm1.1$~\cite{Andreev:2007wg,Czarnecki:2007th}, which is
consistent with the value predicted by heavy-baryon chiral perturbation theory
$g_P^{* \chi{\rm  PT}}=8.26\pm0.16$~\cite{Bernard:2001rs}. However,
it is much smaller than the earlier world average for OMC, $[g_P^{* \rm
OMC}]_{\rm ave}=10.5\pm 1.8$ given in
Ref.~\cite{Gorringe:2002xx}, and the value obtained from a TRIUMF
experiment with radiative muon capture (RMC), $\mu^- p \to \nu_\mu
n \gamma$, which gave $g_P^{* \rm RMC}=12.4 \pm
1.0$~\cite{Jonkmans:1996my}.  After reanalyzing the TRIUMF data, Clark
et~al.~\cite{Clark:2005as} found $g_P^* = 10.6 \pm 1.1$.
When combined with the new MuCap result, the world average is $8.7 \pm
1.0$~\cite{Czarnecki:2007th}.

\begin{figure}
\begin{center}
\includegraphics[height=.2\textheight]{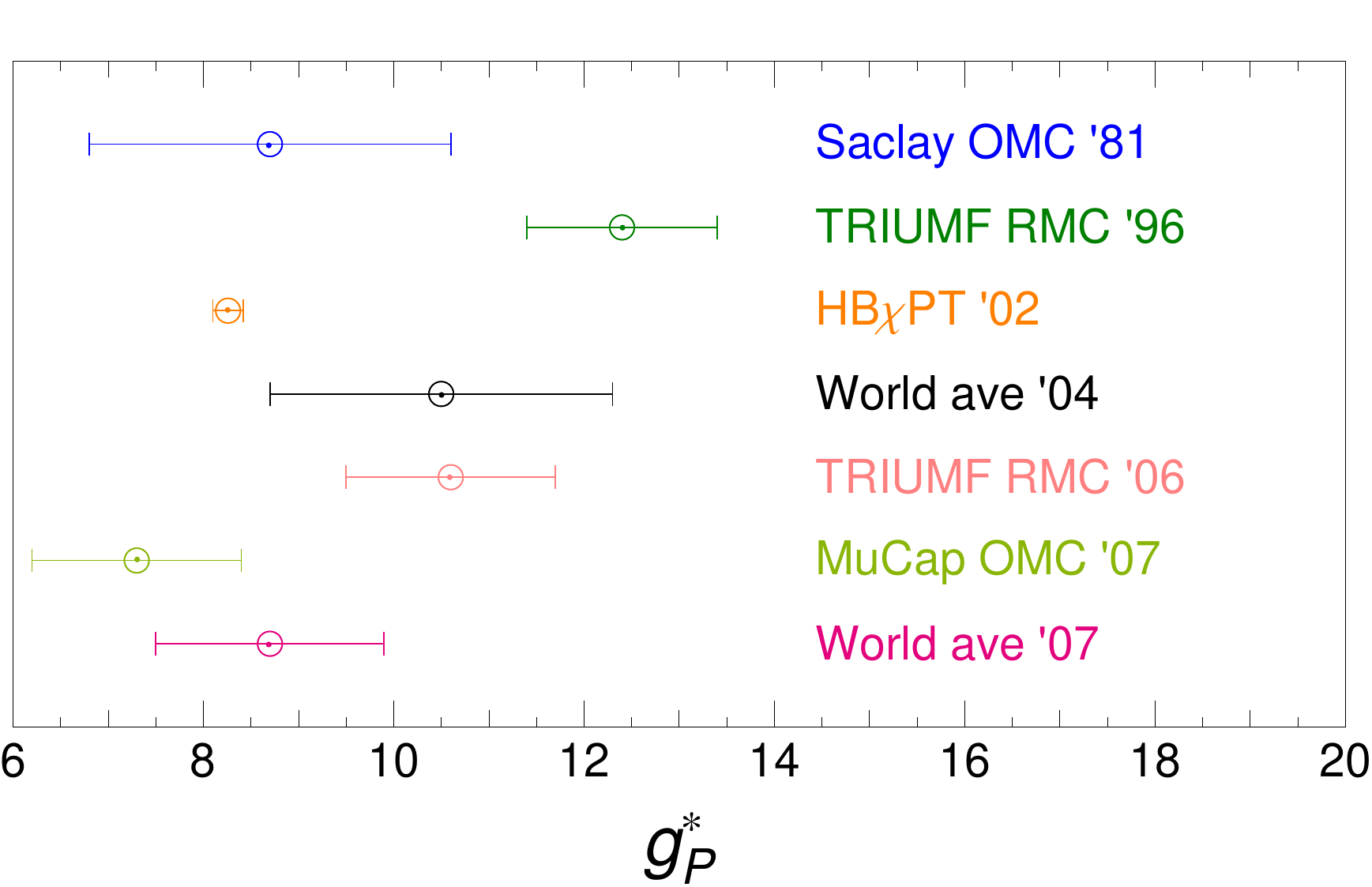}
\hspace{0.01\textwidth}
\includegraphics[height=.2\textheight]{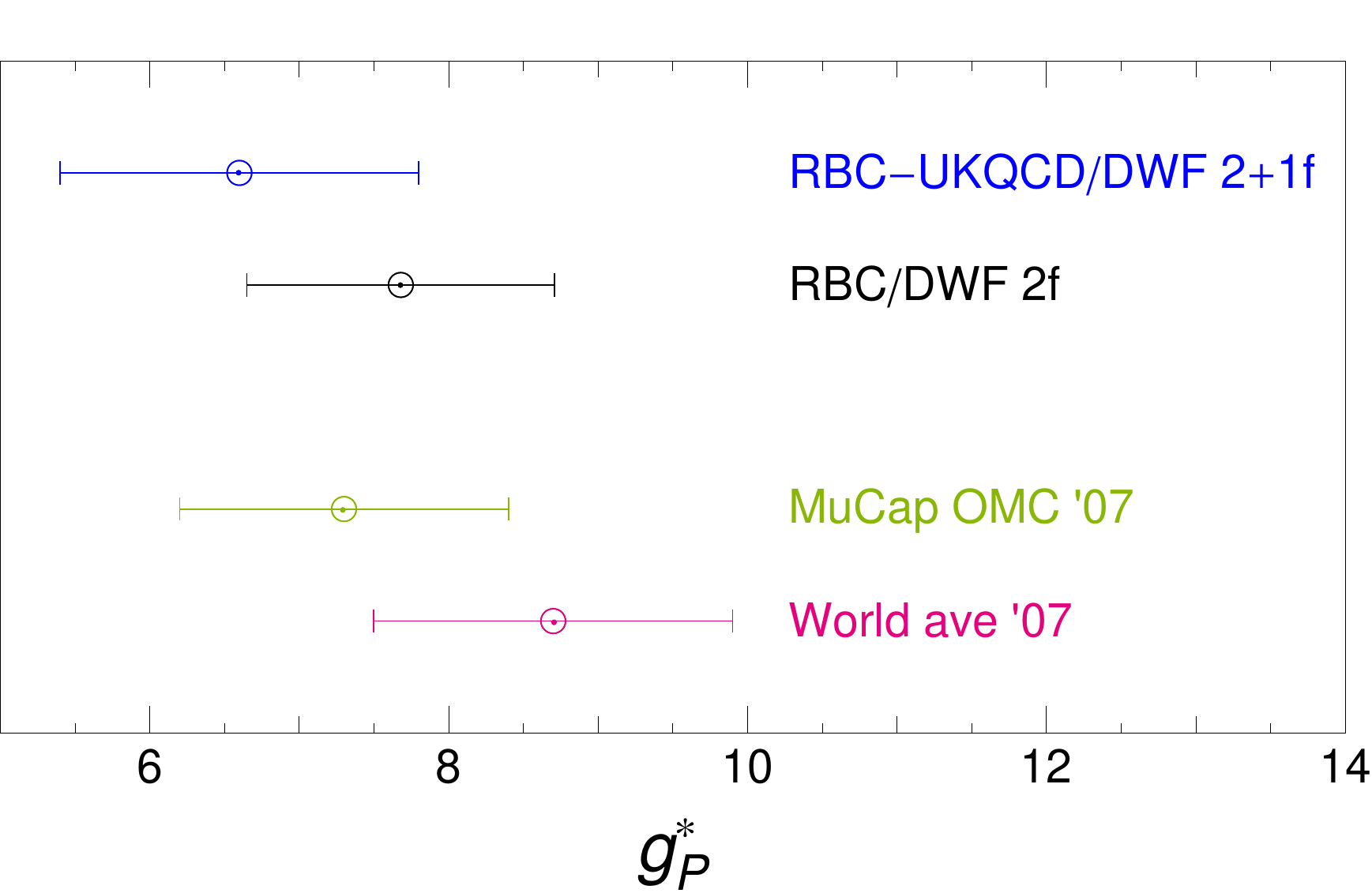}
\end{center}
\caption{(Left) The induced-pseudoscalar charge of the nucleon $g_P^*$ from experimental 
measurements~\cite{Jonkmans:1996my,Gorringe:2002xx,Clark:2005as,Andreev:2007wg,Czarnecki:2007th} 
and an earlier estimation from HB$\chi$PT~\cite{Bernard:2001rs}.
(Right) Comparison of lattice estimates of $g_P^*$ using the DWF fermion
action~\cite{Lin:2008uz,Sasaki:2003jh,Yamazaki:2009zq} with MuCap data.
\label{fig:gP}
}
\end{figure}

There have been few calculations of the induced charge $g_P^*$ in
lattice QCD. Unlike $g_A$, we need to calculate the form-factor at several $q^2$ to
extrapolate $\tilde{g}_P(q^2)$ to the same $q^2$ as those probed in experiments.
The $N_f=2$~\cite{Lin:2008uz} and $2+1$-flavor~\cite{Yamazaki:2009zq}
DWF calculations (Table~\ref{tab:LQCDsummary}) of $g_P^*$ evaluated at 
$(q^2=0.88 m_\mu^2)$ by studying the momentum dependence of the axial
matrix elements give $7.7(1.0)$ and $6.6(1.2)$, respectively. These central
values are about $1 \sigma$ smaller than the world-averaged MuCap
estimate, as shown on the right-hand side of Fig.~\ref{fig:gP}.

Direct calculations of the pseudoscalar charge $g_P$ defined in
Eq.~\ref{eq:defgP} have not been done using LQCD due to the lack
of experimental motivation.  One technical challenge has been removing
the contribution of the pion pole to the amputated vertex in the
calculation of $Z_P$ in RI-MOM schemes. This has recently been
overcome by using non-exceptional momenta in the external quark legs.
We, therefore, expect to provide estimates for $g_P$ at the same level
of precision as $g_T$.

\subsection{Nucleon tensor charge \texorpdfstring{$g_T$}{g\string_T}}
\label{ssec:gT}

The tensor charge $g_T$ is the zeroth moment of transversity, and can
be studied through processes such as SIDIS (semi-inclusive deep
inelastic scattering). The HERMES and COMPASS
experiments~\cite{Anselmino:2008jk,:2008dn,Diefenthaler:2007rj}
presented their first estimates of $g_T$ from data collected at
$Q^2 = 2.4\mbox{ GeV}^2$. Experimentally, to estimate $g_T$ one
first extracts the contribution of individual quarks as a function of
the quark momentum fraction $x$ at a particular $Q^2$. To obtain the
contribution of each quark, the results, estimated from measurements
at a finite number of values of $x$, are integrated over the full
range $0 \le x \le 1$. The isovector tensor charge is then given by
the difference between the up and down quark contributions. Since this
analysis requires data over the full range of $x$, and the low-$x$ and
high-$x$ values are not well known, improvements in precision await
future experiments. Current extracted numbers are highly
model-dependent.  Combining SIDIS (HERMES and COMPASS) results with Belle
$e^+e^-$ analysis~\cite{Seidl:2008xc,Anselmino:2008jk} of data
collected at $Q^2 = 110\ \mbox{GeV}^2$, the best experimental estimate
of $g_T$ at $Q^2=0.8\mbox{ GeV}^2$ (instead of $Q^2=0$)
is $0.77^{+0.18}_{-0.36}$.

There are also estimates from purely theoretical models. 
These include the Nambu--Jona-Lasinio
model~\cite{Cloet:2007em} and the chiral-quark soliton
model~\cite{Wakamatsu:2007nc}; unfortunately, they are not consistent
with each other. Estimates from QCD sum rules~\cite{He:1994gz} have a
large uncertainty.

\begin{figure}
\begin{center}
\leavevmode\raise15pt\hbox{\includegraphics[height=.4\textheight]{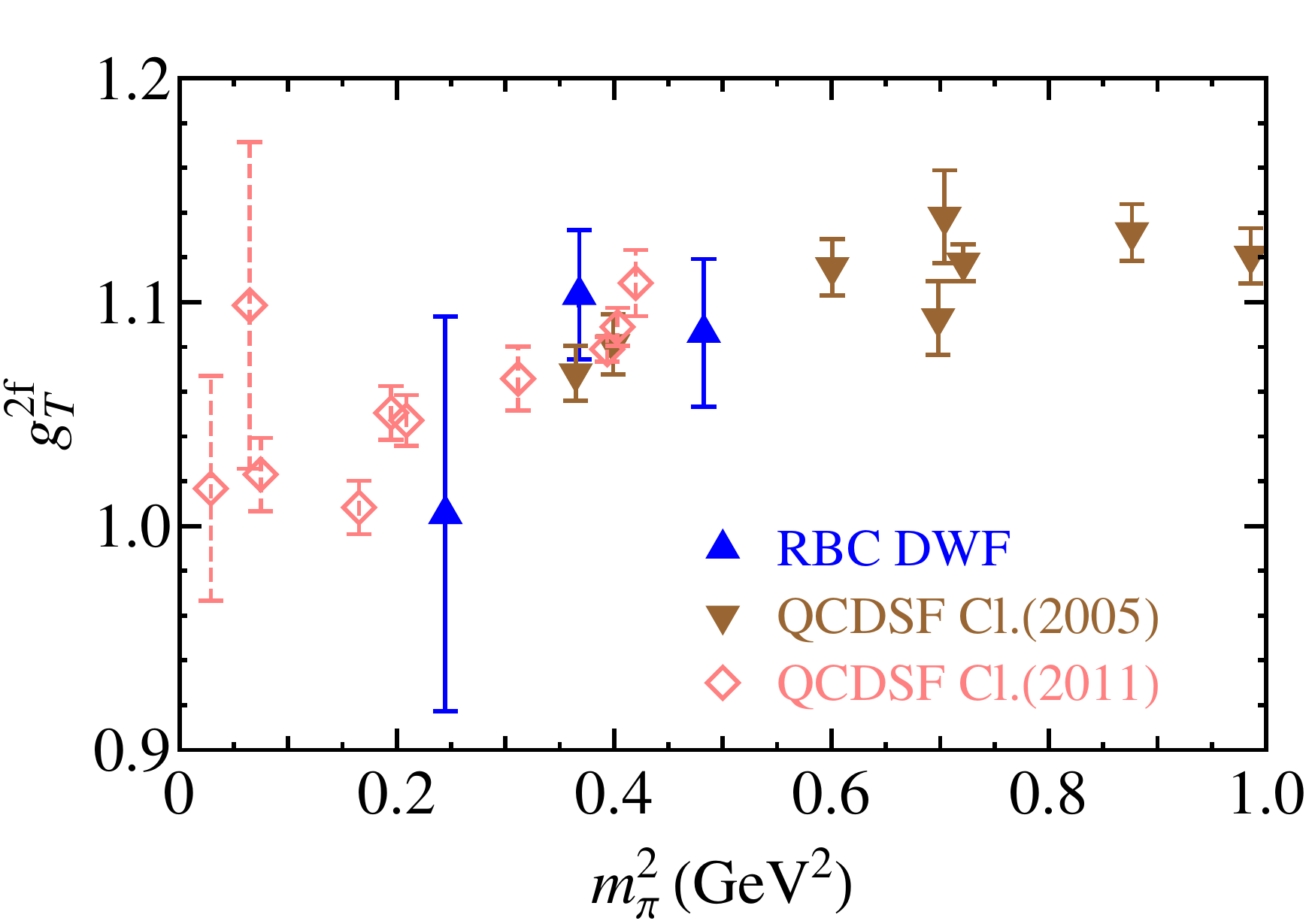}}
\includegraphics[viewport=0 15 475 264,
                 height=.3\textheight]{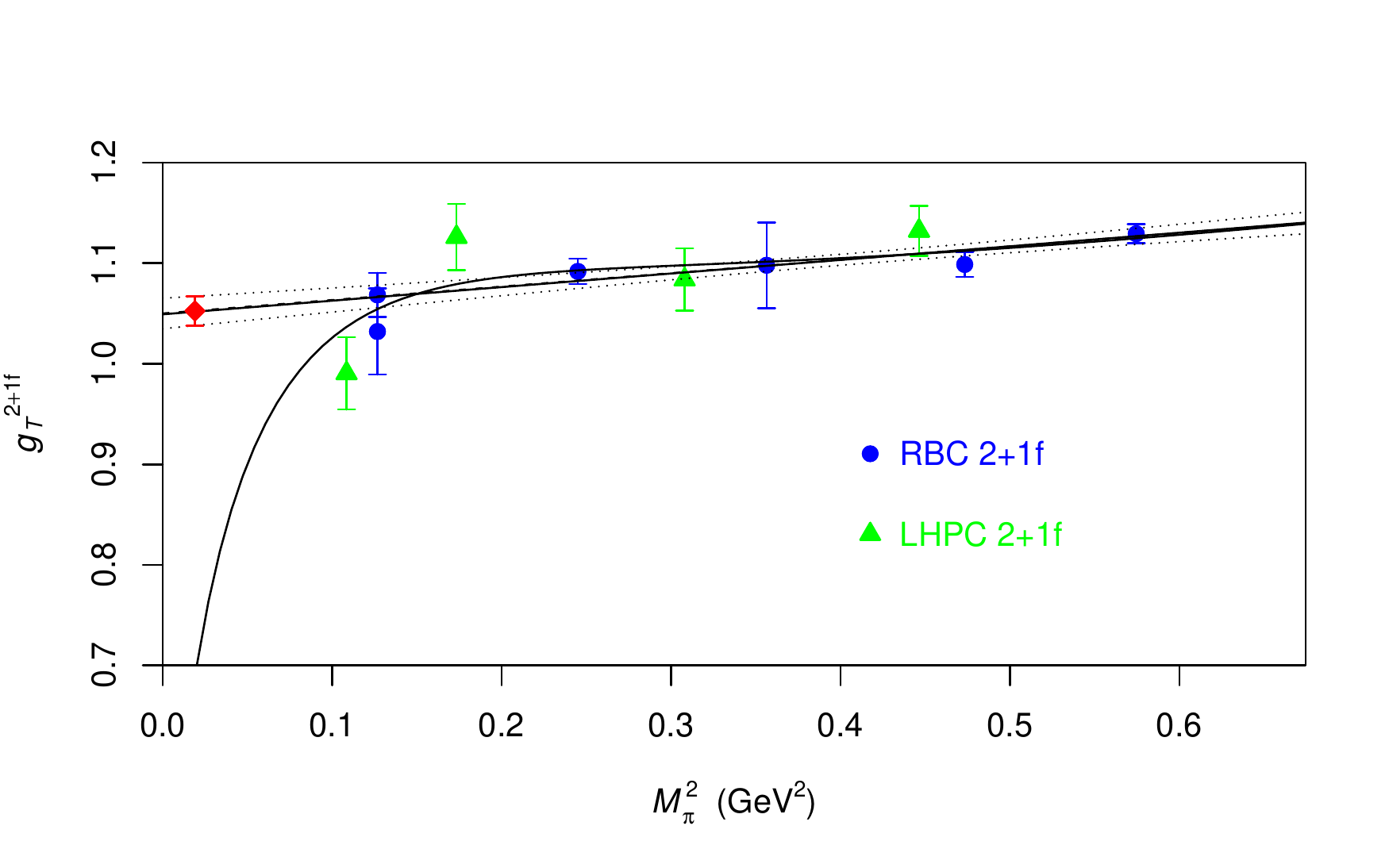}
\end{center}
\caption{
\label{fig:gT}\label{fig:gT-extrap}
Summary of LQCD estimates of $g_{\rm T}$ using
$N_f=2$~\cite{Gockeler:2005cj,Pleiter:2011gw,Lin:2008uz} (top) and
$N_f=2+1$~\cite{Aoki:2010xg,Bratt:2010jn} (bottom) $O(a)$-improved
fermion actions.  Two chiral extrapolations of $g_T$ are shown using
the combined RBC/UKQCD DWF data on their 2.7-fm
ensemble~\cite{Aoki:2010xg} and the LHPC mixed-action
data~\cite{Bratt:2010jn}. The value at the physical pion mass from the
linear fit is shown by the red diamond. Fits using the HB$\chi$PT
ansatz are very sensitive to removing
points at large $M_\pi$ so no error band is shown. For the $g_T^{2 \rm{f}}$
data, the filled symbols and solid error bars (open symbols and dashed
error bars) denote results taken from the published papers (the latest
lattice proceedings).  }
\end{figure}

There are several LQCD estimates of $g_T$, and we review those
listed in Table~\ref{tab:LQCDsummary}.  The QCDSF collaboration's
2-flavor calculations with clover
fermions~\cite{Pleiter:2011gw,Gockeler:2005cj}, over a large range of
pion masses (170--1170~MeV) and 3 lattice
spacings~\cite{Pleiter:2011gw}, show a mild increase in $g_T$ with
$M_\pi^2$.  RBC's 2-flavor DWF calculation~\cite{Lin:2008uz} shows a
similar trend and gave $g_T(\overline{\rm MS},\ 2\ {\rm GeV}) =
0.93(6)$ after extrapolation to the physical pion mass and using $Z_T$
calculated nonperturbatively in the RI-MOM scheme.  These results are
summarized on the top of Fig.~\ref{fig:gT}.

The 2+1-flavor results from the
LHPC~\cite{Edwards:2006qx,Bratt:2010jn} and
RBC/UKQCD~\cite{Aoki:2010xg} collaborations are summarized in the
second of Fig.~\ref{fig:gT}.  RBC/UKQCD used DWF for both dynamical
and valence quarks, while LHPC used the mixed-action approach, DWF on
a 2+1-flavor staggered (asqtad) gauge ensemble.  The lattice spacings
in the two calculations are similar, $0.1224$ fm and $0.114$ fm; thus,
we expect similar lattice-discretization errors.  The range of pion
masses explored is also comparable, 290--760~MeV by LHPC and
330--670~MeV by RBC/UKQCD.  Both find the dependence on the pion-mass
to be small except at the lightest pion-mass points, $290$ and
$330$~MeV, respectively. At these points, the observed downward dip
could be indicative of the onset of chiral logs; however, it is not
yet clear whether these points suffer from finite-volume and
excited-state effects. Further studies at lighter pion masses are
needed to resolve these issues.

To extrapolate the tensor charge to the physical pion mass, we
employed 
the heavy-baryon chiral perturbation theory 
formulation~\cite{Detmold:2001jb,Detmold:2002nf}.  The resulting
formula for $g_T$ contains one low-energy constant 
and two scales
at lowest order in chiral
logs~\cite{Detmold:2002nf}. Analogous formulae for the other twist-two
matrix elements, the quark momentum fraction $\langle x \rangle$ and
helicity distribution function $\langle \Delta x \rangle$, which can
be obtained from $x$-dependent measurements of polarized and
unpolarized form factors, work well in describing the lattice
data. We, therefore, analyzed the combined RBC/UKQCD and LHPC $g_T$
data using the HB$\chi$PT ansatz.  The fits are highly sensitive to
the data points selected, since the chiral log is sensitive to only one
point at the lightest $M_\pi$.  This lack of sensitivity to chiral
logs is illustrated by the linear fit to the $2+1$-flavor data shown
in Fig.~\ref{fig:gT}. It fits all points except the one at lowest
$M_\pi$ and gives $g_T=1.05(2)$.  Given this lack of sensitivity to
the chiral-log term and the high pion masses used relative to the
expected range of validity of this order of HB$\chi$PT, we have little
reason to believe that such an extrapolation is well controlled. We
include in Fig.~\ref{fig:gT} a HB$\chi$PT extrapolation consistent
with the data in order to illustrate the relative size of the chiral
log, which may be quite large and appear at pion masses not much
smaller than those currently available.  For our best estimate we use
$g_T=1.05(35)$ where the central value is from the linear fit and the
uncertainty includes the systematic error associated with the extrapolation
in $M_\pi^2$. Clearly data at smaller $M_\pi^2$ are needed.

\subsection{Nucleon scalar charge \texorpdfstring{$g_S$}{g\string_S}}
\label{ssec:gS}

The nucleon isovector scalar charge $g_S$ has not been analyzed in 
lattice calculations, in contrast to its isoscalar partner, the scalar
density (or the nucleon-$\sigma$ term).  There are no experimental
measurements of this quantity, and theoretical
estimates~\cite{Adler:1975he} (from different model approximations)
give rather loose bounds: $0.25 \leq g_S \leq 1$.
Our preliminary lattice calculations show that $g_S$ has the noisiest
signal compared to the other matrix elements discussed previously and
will, therefore, drive the size of the statistical ensemble required
for precision studies of all the matrix elements.

\begin{figure}
\begin{center}
\includegraphics[clip=true,height=.4\textheight]{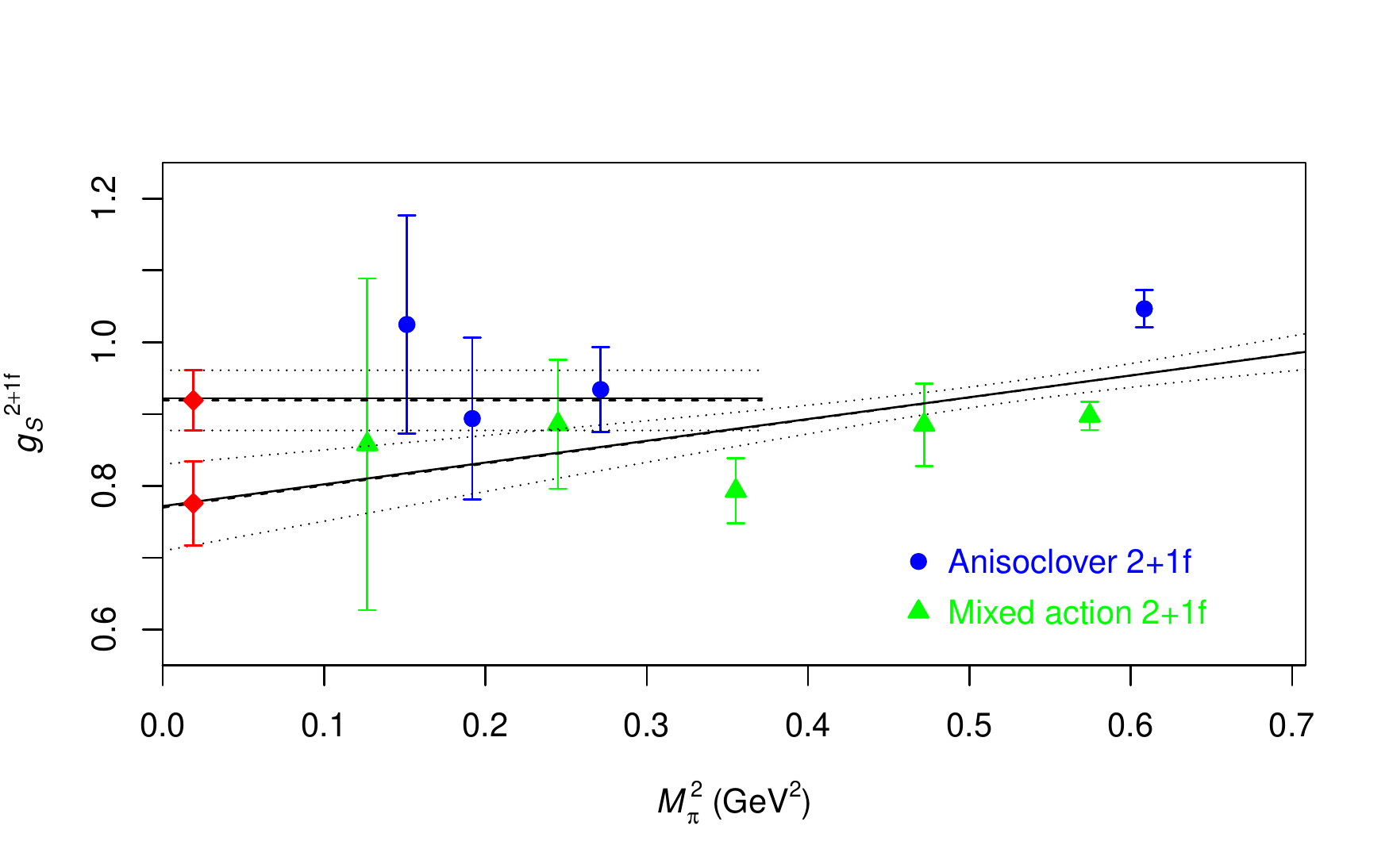}
\end{center}
\caption{
\label{fig:gS}
The scalar charge $g_{\rm S}$ from $N_f=2+1$ anisotropic clover and
DWF/asqtad lattices. We also show a linear fit to the full data set
and a constant fit to the data at the five smallest values of
$M_\pi^2$.  The extrapolated values are shown using red diamonds.  }
\end{figure}

To get a first estimate of $g_S$, we have performed calculations on
two sets of gauge ensembles. The first uses the anisotropic clover lattices
generated by the Hadron Spectrum Collaboration
(HSC)~\cite{Edwards:2008ja,Lin:2008pr} with pion masses ranging from
390 to 780~MeV.   The second uses
$N_f=2+1$ asqtad ensembles but calculates matrix elements with domain-wall valence
quarks with $M_\pi \in \{350,700\}$~MeV. The number of configurations analyzed range between 200 and 650. 
These results are summarized in Fig.~\ref{fig:gS}.
The error bars shown are statistical. 

There is no clear guidance on how to perform a chiral extrapolation to
the physical pion mass since the data show no evidence for chiral
logs.  We, therefore, made fits assuming a behavior linear or constant
in $M_\pi^2$ on the full and different subsets of the data.  In
Fig.~\ref{fig:gS}, we show two fits, a linear one using all the data 
and a constant fit to the five lightest $M_\pi^2$ values. 
The extrapolated value from such fits to different subsets of data
obtained by removing the points corresponding to the heaviest and
lightest $M_\pi$ varies between 0.6--1.0.  We take the mean as the
central value and 0.2 as an estimate of the error associated 
with the mass extraploation. 

In addition to the large statistical error, there is significant
uncertainty in the estimate of the renormalization constant $Z_S$. We
have used the tadpole-improved tree-level value $Z_S = u_0$, where the
tadpole factor $u_0$ is the fourth root of the expectation value of
the $1\times 1$ Wilson loop. For the HSC and DWF ensembles, $u_0 =
0.945$ and $0.938$, respectively.  A recent nonperturbative estimate
of $Z_S$ for the DWF action on lattices with a similar cutoff $a$ as in our calculations, 
converted to the $\overline{\rm MS}$ scheme at 2~GeV, 
gives $~0.65$~\cite{Aoki:2010dy}. We expect a value closer to unity
due to the smearing of links in the formulation of the lattice actions
we use. Nevertheless, based on current nonperturbative estimates with different 
actions and link smearings, our estimate is $Z_S=0.7(2)$ in the 
$\overline{\rm MS}$ scheme at 2~GeV. Using this value 
would lower $g_S$ by about $25\%$. We take the uncertainty in $Z_S$ into account
by doubling the error estimate, and use $g_S= 0.8(4)$.

\subsection{Lattice estimates of tensor and scalar charges for \texorpdfstring{$\epsilon_{S,T}$}{\textepsilon\string_S,T}}

LQCD calculations of $g_S$ and $g_T$ discussed in the previous
sections, while theoretically clean, require reducing a number of
systematic errors.  Our current understanding is that finite-volume
effects are small for $M_\pi L \gsim 4$, and there is little evidence for
discretization errors at current statistics; contributions due to excited states are smaller
than statistical errors once the time separation $t_{\rm sep} > 1.2$~fm 
for the current source operators and lattice parameters; 
and chiral extrapolations gives
rise to the biggest uncertainty in the current data as shown in
Figs.~\ref{fig:gT-extrap} and \ref{fig:gS}.  Thus, we need
high-statistics calculations on large lattices with light-quark masses
close to the physical values. Lastly, nonperturbative calculations of
renormalization constants are essential.

Based on the above analysis, preliminary LQCD estimates are 
\begin{equation}
g_T (\overline{\rm MS}, \mu \mathord{=} 2 \,  {\rm GeV})  = 1.05(35)  \, ,  \  \  \  \ 
g_S  (\overline{\rm MS}, \mu \mathord = 2 \,  {\rm GeV}) = 0.8(4)  .
\label{eq:currentME}
\end{equation}
These are used in the next section to explore bounds on new physics at
the TeV scale. We emphasize that our focus at this point, given the preliminary
nature of the estimates, is on the variation in the bounds under different
scenarios of reduction of errors in lattice calculations.

\section{Impact of lattice results on phenomenology}
\label{sect:latticepheno}

\begin{figure}[!b]
\centering
\includegraphics[width=0.45\textwidth]{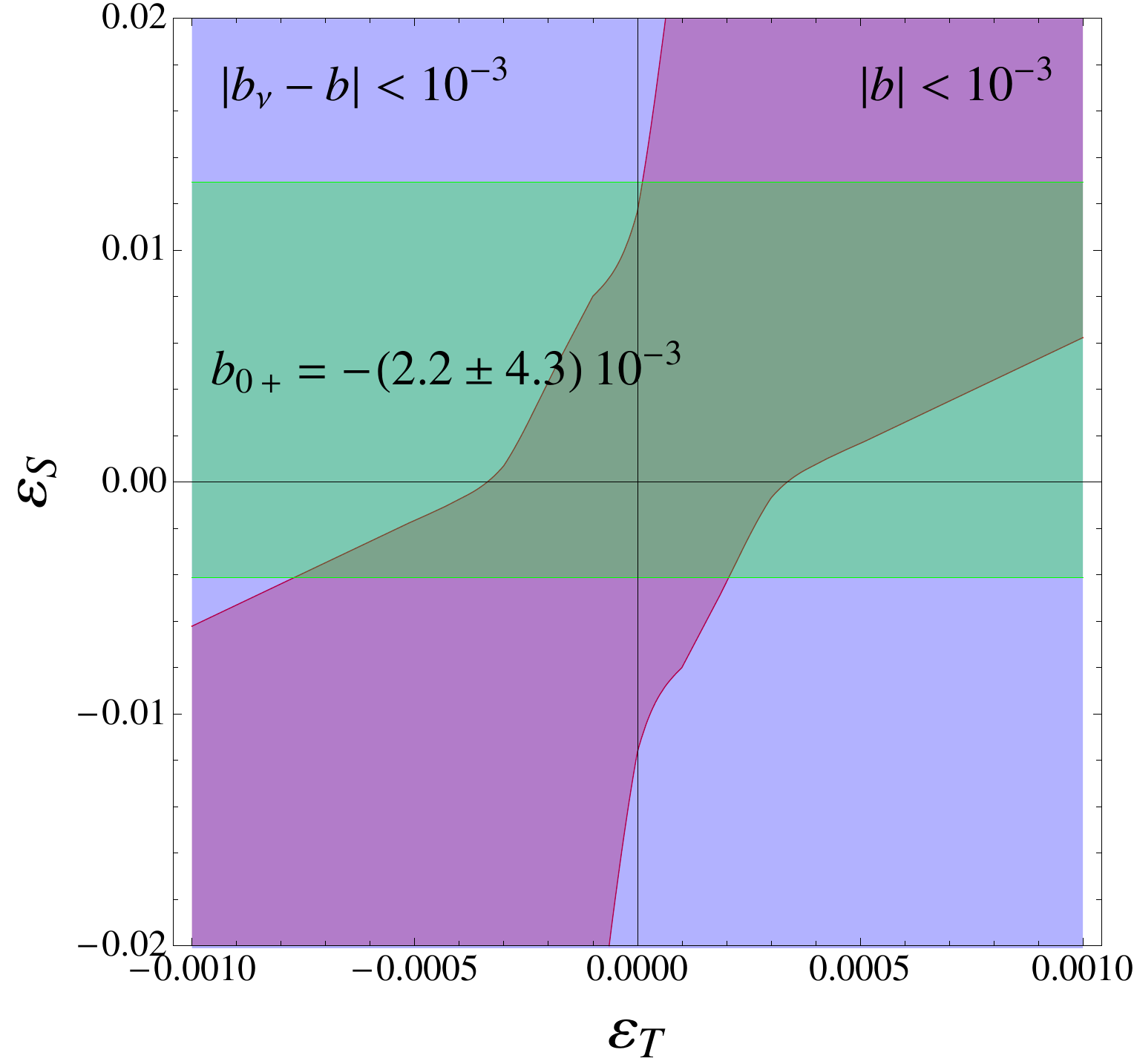}
\hspace{0.2in}
\includegraphics[width=0.45\textwidth]{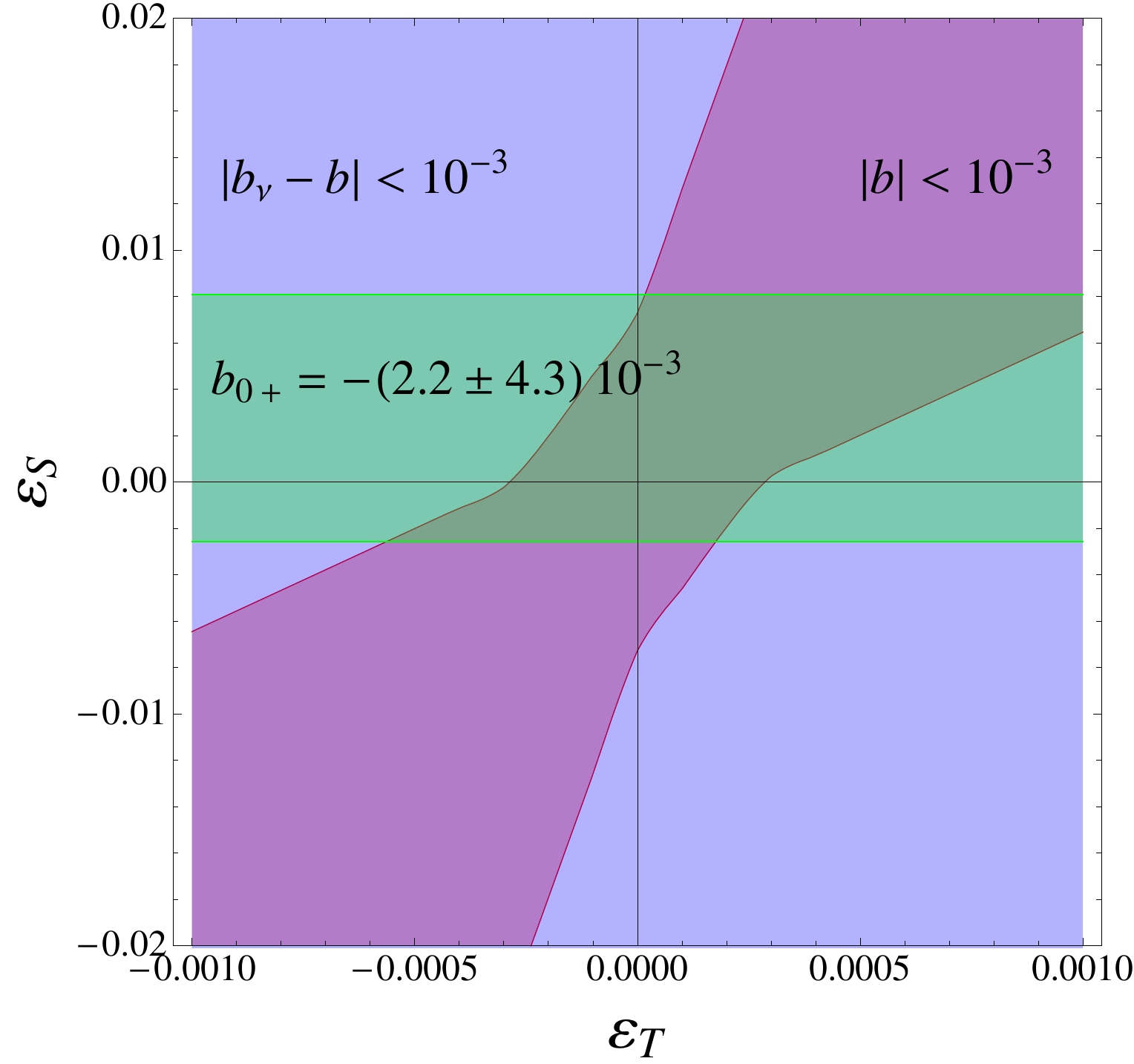}
\caption{
Left panel:  $90\%$ C.L. allowed regions  in 
the  $\epsilon_S$-$\epsilon_T$ plane
implied by (i)  the existing bound on $b_{0^+}$ 
(green horizontal band); 
(ii)  projected measurements of $b$ and $b_\nu-b$ in neutron decay 
(red and blue bow-tie shapes) at the $10^{-3}$ level;  
(iii) hadronic matrix elements taken in the  ranges 
$0.25 < g_S < 1.0$,   $0.6 < g_T < 2.3$~\cite{Herczeg2001vk}. 
Right panel: same as left panel but with scalar and tensor charges taken from 
lattice QCD: 
$g_{S}= 0.8(4)$ and  $g_{T}= 1.05(35)$. 
The effective couplings  $\epsilon_{S,T}$ are defined  in the $\overline{\rm MS}$ scheme at 2~GeV.}
\label{fig:constraints2}
\end{figure}

In Section~\ref{sect:impact1}, while studying the low-energy phenomenology of $\epsilon_{S,T}$, 
we  ignored the uncertainty in the charges $g_{S,T}$. 
Clearly, the impact on $\epsilon_{S,T}$ of future  $10^{-3}$-level  neutron measurements of $b$, $b_\nu$, and 
$b_\nu -b$  depends on how well we know the nucleon matrix elements $g_{S,T}$. 
Since $g_{S,T}$ always multiply factors of the short-distance couplings in physical amplitudes, 
they determine the slope  of the bands on 
the $\epsilon_S$-$\epsilon_T$ plane represented in Figs.~\ref{fig:constraints1} and \ref{fig:constraints1v2}. 
Moreover, if one accounts for the  uncertainty in $g_{S,T}$ the bands Figs.~\ref{fig:constraints1}  and \ref{fig:constraints1v2} 
acquire additional theory-induced thickness and  
their boundaries are mapped  into characteristic ``bow-tie" shapes. 
We illustrate  this  in Fig.~\ref{fig:constraints2}, 
assuming 
experimental sensitivities in $b$ and $b_\nu - b$ at the $10^{-3}$ level. 
For the scalar and tensor charges  we use in the left panel the ranges quoted in  Ref.~\cite{Herczeg2001vk}  (based on earlier 
quark-model estimates): 
$0.25 < g_S < 1.0$,   $0.6 < g_T < 2.3$;  
while in the right panel we use  the lattice estimates  
$g_{S}= 0.8 (4)$ and  $g_{T}= 1.05 (35)$, corresponding to 
$\delta g_{S}/g_S  \sim 50 \% $  and 
$\delta g_{T}/g_T \sim 35\%$.  
Comparing these plots to the ones in Fig.~\ref{fig:constraints1} 
the loss of constraining power  is quite evident. 
Especially in the left panel one sees that the impact of neutron measurements is greatly diluted.

\begin{figure}[t!]
\centering
\includegraphics[width=0.60\textwidth]{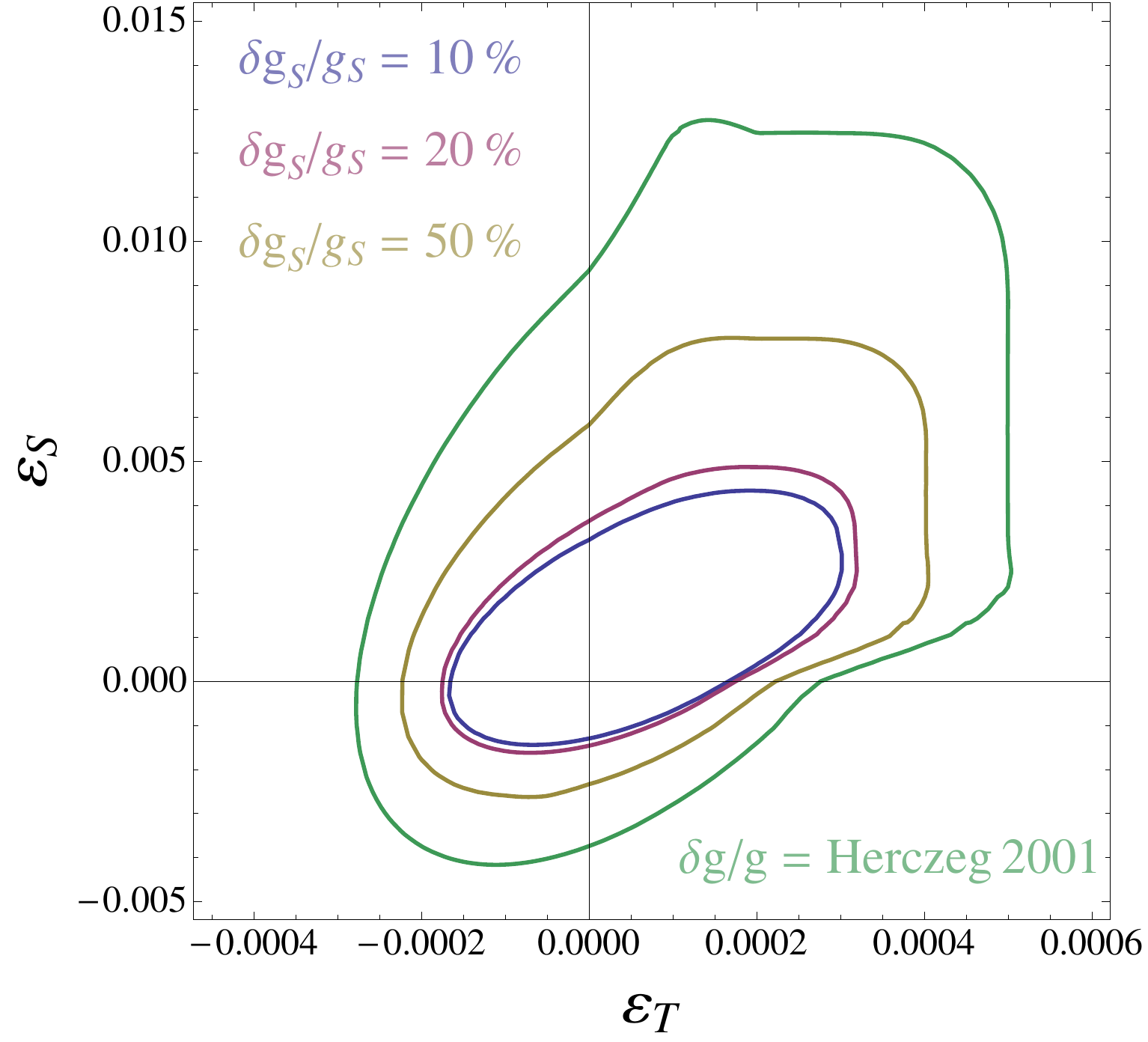}
\caption{
Combined $90\%$ C.L. allowed regions in the 
$\epsilon_S$-$\epsilon_T$ plane based on:
 (i) existing  limit on $b_{0^+}$ from  $0^+ \to 0^+$ nuclear decays;
 (ii) future neutron decay measurements with  projected  sensitivity of $10^{-3}$ in $b$ and $b_\nu-b$. 
The four curves correspond to 
 four different scenarios for  the hadronic 
matrix elements:  
$0.25 < g_S < 1.0$,   $0.6 < g_T < 2.3$ as quoted in Ref.~\cite{Herczeg2001vk};  
lattice results with 
 current central values from Eq.~(\ref{eq:currentME}) and 
 $\delta g_{S}/g_{S} = 50 \%, 20\%, 10\%$ with $\delta g_T/g_T = 2/3 \ 
\delta g_S/g_S$  (this choice assumes that the ratio of fractional uncertainties in $g_S$ and $g_T$ will  
remain approximately  constant as these uncertainties decrease).
The effective couplings  $\epsilon_{S,T}$ are defined in the $\overline{\rm MS}$ scheme at 2~GeV.
} 
\label{fig:constraints3}
\end{figure}

In Fig.~\ref{fig:constraints3} we summarize the low-energy constraints 
on $\epsilon_{S,T}$, taking into account the effects of hadronic uncertainties.  
We plot the combined $90\%$ C.L.  regions in the 
$\epsilon_S$-$\epsilon_T$ plane allowed by the 
 current limit on $b_{0^+}$ and future $10^{-3}$-level measurements of  $b$ and $b_\nu-b$ in neutron decay.  
 The different curves reflect four different scenarios for  the hadronic 
matrix elements: 
the outer-most curve corresponds the range of Ref.~\cite{Herczeg2001vk}, 
while the three inner curves  correspond to lattice results 
with current central values from Eq.~(\ref{eq:currentME}) and  three different uncertainties: 
$\delta g_{S}/g_{S} \in \{ 50 \%, 20\%, 10\%\}$ with $\delta g_T/g_T = 2/3   \  \delta g_S/g_S$ 
(this choice assumes that the ratio of fractional uncertainties in $g_S$ and $g_T$ will  
remain approximately  constant as these uncertainties decrease).

The confidence intervals on $\epsilon_{S,T}$  are obtained using the so-called R-Fit method,  
as described  in Ref.~\cite{Charles:2004jd}.  
In this approach the QCD parameters $g_{S,T}$ are bound to remain within allowed ranges  
determined by the lattice calculations and  estimates of systematic uncertainties 
(in the case  at hand the ranges are $0.4 \leq  g_S  \leq 1.2$ and  $0.7 \leq g_T \leq 1.4$). 
The chi-squared function 
\begin{equation}
\chi^2 (\epsilon_{S},\epsilon_T, g_{S},g_T) =  \sum_{i=1}^{N_{\rm obs}}  
\left(\frac{O_i^{\rm exp}    -   O_i^{\rm th} (\epsilon_{S}, \epsilon_T,   g_{S},g_T)}{\sigma_i^{\rm exp}}\right)^2
\end{equation}
is then minimized with respect to $g_{S,T}$ (varying $g_{S,T}$ in their allowed ranges), 
leading to 
\begin{equation}
\bar{\chi}^2 (\epsilon_{S},\epsilon_T)  =  {\rm min}_{g_{S,T}}  \   
\chi^2 (\epsilon_{S},\epsilon_T, g_{S},g_T)~.
\end{equation} 
Finally, the confidence intervals on $\epsilon_{S,T}$  are deduced applying 
the standard procedure~\cite{Nakamura:2010zzi} to  $\bar{\chi}^2 (\epsilon_{S}, \epsilon_{T})$, 
with an effective number of degrees of freedom 
given by  ${\rm min} (N_{\rm obs} - N_{g}, N_{\epsilon})$, 
where $N_{\rm obs}$ is the number of experimental constraints, 
$N_g = 2$ is the number of QCD parameters ($g_{S,T}$),  
and  $N_\epsilon = 2$ is the number of parameters we wish to constrain ($\epsilon_{S,T}$). 

From Fig.~\ref{fig:constraints3} several clear messages emerge:
\begin{itemize}
\item  Hadronic uncertainties in $g_{S,T}$ 
strongly dilute the significance of new $10^{-3}$-level experiments. 
Experimental progress without theoretical progress will not lead to competitive  
constraints on the short-distance scalar and tensor interactions. 

\item Our preliminary lattice results (curve labeled by $\delta g_S/g_S = 50 \%$) 
already provide  a significant improvement over previous knowledge of $g_{S,T}$ 
summarized  in Ref.~\cite{Herczeg2001vk}. 

\item In order to fully exploit the constraining power of planned   $10^{-3}$
measurements of $b$ and $b_\nu$, 
the uncertainty on $g_{S}$ should be reduced to $20\%$.  
Improvement beyond this level would not significantly increase the constraining power 
(see difference between the curves labeled as $\delta g_S/g_S = 20\%$ 
and $\delta g_S/g_S = 10\%$). 
\end{itemize}

    \section{Collider limits\label{sec:4}}

The contact interactions probed at low energy can also be directly probed at high-energy colliders. The rate, however, depends on whether the particles that generate the 4-fermi interaction are kinematically accessible at the collider energies. We begin in Section \ref{sec:Model-Independent Limits} under the assumption that the scalar and tensor interactions remain point-like at TeV scale energies. Then in Section \ref{sec:resonance} we derive a relation between  $\epsilon_S$ and 
the production cross-section, Eq. \ref{lowerlimit-epsS}~, when the scalar interaction is generated by the exchange of a resonance that is  kinematically accessible at the LHC.

\subsection{Model-independent limits} 
\label{sec:Model-Independent Limits}

Assuming that the scalar and tensor interactions remain point-like at TeV-scale energies, 
we can employ the operator formalism to put bounds on $\epsilon_{S,T,P}$ from collider physics.  
$SU(2)$ gauge invariance implies that  $\epsilon_{S,T,P}$ control  not only charged-currrent processes but 
also the corresponding neutral-current versions, as the 
weak-scale effective Lagrangian 
includes terms proportional to 
$(\epsilon_S - \epsilon_P) \bar{e}_R e_L \bar{d}_L  d_R$, 
$(\epsilon_S + \epsilon_P) \bar{e}_R e_L \bar{u}_R u_L$, 
and $ \epsilon_T   \bar{e}_R \sigma^{\mu \nu}  e_L \bar{u}_R \sigma_{\mu \nu} u_L$.
Exploiting this property,
from an early CDF analysis~\cite{Abe:1997gt} 
of contact interactions in $p \bar{p} \to e^+ e^- + X$,  
after matching the different conventions for the effective couplings, 
we obtain the 90$\%$ C.L.  limit $|\epsilon_S |< 0.135$.
There are a number of LHC searches for contact interactions, specifically in dijet  \cite{Khachatryan:2010te,Aad:2011aj,Khachatryan:2011as} and dimuon \cite{Collaboration:2011tt} final states. All of these studies, however,  focus only on specific vector-like interactions and do not consider scalar (i.e, helicity flipping) contact interactions.

\begin{figure}
\begin{center}
\includegraphics[width=0.65\hsize]{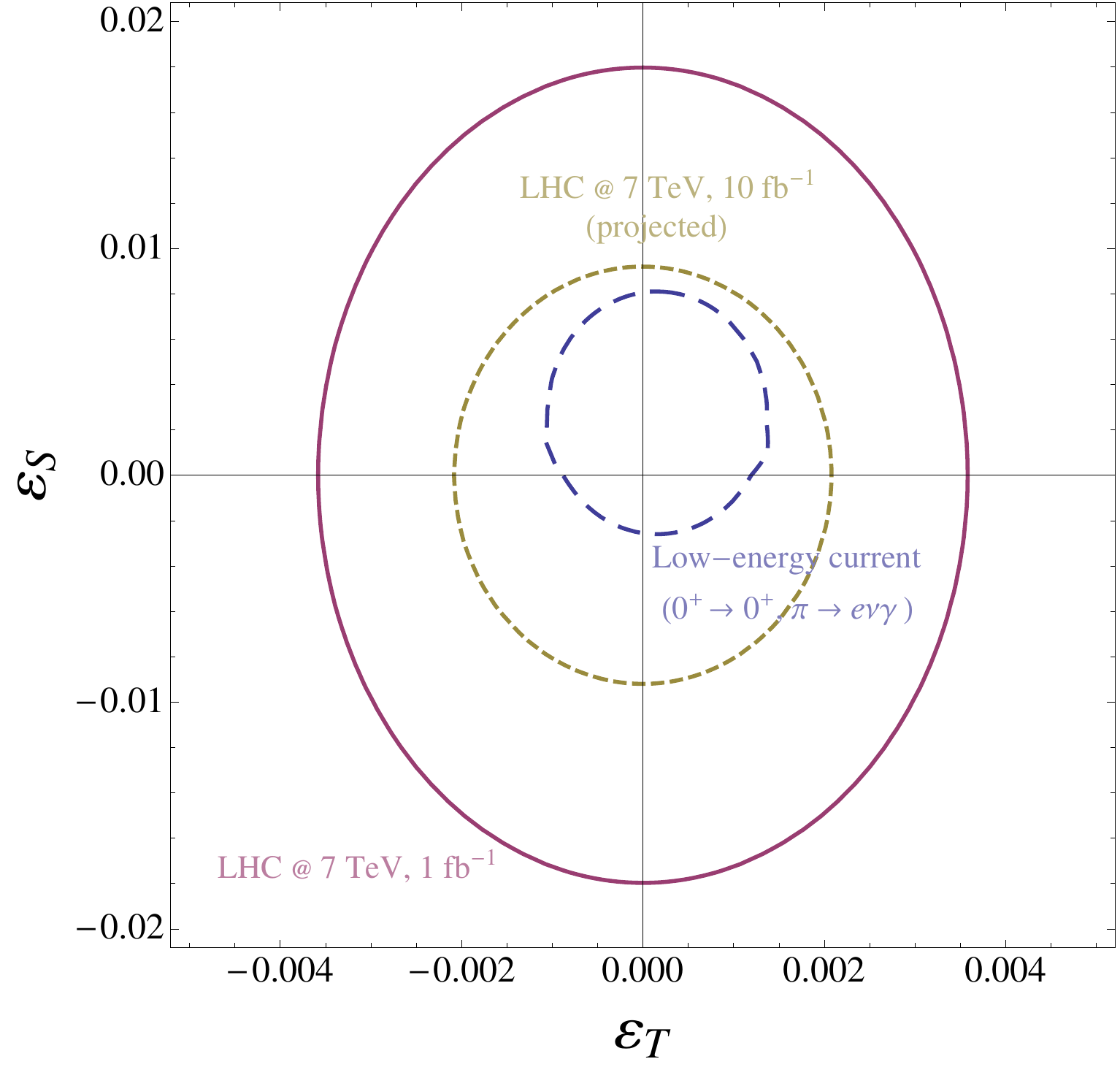}
\caption{\small 
Joint $90 \%$ CL limit on $\epsilon_S$ and $\epsilon_T$ implied by:
(i) current bounds from nuclear $\beta$ decay $0^+ \rightarrow 0^+$ and radiative pion decay (blue, dashed); 
(ii)  CMS search~\cite{CMSWprime} in the channel $pp  \rightarrow e \nu +X$ at $\sqrt{s}=7$~TeV  with $1.03$~fb$^{-1}$ of data.  The limit is obtained  by requiring less than 3.7 $e \nu$-produced events  having $m_T>1$~TeV (red, solid).  
LO MSTW 2008  \cite{Martin:2009iq} parton distribution functions are used; 
(iii)  projected LHC searches in the channel  $pp  \rightarrow e \nu +X$
at $\sqrt{s}=7$ TeV with  $10$ fb$^{-1}$ of data  (gold, dotted).  
The limit is obtained by requiring less than 3 $e \nu$-produced signal events with  $m_T>$ 1.5~TeV 
and assuming that no events are observed.  
The cut is  chosen to reduce the expected leading background to be below 1 event.
The effective couplings  $\epsilon_{S,T}$ are defined  in the $\overline{\rm MS}$ scheme at 2~GeV.
\newline
\label{LHCbounda}}
\end{center}
\end{figure}

\begin{figure}
\begin{center}
\includegraphics[width=0.65\hsize]{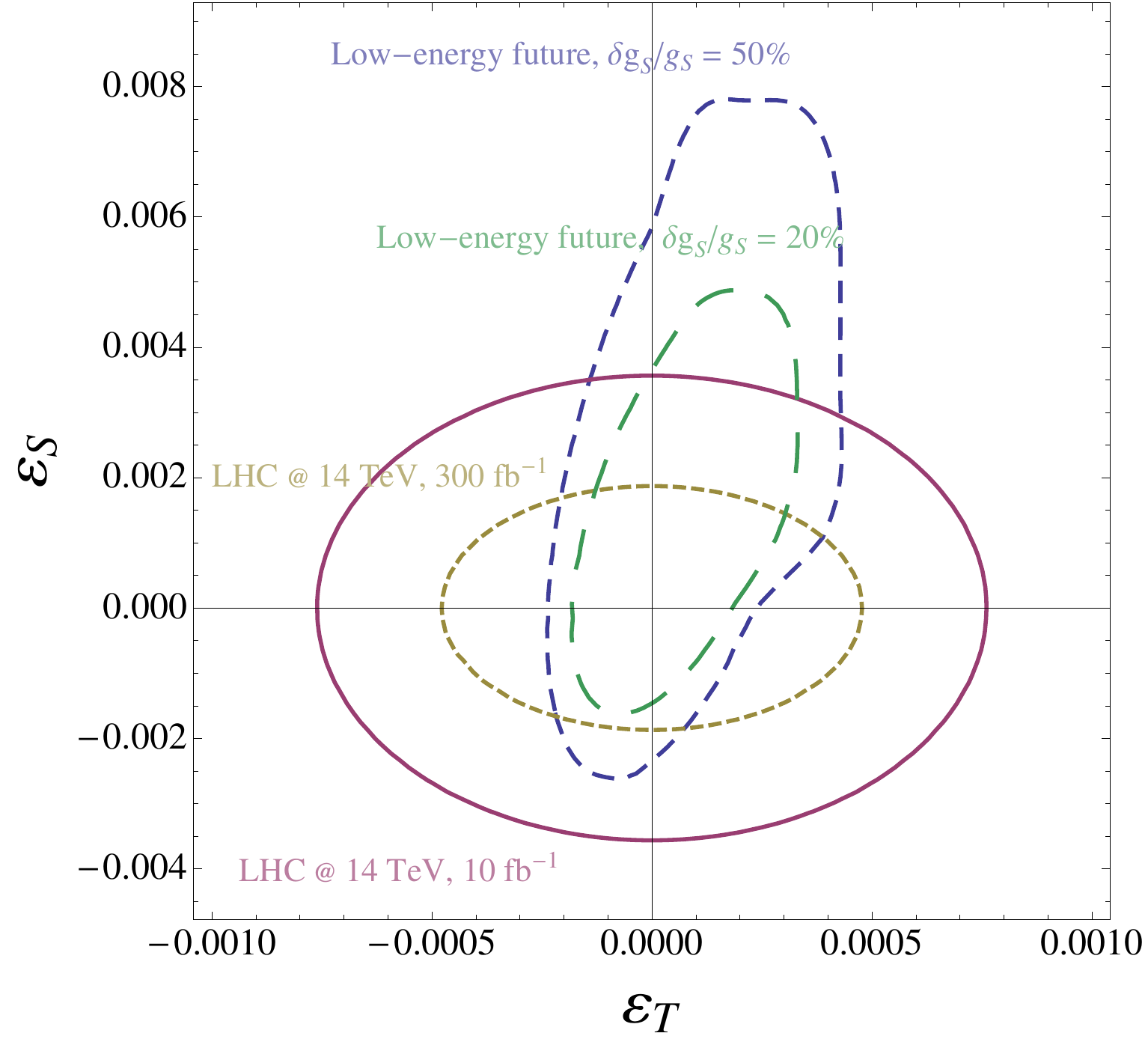}
\caption{\small  Projected joint $90 \%$ credibility level limit  on $\epsilon_S$ and $\epsilon_T$ from the LHC at $\sqrt{s}=14$ TeV, 
obtained from requiring less than 3 $e \nu$-produced signal events with: (i) $m_T>$ 2.5 TeV and 
$10$ fb$^{-1}$ of integrated luminosity (solid, red ellipse); and (ii) $m_T>4$ TeV and 300 fb$^{-1}$ (dashed, yellow ellipse). 
Cuts are chosen to reduce the expected leading background to be below 1 event. To obtain the projection it is assumed no events are found. Same PDFs are used as in Fig. \ref{LHCbounda}. Note the change in scale between these two figures. Anticipated bounds from low-energy experiments and reduced LQCD uncertainties, redrawn from Fig. \ref{fig:constraints3}, are shown for comparison. 
The effective couplings  $\epsilon_{S,T}$ are defined  in the $\overline{\rm MS}$ scheme at 2~GeV.
\label{LHCboundb}}
\end{center}
\end{figure}

Here we focus for definiteness on the charged-current part of the scalar and tensor effective operators. 
These contact interactions fall into the signature class of collider searches for an exotic  $ W^\prime$ gauge boson, since they both can contribute to the signature  $pp \rightarrow e \nu +X$.  
We will use the analyses and results of  searches for this process to obtain bounds on $\epsilon_{S}$ and $\epsilon_{T}$. 
In the limit $m_l=0$ the analysis is simplified, since these operators do not interfere with SM processes. We do include the interference between the scalar and tensor interactions, which does not vanish in the chiral limit. 
The relevant part of the effective Lagrangian is given by
\ba 
{\cal L} &=&- \frac{\eta_S}{\Lambda^2_S} V_{ud}(\overline{u} d)   (\overline{e} P_L \nu_e) 
- \frac{\eta_T}{\Lambda^2_T} V_{ud} (\overline{u} \sigma^{\mu \nu} P_L d)   (\overline{e} \sigma_{\mu \nu}P_L \nu_e)
\label{ST-Lag}
\ea 
where $\sigma^{\mu \nu} = i [\gamma^\mu, \gamma^\nu]/2$,  and
$\eta_S,\eta_T=\pm$ denotes the sign of the coefficients of the scalar and tensor operators. 
The relations between $\Lambda_{S,T}$ and the effective couplings $\epsilon_{S,T}$ at  $\mu=1~{\rm TeV}$ 
are given by 
$\epsilon_S \equiv 2 \eta_S v^2/\Lambda^2_S$ and $\epsilon_T \equiv \eta_T v^2/\Lambda^2_T$.
Note  that since collider searches set  limits on the effective couplings $\epsilon_{S,T}$ at the 
high renormalization scale $\mu = 1$~TeV,  a direct comparison with the low-energy constraints 
requires an appropriate rescaling down to the hadronic scale. Using the one-loop anomalous dimensions 
for scalar and tensor operators (see~\cite{Broadhurst:1994se} and references therein),  
the one-loop beta function for the strong coupling constant,  and including the 
appropriate heavy quark thresholds, we find in the $\overline{\rm MS}$ scheme 
$\epsilon_S (1~{\rm TeV})/\epsilon_S (2~{\rm GeV}) = 0.56$ and 
$\epsilon_T (1~{\rm TeV})/\epsilon_T (2~{\rm GeV}) = 1.21$.  We will use these factors to rescale the collider 
limits and  compare them to low-energy limits in Figs.~\ref{LHCbounda} and  \ref{LHCboundb}.

To determine the transverse mass distribution of the electron--neutrino pair we start with 
 \cite{EllisSterlingWebber} 
\ba 
\label{masterdsigma}
\frac{d^3 \sigma}{dy ~dy^\prime ~d m_T^2} &=& \frac{1}{64 \pi s^2} \sum_{ij} \frac{f_i(x_1)}{x_1} \frac{f_j(x_2)}{x_2}  \langle | {\cal M} |^2 \rangle
\ea
where $i$ and $j$ are summed over the initial partons (with parton distribution functions (PDF) $f_{i,j}$ and momentum fractions $x_{1,2}$),   
and $y,y^\prime$ are the rapidities of the electron and neutrino. One finds 
 $x_1 = m_T (e^y + e^{y^\prime})/2\sqrt{s} $, $x_2 =  m_T (e^{-y} + e^{-y^\prime})/2\sqrt{s} $. We also used the observation that the transverse mass of the electron and neutrino, $m_T \equiv \sqrt{2 E^e_T E^\nu_T(1- \cos \Delta \phi_{e \nu})}$ (where $E^{e,\nu}_T$ is the transverse energy of the electron or neutrino, and $\Delta \phi_{e \nu}$ is the azimuthal angle between the two leptons),  is simply $m_T=2 p_T$ at leading order, where $p_T$ is the transverse momentum of the electron. 

To leading order (LO), the contributions of the color- and spin-averaged scalar and tensor matrix elements to $pp \rightarrow e \nu +X$, including the interference term,  is
\ba 
\frac{\langle | {\cal M} |^2 \rangle}{ |V_{ud}|^2} &=& \frac{2}{3}\frac{1}{\Lambda^4_S} (p \cdot p^\prime) (k \cdot k^\prime)
 -\frac{8}{3} \frac{\eta_S \eta_T }{\Lambda^2_S \Lambda^2_T} \left[ (p \cdot k) (p^\prime \cdot k^\prime) - (p \cdot k^\prime) (p ^\prime \cdot k ) \right]
 \nonumber  \\
          & & + \frac{16}{3} \frac{1}{\Lambda^4_T} \left[ 2 (p \cdot k) (p^\prime \cdot k^\prime) +2 (p \cdot k^\prime) (p^\prime \cdot k) - (p \cdot p^\prime)(k \cdot k^\prime) \right] 
\label{intterms}
\ea 
where $p,p^\prime$ are the momenta of the incoming partons, and $k,k^\prime$ are the momenta of the electron and neutrino. The interference term is antisymmetric under $k \longleftrightarrow k^\prime$, so it does not contribute to the transverse mass distribution obtained by integrating over $y$ and $y^\prime$. 
After some substitutions this expression becomes
\ba 
\label{finalmatrixelement}
\frac{\langle | {\cal M} |^2 \rangle}{ |V_{ud}|^2} &=&\frac{1}{6}\frac{\hat{s}^2}{ \Lambda^4_S} 
 - \frac{\eta_S \eta_T}{3} \frac{m^2_T \hat{s}}{\Lambda^2_S \Lambda^2_T}  \sinh(y-y^\prime)  + \frac{4}{3} 
 \frac{\hat{s}^2}{\Lambda^4_T} \left(1-\frac{m^2_T}{\hat{s}} \right)
\ea 
with $\hat{s}=x_1 x_2 s$ and 
$m_T $ is the transverse mass of the lepton-neutrino pair. 

Next we need the cross-section  with  $m_{T}$ greater than a threshold $m_{T,{\rm cut}} $. Using (\ref{masterdsigma}) and  (\ref{finalmatrixelement}), one finds 
\ba 
\label{sigmaMTcut}
\sigma(m_T > m_{T, {\rm cut}}) \! \! \! &=&  \! \! \! \! 
\frac{s}{48 \pi}\int^1 _{m^2_{T, {\rm cut}}/s} \! \! \! \! d \tau \sqrt{\tau}  \left[ \frac{|V_{ud}|^2}{ \Lambda^4_S} \sqrt{\tau - m^2_{T, {\rm cut}}/s}
+ \frac{8}{3} \frac{|V_{ud}|^2}{\tau  \Lambda^4_T} \left(\tau- \frac{m^2_{T, {\rm cut}}}{s} \right)^{3/2} \right]   \nonumber \\
 &\times & \int ^{-\frac{1}{2} \ln \tau}_{\frac{1}{2} \ln \tau} d y_P  \left[ f_{\overline{u}}(\sqrt{\tau} e^{y_P}) f_d (\sqrt{\tau} e^{-y_P})
+ (\overline{u},d) \rightarrow (u,\overline{d}) \right] ~,
 \ea
where the sum over $i,j=\overline{u},d$ and $i,j=u, \overline{d}$ has been done, $\tau=x_1 x_2$, and $y_P=0.5 \ln(x_1/x_2)$.

The ATLAS and CMS experiments have searched for new physics in  $pp \rightarrow e \nu +X$ by looking for an excess of events at large transverse mass 
\cite{Aad:2011yg,CMSWprime}.  The CMS study analyzes 1.03 fb$^{-1}$ of data for the electron final state, and 
we begin by  following their analysis in setting limits on the scalar and tensor interactions.  
The CMS search window is defined by specifying a cut on $m_T$ and counting the number of events detected with transverse mass larger than the cut. 
Specifically, they looked for the production of a heavy $W^\prime$ with decay $W^\prime \rightarrow e \nu$ by  
searching for events having transverse mass  above a variable threshold  $m_{T, {\rm cut}}$, finding 1 event for $m_{T, {\rm cut}}= 1$ TeV and 1 event for   $m_{T, {\rm cut}} = 1.1$ TeV.  In general the limit on the number of expected signal events depends on the expected background, 
$n_{\rm{b}}$, which for this search  
 is quoted to be  $n_{\rm{b}}=2.2 \pm 1.1$ events for $m_{T, {\rm cut}} = 1$ TeV and  $n_{\rm{b}}=1.4 \pm 0.80$ events for $m_{T, {\rm cut}} =  1.1$ TeV~\footnote{
These $n_{\rm b}$ values are taken from Table 1 of  Ref.~\cite{CMSWprime}.
Different central values for $n_{\rm b}$ appear in  Figure 2 of Ref.~\cite{CMSWprime}: 
for example   $n_{\rm{b}}=1.15$   for $m_{T, {\rm cut}} = 1$~TeV. 
The two sets of $n_{\rm b}$ are  consistent within the quoted error bars and lead to minor $(5\%)$ differences 
in the bounds on $\epsilon_{S,T}$.}. 
 
 To set a limit we follow Ref. \cite{CMSWprime} and use Bayesian statistics with a flat prior in the signal $n_{\rm{s}}$. The likelihood function 
 $L(n|n_{\rm{s}})$ is given by the Poisson
 distribution for $n$ detected events with $n_{\rm{s}}$ signal and $n_{\rm{b}}$ background events expected. 
 The expected number of signal events is given by $n_{\rm{s}}=\epsilon \sigma  {\cal L}$,  where $\sigma$ is given by (\ref{sigmaMTcut}), $ {\cal L} $ is the integrated luminosity, and $\epsilon$ is the detection efficiency times the geometric acceptance. Ref. \cite{CMSWprime} quotes the signal efficiency for a $W^\prime$ to be $80\%$. Their earlier analysis (Ref. \cite{Khachatryan:2010fa}), 
 based on 36 pb$^{-1}$ of data, quotes the product of the geometric acceptance and detection efficiency as being greater than 64$\%$ in the $W^\prime$ mass range of interest. In the absence of a detector simulation for our signal, in what follows we will assume our signal has a $80\%$ detection times geometric acceptance efficiency. 
 
 The credibility level $1-\alpha$ for a flat prior in the signal  is then derived from  \cite{Nakamura:2010zzi} 
 \beq 
 1- \alpha= \frac{\int^{s_{\rm{up}}}_0 d n_{\rm{s}} ~L(n|n_{\rm{s}}) } {\int^{\infty}_0 d n_{\rm{s}} ~L(n|n_{\rm{s}}) }
 \label{cl1}
 \eeq 
 which is equivalent to \cite{Nakamura:2010zzi} 
 \beq 
 \alpha = e^{-s_{\rm{up}}} \frac{\sum_{m=0} ^{n} \frac{1}{m!}(s_{\rm{up}}+n_{\rm{b}})^m}{\sum_{m=0} ^{n} \frac{1}{m!}n_{\rm{b}}^m}
\label{cl2}
\eeq 
To set a limit on $s_{\rm{up}}$, we choose the lower value of $m_{T,\rm{cut}}=1$ TeV in order to maximize the signal rate. 
Then for $n_{\rm{b}}=2.2$ expected background events and $n=1$ event detected, one finds that $s_{\rm{up}}=3.0$ at the 90$\%$ credibility level. 
Dividing by $\epsilon$, we obtain a 90$\%$ upper credibility limit of $3.7$ produced signal events. 

In Fig. \ref{LHCbounda}  we show the corresponding limits on $\epsilon_S$ and $\epsilon_T$ (red,  solid curve), using Eq. (\ref{sigmaMTcut}) with  $1.03$ fb$^{-1}$ of integrated luminosity  at $\sqrt{s}=7$ TeV.  LO MSTW 2008 \cite{Martin:2009iq} 
PDFs are used and evaluated at $Q^2=1$ TeV$^2$.  We also checked that limits obtained using the CTEQ6 PDF set \cite{Pumplin:2002vw} are quantitatively in good agreement. 
When only one of these operators is present, the bounds on $\epsilon_{S,T}$ correspond to 
$\Lambda_S >  2.5$ TeV and $\Lambda_T >  2.7$ TeV.  As illustrated by~Fig.~\ref{LHCbounda}, 
our LHC bound on $\epsilon_S$ is a factor of 7 stronger than the old CDF limit, but  about  4 
times weaker than the bound from nuclear beta decay.
Similarly, our new collider bound on  $\epsilon_T$  is a  factor of 3 
weaker than the bound from radiative pion decay.  
 
We have performed a parallel analysis using the ATLAS results~\cite{Aad:2011yg} on $W'$ search. 
Use of the ATLAS results requires an extra step, since their quoted efficiency for a given $m_{T,{\rm cut}}$ 
includes the fraction of total $W' \to e \nu$ events with   $m_T > m_{T,{\rm cut}}$. 
After determining  this fraction with a leading-order calculation, we  infer the 
ATLAS detection times geometric acceptance efficiency 
for $m_{T,{\rm cut}} = 1$~TeV to be 80\%,  the same as quoted by CMS for their experiment. 
Using then the fact that for $m_T > 1$~TeV ATLAS observes $n=1$ event 
with an expected number of  background events $n_{\rm b} = 0.89(20)$,  
we find that the ATLAS limits on $\epsilon_{S,T}$ 
differ from the CMS ones  only at the 5\% level, well within the uncertainties of  our  leading order calculation.
We also estimate  that the  bounds on $\epsilon_{S,T}$ can be reduced by at least a factor of 2 once the full data set collected at $\sqrt{s}=7$~TeV is analyzed 
(see dotted, gold line in Fig.~\ref{LHCbounda}).
We expect that stronger limits can be obtained by a combined analysis of the ATLAS and CMS data, 
which  goes beyond the scope of this work.

To obtain projected limits at higher luminosities and $\sqrt{s}=7$~TeV or $\sqrt{s}=14$~TeV, we repeat the same LO analysis, assuming the same 80$\%$ detection times acceptance efficiency for the signal as before.
We  choose an aggressive cut to make the expected background small. 
The location of the cut on the transverse mass $m_{T,\rm{cut}} \sim $~TeV is estimated by computing at tree-level the transverse mass distribution of the dominant SM physics background, due to the production of a  high-$p_T$ lepton from an off-shell $W$, and finding the value above which the expected background is less than 1 event. 
At $\sqrt{s}=7$~TeV and with an integrated luminosity of 10 fb$^{-1}$,  we find that the number of background 
events drops below one for $m_{T,\rm{cut}} = 1.5$ TeV. 
At $\sqrt{s}=14$~TeV,   with $m_{T,\rm{cut}} = 2.5$ TeV and an integrated luminosity of 10 fb$^{-1}$ we find 0.5 background events expected above the cut.  
We also consider an ultimate luminosity of $300$ fb$^{-1}$, finding that for $m_T>4$ TeV there are  0.3 expected background events. 
We therefore impose $m_{T,\rm{cut}}=2.5$ TeV (4~TeV), and assume an integrated luminosity of $10$ fb$^{-1}$ (300 fb$^{-1}$).  To set a limit we will assume that no events are found, which is consistent with less than 1 background event expected. From Eqs. (\ref{cl1}) and (\ref{cl2}) we then obtain a 90$\%$ Credibility Limit of  3 produced events. The anticipated joint $90\%$ credibility level limits on $\epsilon_S$ and $\epsilon_T$ from LHC at $\sqrt{s} = 14$~TeV  are shown in Fig. \ref{LHCboundb}. 

From an inspection of Fig. \ref{LHCboundb} we find that  at high luminosity and center-of-mass energy the expected improvement in the limits are nearly an order of magnitude compared to the existing collider limits.  Even with only $10$ fb$^{-1}$ taken at 14 TeV we expect the limits to improve substantially from the current collider limit. At these energies and luminosities the bound on $\epsilon_S$ from the LHC will become stronger than anticipated future bounds from low-energy experiments. This conclusion is illustrated in Fig. \ref{LHCboundb}, where we also overlay the projected low-energy bounds presented in Fig. \ref{fig:constraints3}.

We expect  these projected limits can be tightened, since we have chosen hard cuts to reduce the expected leading background to below one event, at the cost of significantly cutting into the signal.  Optimizing the choice of the cut to maximize the sensitivity to the contact interactions will require including additional backgrounds, such as QCD and top quarks, and more generally, a better understanding of the systematic errors involved.  

Our analyses can certainly be improved. Our estimate of the detection times acceptance efficiency was borrowed from the estimate for a $W^\prime$ signal from \cite{CMSWprime}.  Obviously a detector simulation of the signal will provide a better estimate of this factor. The theoretical error on the signal rate will be further reduced once the 
next-leading-order QCD contributions are known. 

Finally, the interaction Lagrangian (\ref{ST-Lag}) can be generalized to include interactions of the electron with neutrinos of all flavors, with a corresponding generalization of $\epsilon_{S,T} \rightarrow \epsilon^{\alpha}_{S,T}$ where $\alpha \in \{e,\mu,\tau \}$. Because the final states with different neutrino flavors do not interfere and neither do the scalar and tensor interactions after integrating over the rapidity distributions, 
the derived and projected bounds shown in Figs. \ref{LHCbounda} and \ref{LHCboundb} now apply to the quantities$\sqrt{\sum_{\alpha} (\epsilon^{\alpha}_{S,T})^2}$.  

\subsection{Scalar resonance} 
\label{sec:resonance} 

A larger signal rate is obtained if the particle that generates the scalar interaction is kinematically accessible at the LHC.  In this case there can be a direct relationship between $\epsilon_S$ and the production cross-section and mass of the resonance, as we now demonstrate.

We assume that after electroweak symmetry breaking there is a charged scalar $\phi^+$ of mass $m$, with the following couplings to first-generation quarks and leptons:  
\begin{eqnarray} 
{\cal L} &=&  \lambda_S  V_{ud} \phi^+ \overline{u} d + \lambda_{P} V_{ud} \phi^+ \overline{u}\gamma_5 d+ \lambda_l \phi^- \overline{e} P_L \nu_e +\hbox{h.c.}~,
\label{S-Lagrangian}
\end{eqnarray} 
where $\phi^- \equiv (\phi^+)^*$.  
At low energies 
the exchange of $\phi^+$ generates  a scalar operator with
\begin{eqnarray} 
\epsilon_S & = & 2  \lambda_S \lambda_l \frac{v^2}{m^2}
\label{epsS}
\end{eqnarray} 
and a pseudoscalar operator with
\begin{eqnarray} 
 \epsilon_P & = & 2  \lambda_{P}  \lambda_l \frac{v^2}{m^2} ~.
 \end{eqnarray}

\begin{figure}
\begin{center}
\includegraphics[width=0.75\hsize] {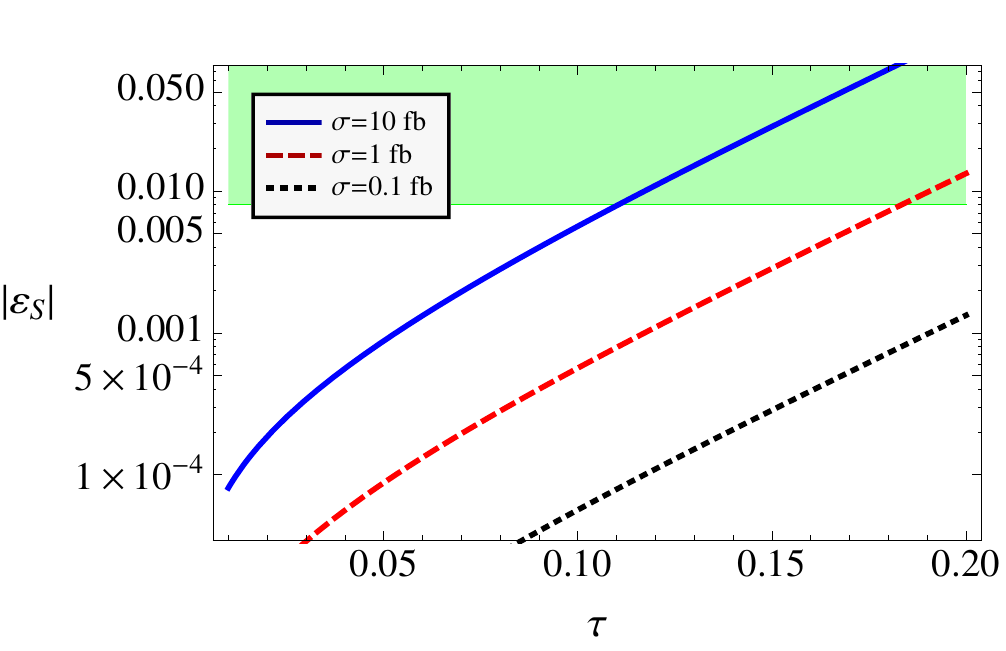}
\end{center}
\vspace{-0.75cm}
\caption{Projected lower bound on $|\epsilon_S|$ 
(at $\mu=2$~GeV), 
for a discovery cross-section of $\sigma(pp \rightarrow e +~{ \rm MET }+X)=10$  fb (blue, solid),
$1$ fb (red, dashed) and 
$0.1$ fb (black, dotted), 
as a function of $\tau =m^2/s$. Shaded region (green) shows the current experimental exclusion on $\epsilon_S$ from $0^+ \rightarrow 0^+$ nuclear $\beta$ decay.
\label{plot:eps-tau}}
\end{figure}

\begin{figure}
\begin{center}
\includegraphics[width=0.75\hsize] {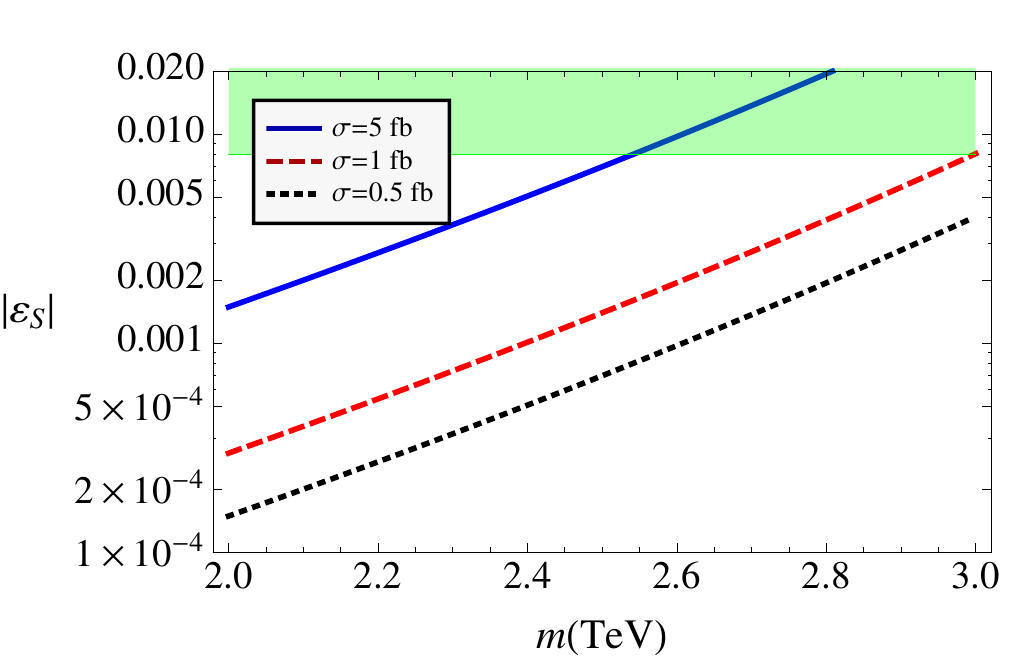}
\end{center}
\vspace{-0.75cm}
\caption{Projected lower bound on $|\epsilon_S|$ 
(at $\mu=2$~GeV)
for $\sqrt{s}=7$ TeV and a discovery cross-section of $\sigma(pp \rightarrow e +~{ \rm MET } +X)=5$ fb (blue, solid), 
$1$ fb (red, dashed) and $0.5$ fb (black, dotted).  Shaded region (green) shows the current experimental exclusion on $\epsilon_S$ from $0^+ \rightarrow 0^+$ nuclear $\beta$ decay.
The bound scales linearly with $\sigma \cdot \rm{BR}$.  
\label{plot:eps-bound1}}
\end{figure}

\begin{figure}
\begin{center}
\includegraphics[width=0.75\hsize] {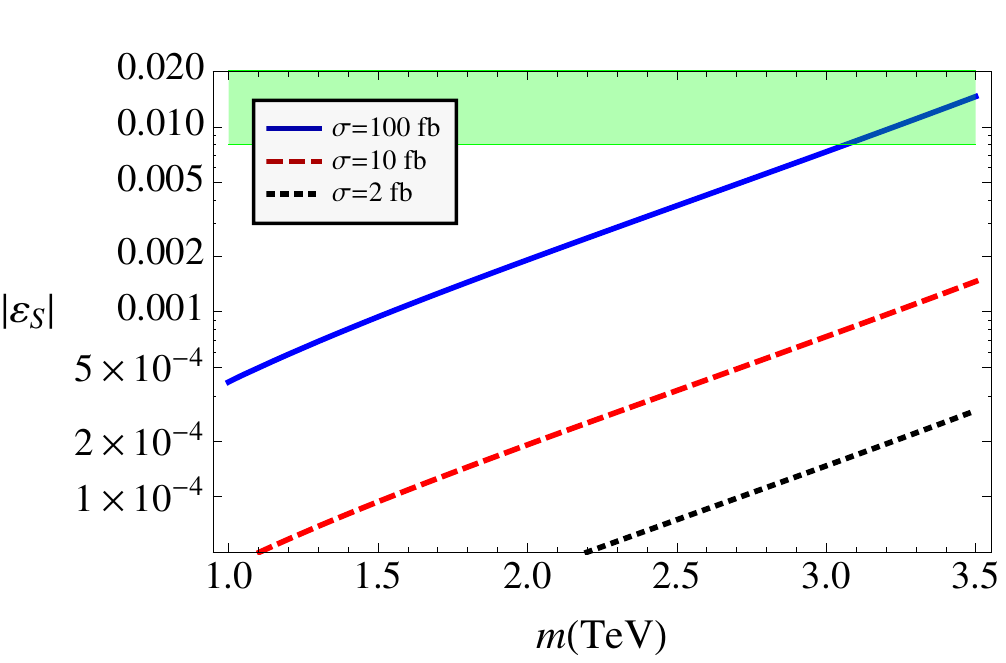}
\end{center}
\vspace{-0.75cm}
\caption{Same as Fig. \ref{plot:eps-bound1} but for $\sqrt{s}=14$ TeV and $\sigma \cdot \rm{BR} = 100$~fb (blue, solid), 10~fb (red, dashed)
and 2~fb (black, dotted). 
\label{plot:eps-bound2}}
\end{figure}

To proceed, at leading order the cross-section for the on-shell production of $\phi$, which then decays to $l \nu$  (of a given sign), is given in the narrow-width approximation by  
\begin{eqnarray} 
\sigma \cdot \rm{BR}   & = &  \lambda^2_l \left(\lambda^2_S+\lambda^2_{P} \right)  |V_{ud}|^2   \frac{m}{48  s \Gamma_\phi} L(\tau) 
\end{eqnarray}
with $\tau = m^2/s$,  $ L(\tau)=  \int^1 _\tau dx f_q(x) f_q^\prime(\tau/x)/x $, and where $\Gamma_\phi$ is the total decay width of $\phi$. 
Next, note that since $\phi$ may decay to other particles (not just to $l \nu$ and $u d$) , 
\begin{eqnarray} 
\Gamma_\phi \geq \Gamma_l + \Gamma_q &=& \left(\lambda_l^2 + 2 N_c (\lambda^2_S+\lambda^2_{P})|V_{ud}|^2 \right) \frac{m}{ 16 \pi}  
\label{naive}
\end{eqnarray} 
with $N_c=3$. 
Next note that 
$m/\Gamma_\phi  \leq  16 \pi /(\lambda_l^2 + 2 N_c (\lambda^2_S+\lambda^2_{P})|V_{ud}|^2)$, 
and then use the arithmetic-geometric inequality $\sqrt{2 N_c} \lambda_\phi \lambda_l < \frac{1}{2} (\lambda^2_l+ 2 N_c (\lambda^2_S+\lambda^2_{P})|V_{ud}|^2) $, 
where $\lambda_\phi =\sqrt{\lambda^2_S+\lambda^2_{P}}| V_{ud}|$, to finally obtain our main result of this subsection,
\ba 
\sigma \cdot \rm{BR}  & \leq & \frac{|V_{ud}|}{12  v^2} \frac{ \pi}{\sqrt{2 N_c}} \left(\sqrt{\epsilon^2_S +\epsilon^2_{P}} \right) \tau L(\tau) 
\ea
Because of the severe constraint imposed by $\pi \rightarrow e \nu$, 
the coupling $\lambda_{P}$ of $\phi$ to the pseudoscalar quark scalar density must be significantly suppressed. In the limit  $\epsilon_{P} \ll \epsilon_S$ one then has
\ba 
\label{lowerlimit-epsS}
\sigma \cdot \rm{BR}  & \leq & \frac{|V_{ud}|}{12  v^2} \frac{ \pi}{\sqrt{2 N_c}} |\epsilon_S|  \tau L(\tau) 
\ea
This expression can be rearranged to obtain a lower bound on $\epsilon_S$, that is stronger after summing in $L$ over both charged-particle final states. 
The bound  depends only on $\tau$ and $\sigma \cdot \rm{BR}$. Figure \ref{plot:eps-tau} shows the bound as a function of $\tau$ 
for several choices of $ \sigma \cdot \rm{BR}$ that will be probed by the LHC.  
Equivalent limits are shown in Figs. \ref{plot:eps-bound1} and 
\ref{plot:eps-bound2}, where we show the dependence of the bound on the mass of the resonance, 
for several values of $\sigma \cdot \rm{BR}$ and at $\sqrt{s}=7$ and 14 TeV. 
LO CTEQ6 \cite{Pumplin:2002vw} parton distribution functions are used for all these figures. 

We have shown that if a signal is observed in $pp \rightarrow e + {\rm{missing}}~{\rm{energy}}~{\rm{(MET)}}+X$, then a lower bound on $\epsilon_S$ can be obtained, provided the signal is due to the on-shell production of a scalar, which decays to an electron and missing energy provided by an electron neutrino, and whose pseudoscalar coupling to quarks is suppressed compared to its scalar coupling. That the resonance couples to an electron neutrino is important in deriving the above relation to $\epsilon_S$, since 
at linear order in the $\epsilon$'s, neutron-decay experiments do not probe 
couplings to other neutrino flavors $\epsilon_S^{\alpha \in \{\mu,\tau \}}$. 

Further confidence that the signal is due to the production of an on-shell particle can be established by the detection of an edge in the transverse electron-neutrino mass distribution, and through the detection of a resonance in dijets. 
The only additional theoretical assumption used to obtain the lower limit (\ref{lowerlimit-epsS}) is that the charged resonance $\phi$ is interpreted as a scalar and not as a vector or a tensor. Measurements of the rapidity distribution of the electron should determine the spin of $\phi$. 

If a signal is discovered in $pp \rightarrow e + {\rm{MET}} +X$ then  neutron and nuclear $\beta$-experiments will be crucial in order to pin down the properties of the resonance and of the MET. As an illustration, suppose the measured cross-section and mass  imply an $\epsilon_S$ in excess of existing neutron decay bounds, then either: (i) the resonance does not have spin 0; or (ii) as already described above, since the outgoing neutrino flavor is not identified, the relationship between the cross-section and $\epsilon_S$ can be undone simply if the scalar $\phi$ couples preferentially to muon and tau neutrinos rather than to the electron neutrino; 
or (iii) there are additional scalars at the TeV scale or a scalar contact interaction, such that partial cancellations occur in summing the multiple contributions to $\epsilon_S$.

    \section{Discussion\label{sec:discuss}}
       It is anticipated that the next generation of neutron $\beta$-decay
experiments will increase their sensitivity to BSM scalar and tensor
interactions by an order of magnitude, through improved measurements of the 
neutrino asymmetry parameter $B$ and the  Fierz interference term $b$ (see 
 Figs.~\ref{fig:constraints1}, \ref{fig:constraints1v2}, and  \ref{fig:constraints3}). 
In order to assess the impact of these future experiments, we have performed  
a  comprehensive analysis of constraints on scalar and tensor BSM interactions 
from  a broad range of  low-energy probes (neutron decay, nuclear decays, pion decays) as well as collider 
searches. 

Extracting bounds on scalar and tensor BSM  couplings from neutron and nuclear beta decays 
requires knowledge of the nucleon scalar and tensor form factors 
at zero momentum transfer.  
In this paper we have provided  the first lattice-QCD estimate of the scalar form-factor, 
 $g_{S} = 0.8 (4)$,  and a new average of existing tensor form-factor results, 
 $g_T = 1.05 (35)$.
We find that to fully exploit the increased experimental sensitivity will  
require understanding the lattice-QCD
estimates of the proton-to-neutron matrix elements at the level of
10--20\% (see Fig.~\ref{fig:constraints3}). 
To do that will require analyzing a few thousand samples at
each value of the simulation parameters using a combination of decorrelated lattices and multiple
source points on each lattice, improvements in source and sink
interpolating operators for nucleons, and simulations close to
physical light-quark masses. With the anticipated increase in computing power and 
resources,  we estimate calculations will reach this precision in 2--4 years. 

In our survey of probes of BSM scalar and tensor interactions, 
we have found that the
currently strongest bounds arise from nuclear $\beta$ decay $(\epsilon_S)$ and
radiative pion decay $(\epsilon_T)$,  probing effective scales 
$\Lambda_S > 4.7$~TeV  and $\Lambda_T >  5$~TeV, respectively.
We also find that within a specific model
for the lepton flavor structure of the scalar and tensor interactions,
significantly stronger bounds arise from $\pi \rightarrow e \nu$ decay, a
conclusion in agreement with previous literature.

We have used LHC data to obtain constraints on the scalar and tensor
interactions, finding bounds within sight of current limits obtained from
low-energy measurements (see Fig.~\ref{LHCbounda}). We have also provided a preliminary estimate of
expected future bounds from the LHC, finding that an order of magnitude
improvement should  ultimately  be achievable and that the future collider constraints 
(associated with effective scales $\Lambda_{S,T} \sim 7$~TeV)
will compete with improved neutron-decay constraints 
based on experimental sensitivities  $\delta b, \delta b_\nu  \sim 10^{-3}$ 
 (see Fig.~\ref{LHCboundb}).
Finally,  if a charged resonance decaying to
an electron plus missing energy is discovered at the LHC, we  have shown how, with
some theoretical assumptions, the production cross-section provides a
lower bound on the scalar interaction probed at low energy 
(see  Figs.~\ref{plot:eps-tau}, \ref{plot:eps-bound1}, 
and \ref{plot:eps-bound2}).

Our analysis shows that in order to compete with upcoming collider bounds on scalar and tensor interactions,  future neutron-decay experiments
should aim at the very least to sensitivities  $\delta b, \delta b_\nu \sim 10^{-3}$
in the Fierz interference term and neutrino asymmetry.
Moreover, experiments aiming for
$\delta b, \delta b_\nu \sim 10^{-4}$ would provide an unmatched
discovery potential for new scalar and tensor interactions, and therefore
should be vigorously pursued.

    \phantomsection
    \addcontentsline{toc}{section}{Acknowledgments}
    \section*{Acknowledgments}
       The lattice tensor- and scalar-charge calculations were done on Hyak cluster (NSF MRI PHY-0922770) at University of Washington. (We thank MILC and HSC for their dynamical lattices,
and LHPC and NPLQCD collaborations for the light DWF propagators).
TB and RG are supported in part by DOE grant No. DE-KA-1401020.  HWL and SDC
are supported in part by the U.S. Dept. of Energy under grant
No. DE-FG02-97ER4014. SDC was also supported by DOE grant
No. DE-FG02-91ER40676 and DE-FC02-06ER41440 and NSF grant
No. 0749300.  TB, VC, MG, and RG are supported by DOE grant No.\
DE-AC52-06NA25396 and by  the LDRD program at LANL.   
MGA was supported by the U.S. DOE contract DE-FG02-08ER41531 and by the Wisconsin Alumni Research Foundation.
The work of AF  is supported
by an FPU Grant (MICINN, Spain). This work has been supported in part
by MEC (Spain) under Grant FPA2007-60323, by the Spanish Consolider
Ingenio 2010 Programme CPAN (CSD2007-00042) and by CSIC under grant
PII-200750I026.  
We thank Joe Carlson, 
Sacha Davidson,  
Anosh Joseph, 
Greg Landsberg, 
and   Michael Ramsey-Musolf  for discussions and correspondence.

\appendix 
\section[Appendix: Operators]{\(SU(2)\times U(1)\) invariant operators contributing to charged-current processes\label{sect:operators}}
       The building blocks to construct gauge-invariant  local operators are
the  gauge fields $G_\mu^A,  \, W_\mu^a, \, B_\mu$,
corresponding to $SU(3)\times SU(2)_L \times U(1)_Y$,
the five  fermionic  gauge multiplets,
\beq
l^i =
\left(
\begin{array}{c}
\nu_L^i\\
e_L^i
\end{array}
\right)
\qquad e^i = e_R^i
\qquad q^i =
\left(
\begin{array}{c}
u_L^i \\
d_L^i
\end{array}
\right)  \qquad
u^i = u_R^i \qquad
d^i = d_R^i~,
\label{eq:fermions}
\eeq
the Higgs doublet $\varphi$
\beq
\varphi =
\left(
\begin{array}{c}
\varphi^+ \\
\varphi^0
\end{array}
\right)~,
\eeq
and the covariant derivative
\beq
D_{\mu} =   I \, \partial_\mu   \, - \, i g_s \frac{\lambda^A}{2} G_\mu^A \,
- \, i g \frac{\sigma^a}{2} W_\mu^a   \, - \, i g'  Y B_\mu~.
\eeq
In the above expression  $\lambda^A$ are the $SU(3)$ Gell-Mann matrices,
$\sigma^a$ are the $SU(2)$  Pauli matrices,  $g_s, g, g'$ are the gauge couplings
and $Y$ is the hypercharge of a given multiplet.

The minimal set of operators contributing to low-energy charged current processes  can 
be divided into two groups: four-fermion operators
\begin{subequations}
\bea
O_{l q}^{(3)}&=& (\overline{l} \gamma^\mu \sigma^a l) (\overline{q} \gamma_\mu \sigma^a q)
\\
O_{qde} &=& (\overline{\ell} e) (\overline{d} q)+ {\rm h.c.}
\\
O_{l q} &=& (\bar{l}_a e)\epsilon^{ab}(\bar{q}_b u)+ {\rm h.c.}
\\
O^t_{l q} &=& (\bar{l}_a\sigma^{\mu\nu}e)\epsilon^{ab}(\bar{q}_b\sigma_{\mu\nu}u)+ {\rm h.c.}
\eea
\end{subequations}
and vertex corrections
\begin{subequations}
\bea
O_{\varphi \varphi} &=& i(\varphi^T \epsilon  D_\mu \varphi) (\overline{u}\gamma^\mu d)+ {\rm h.c.}~,
\\
O_{\varphi q}^{(3)} &=&  \! i (\varphi^\dagger D^\mu \sigma^a \varphi)(\overline{q} \gamma_\mu \sigma^a q)+\!{\rm h.c.}.
\eea
\end{subequations}
Moreover,  the extraction of the Fermi constant from muon decay (needed for weak universality tests) 
is affected by two more operators (four-lepton and lepton-gauge vertex correction): 
\begin{subequations}
\bea
O_{ll}^{(3)} &=& \frac{1}{2} (\overline{l} \gamma^\mu \sigma^a l) (\overline{l} \gamma_\mu \sigma^a l)    
\\
O_{\varphi l}^{(3)} &=& \! i (\varphi^\dagger D^\mu \sigma^a \varphi)(\overline{l} \gamma_\mu \sigma^a l)+\!{\rm h.c.}~.
\eea
\end{subequations}

In terms of the coefficients of the above operators, the low-energy effective couplings appearing in ${\cal L}_{\rm CC}$ 
(see Eq.~\ref{eq:leffq}) are given by
\begin{subequations}
\bea
V_{ij} \cdot
 \left[v_{L}\right]_{\ell \ell i j}
&=& 2 \, V_{ij}  \,  \left[\hat{\alpha}_{\varphi l}^{(3)}\right]_{\ell\ell}   +   2 \, V_{im} \left[\hat{\alpha}_{\varphi q}^{(3)}\right]_{jm}^*
-    2\, V_{im} \left[\hat{\alpha}_{l q}^{(3)}\right]_{\ell\ell mj}   \\
V_{ij} \cdot  \left[v_R\right]_{\ell \ell ij } &=& - \left[\hat{\alpha}_{\varphi \varphi}\right]_{ij} \\
V_{ij} \cdot  \left[s_L\right]_{\ell \ell ij } &=& - \left[\hat{\alpha}_{l q}\right]_{\ell\ell ji}^* \\
V_{ij} \cdot \left[s_R\right]_{\ell \ell ij} &=& -  V_{im}\left[\hat{\alpha}_{qde}\right]_{\ell\ell jm}^*  \\
V_{ij} \cdot  \left[t_L\right]_{\ell \ell ij } &=& - \left[\hat{\alpha}^t_{l q} \right]_{\ell\ell ji}^* ~.
\eea
\end{subequations}

\section[Appendix: Details of neutron decay distribution]{Details of neutron decay distribution\label{sect:ndecaydetails}}
       The effective  Fierz interference term $\bar{b}$  and effective energy-dependent correlation coefficients $\bar{a} (E_e)$, $\bar{A} (E_e)$, $\bar{B} (E_e)$ and $\bar{C}_{(aa,aA,aB)} (E_e)$  introduced in Eq.~\ref{eq:diffn} are~\cite{Ando:2004rk,Gudkov:2005bu}:
\begin{subequations}
\bea
\bar{b}  &=&   b^{\rm SM} +   b^{\rm BSM}
\\
\bar{a} (E_e) &=&  \left( a_{\rm LO} (\tilde{\lambda}) + c_0^{(a)} +  c_1^{(a)} \frac{E_e}{M_N}  \right)  \left(1 + \frac{\alpha}{2 \pi} \delta_\alpha^{(2)} (E_e) \right)
\\
\bar{A} (E_e) &=&  \left( A_{\rm LO}(\tilde{\lambda})  + c_0^{(A)} +  c_1^{(A)} \frac{E_e}{M_N} \right)  \left(1 + \frac{\alpha}{2 \pi} \delta_\alpha^{(2)} (E_e) \right) 
\\
\bar{B} (E_e) &=&  B_{\rm LO} (\tilde{\lambda}) + c_0^{(B)} +  c_1^{(B)} \frac{E_e}{M_N}  +  \frac{m_e}{E_e}  \left( b_\nu^{\rm SM} +   b_\nu^{\rm BSM}  \right)
\\
\bar{C}_{(aa)} (E_e) &=& c_1^{(aa)}   \frac{E_e}{M_N}
\\
\bar{C}_{(aA)} (E_e) &=& c_1^{(aA)}   \frac{E_e}{M_N}
\\
\bar{C}_{(aB)} (E_e) &=& \left( c_0^{(aB)} + c_1^{(aB)}  \frac{E_e}{M_N} \right)~.
\eea
\end{subequations}
In these expressions the subscript LO indicates the well-known leading-order contributions that survive if we neglect the radiative corrections, recoil effects and new-physics contributions\footnote{In that limit, of course $\tilde{\lambda}\to \lambda$ in $a_{\rm LO},A_{\rm LO}$ and $B_{\rm LO}$.}
\bea
a_{\rm LO}(\lambda)  =  \frac{1 - \lambda^2}{1 + 3 \lambda^2} ~,
~~~~~~ A_{\rm LO} (\lambda)  =  \frac{2 \lambda (1 - \lambda)}{1 + 3 \lambda^2} ~,
~~~~~~ B_{\rm LO} (\lambda)  =   \frac{2 \lambda ( 1 + \lambda)}{1 + 3 \lambda^2}~.
\eea
As discussed in the main text,  the linear new-physics effect due to the $\epsilon_R$ coupling has been included in the replacement $\lambda \to \tilde{\lambda}= \lambda (1 - 2 \epsilon_R)$. The only other linear BSM effects in the differential distribution
 are $b^{\rm BSM}$ and $b_\nu^{\rm BSM}$, whose expressions are shown in the main text, Eqs. \eqref{eq:bbsm}.

Radiative corrections are encoded in the function $ \delta_\alpha^{(2)} (E_e)$~\cite{Ando:2004rk}, 
while recoil corrections are encoded in the coefficients $c_{0,1}^{a,A,B,aa,aA,aB}$, $b^{\rm SM}$ and $b_\nu^{\rm SM}$, whose  explicit expressions 
are~\cite{Gudkov:2005bu}
\bea
c_0  &=&  -  \frac{2 \lambda (\lambda + \mu_V)}{1 + 3 \lambda^2}  
\  \frac{E_0}{M_N} 
 \\
c_1 & = &    \frac{3 + 4 \lambda \mu_V + 9 \lambda^2}{1 + 3 \lambda^2}  
\\
c_0^{(a)} & = & 
 \frac{2 \lambda (\lambda + \mu_V)}{1 + 3 \lambda^2}  
\  \frac{E_0}{M_N} 
\\
c_1^{(a)} &=& 
 - \frac{4 \lambda (3 \lambda + \mu_V)}{1 + 3 \lambda^2}  
 \\
 c_0^{(A)} & = & 
 \frac{( \lambda - 1)  (\lambda + \mu_V)}{1 + 3 \lambda^2}  
\  \frac{E_0}{M_N} 
\\
c_1^{(A)} &=& 
 \frac{ \mu_V (1 - 3 \lambda) + \lambda (7 - 5 \lambda)}{1 + 3 \lambda^2}  
 \\
  c_0^{(B)} & = & 
-  \frac{ 2 \lambda  (\lambda + \mu_V)}{1 + 3 \lambda^2}  
\  \frac{E_0}{M_N} 
\\
c_1^{(B)} &=& 
 \frac{ \mu_V (1 + 3 \lambda) + \lambda (5 + 7 \lambda)}{1 + 3 \lambda^2}  
\\
  c_0^{(aB)} & = & 
 \frac{ (1+ \lambda)  (\lambda + \mu_V)}{1 + 3 \lambda^2}  
\  \frac{E_0}{M_N} 
\\
c_1^{(aB)} &=& 
-  \frac{( \mu_V  + 7 \lambda)  (1 + \lambda)}{1 + 3 \lambda^2}  
\\
c_1^{(aa)} &=& 
-  \frac{3 (1 - \lambda^2)}{1 + 3 \lambda^2}  
\\
c_1^{(aA)} &=& 
 \frac{(\lambda - 1)  (\mu_V + 5  \lambda)}{1 + 3 \lambda^2}  
\\
b^{\rm SM} &=&  -  \frac{m_e}{M_N} \frac{1 +  2 \, \mu_V \lambda + \lambda^2}{1 + 3 \lambda^2}  
\\
b_\nu ^{\rm SM} &=&  -  \frac{m_e}{M_N} \frac{ (1 + \lambda)  (\mu_V + \lambda)}{1 + 3 \lambda^2} 
\eea
In the above relations  $\mu_V$ represents the difference between the proton and neutron magnetic moments.
Numerically,  one has  $b^{\rm SM} = - 1.35(1) \times 10^{-3}$ and 
$b_\nu^{\rm SM} = - 1.27(1) \times 10^{-3}$.

    \phantomsection
    \addcontentsline{toc}{section}{Bibliography}
    \bibliographystyle{doiplain}
    \let\oldnewblock=\newblock
    \newcommand\dispatcholdnewblock[1]{\oldnewblock{#1}}
    \renewcommand\newblock{\spaceskip=0.3emplus0.3emminus0.2em\relax
                           \xspaceskip=0.3emplus0.6emminus0.1em\relax
                           \hskip0ptplus0.5emminus0.2em\relax
                           {\catcode`\.=\active
                           \expandafter}\dispatcholdnewblock}
    \bibliography{Vincenzo,Huey-Wen,Michael}{}

\begin{thebibliography}{100}

\bibitem{Weinberg:2009zz}
Steven Weinberg.
\newblock {V-A was the key}.
\newblock {\em J.Phys.Conf.Ser.}, 196:012002, 2009,
  \doi{10.1088/1742-6596/196/1/012002}.

\bibitem{Severijns2006dr}
Nathal Severijns, Marcus Beck, and Oscar Naviliat-Cuncic.
\newblock {Tests of the standard electroweak model in beta decay}.
\newblock {\em Rev. Mod. Phys.}, 78:991\unskip--\ignorespaces 1040, 2006,
  \doi{10.1103/RevModPhys.78.991}, \eprint{arXiv}{nucl-ex/0605029}.

\bibitem{Abele:2008zz}
Hartmut Abele.
\newblock {The neutron. Its properties and basic interactions}.
\newblock {\em Prog.Part.Nucl.Phys.}, 60:1\unskip--\ignorespaces 81, 2008,
  \doi{10.1016/j.ppnp.2007.05.002}.

\bibitem{Dubbers:2011ns}
Dirk Dubbers and Michael~G. Schmidt.
\newblock {The neutron and its role in cosmology and particle physics}.
\newblock 2011, \eprint{arXiv}{1105.3694}.

\bibitem{Dewey2009189}
M.~Dewey, K.~Coakley, D.~Gilliam, G.~Greene, A.~Laptev, J.~Nico, W.~Snow,
  F.~Wietfeldt, and A.~Yue.
\newblock Prospects for a new cold neutron beam measurement of the neutron
  lifetime.
\newblock {\em Nucl.Instr.Meth.}, A611(2\unskip--\ignorespaces 3):189
  \unskip--\ignorespaces  192, 2009, \doi{10.1016/j.nima.2009.07.069}.

\bibitem{Arzumanov2009186}
S.~Arzumanov, L.~Bondarenko, P.~Geltenbort, V.~Morozov, V.V. Nesvizhevsky, Yu.
  Panin, and A.~Strepetov.
\newblock A new project to measure the neutron lifetime using storage of
  ultracold neutrons and detection of inelastically scattered neutrons.
\newblock {\em Nucl.Instr.Meth.}, A611(2\unskip--\ignorespaces 3):186
  \unskip--\ignorespaces  188, 2009, \doi{10.1016/j.nima.2009.07.070}.

\bibitem{Walstrom200982}
P.L. Walstrom, J.D. Bowman, S.I. Penttila, C.~Morris, and A.~Saunders.
\newblock A magneto-gravitational trap for absolute measurement of the
  ultra-cold neutron lifetime.
\newblock {\em Nucl.Instr.Meth.}, A599(1):82 \unskip--\ignorespaces  92, 2009,
  \doi{10.1016/j.nima.2008.11.010}.

\bibitem{Materne2009176}
S.~Materne, R.~Picker, I.~Altarev, H.~Angerer, B.~Franke, E.~Gutsmiedl, F.J.
  Hartmann, A.R. M{\"u}ller, S.~Paul, and R.~Stoepler.
\newblock {PENeLOPE}---on the way towards a new neutron lifetime experiment
  with magnetic storage of ultra-cold neutrons and proton extraction.
\newblock {\em Nucl.Instr.Meth.}, A611(2\unskip--\ignorespaces 3):176
  \unskip--\ignorespaces  180, 2009, \doi{10.1016/j.nima.2009.07.055}.

\bibitem{Leung2009181}
K.K.H. Leung and O.~Zimmer.
\newblock Proposed neutron lifetime measurement using a hybrid magnetic trap
  for ultra-cold neutrons.
\newblock {\em Nucl.Instr.Meth.}, A611(2\unskip--\ignorespaces 3):181
  \unskip--\ignorespaces  185, 2009, \doi{10.1016/j.nima.2009.07.087}.

\bibitem{PERKEOIII:2009}
B.~Markisch et~al.
\newblock {The new neutron decay spectrometer {P\sc erkeo III}}.
\newblock {\em Nucl.Instrum.Meth.}, A611:216\unskip--\ignorespaces 218, 2009,
  \doi{10.1016/j.nima.2009.07.066}.

\bibitem{Plaster:2008si}
B.~Plaster, R.~Carr, B.W. Filippone, D.~Harrison, J.~Hsiao, et~al.
\newblock {A Solenoidal electron spectrometer for a precision measurement of
  the neutron {$\beta$}-asymmetry with ultracold neutrons}.
\newblock {\em Nucl.Instrum.Meth.}, A595:587\unskip--\ignorespaces 598, 2008,
  \doi{10.1016/j.nima.2008.07.143}, \eprint{arXiv}{0806.2097}.

\bibitem{abBA}
R.~Alarcon et~al.
\newblock {Precise Measurement of Neutron Decay Parameters}, 2007,
  \url{http://nab.phys.virginia.edu/abba\_proposal\_2007.pdf}.

\bibitem{Dubbers:2007st}
D.~Dubbers, H.~Abele, S.~Baessler, B.~Maerkisch, M.~Schumann, et~al.
\newblock {A Clean, bright, and versatile source of neutron decay products}.
\newblock {\em Nucl.Instrum.Meth.}, A596:238\unskip--\ignorespaces 247, 2008,
  \doi{10.1016/j.nima.2008.07.157}, \eprint{arXiv}{0709.4440}.

\bibitem{WilburnUCNB}
W.S. Wilburn et~al.
\newblock {Measurement of the neutrino-spin correlation Parameter {$b$} in
  neutron decay using ultracold neutrons}.
\newblock {\em Rev. Mex. Fis.}, Suppl. 55(2):119, 2009.

\bibitem{Pocanic:2008pu}
Dinko Pocanic et~al., Nab Collaboration.
\newblock {Nab: Measurement Principles, Apparatus and Uncertainties}.
\newblock {\em Nucl.Instrum.Meth.}, A611:211\unskip--\ignorespaces 215, 2009,
  \doi{10.1016/j.nima.2009.07.065}, \eprint{arXiv}{0810.0251}.

\bibitem{aSPECT:2008}
S.~S.~Bae{\ss}ler et~al.
\newblock {First measurements with the neutron decay spectrometer aSPECT}.
\newblock {\em Eur.Phys.J.}, A38:17\unskip--\ignorespaces 26, 2008,
  \doi{10.1140/epja/i2008-10660-0}.

\bibitem{Wietfeldt:2005wz}
F.E. Wietfeldt, B.M. Fisher, C.~Trull, G.L. Jones, B.~Collet, et~al.
\newblock {A method for an improved measurement of the electron--antineutrino
  correlation in free neutron beta decay}.
\newblock {\em Nucl.Instrum.Meth.}, A545:181\unskip--\ignorespaces 193, 2005,
  \doi{10.1016/j.nima.2005.01.339}.

\bibitem{UCNb}
K.~P. Hickerson.
\newblock {The Fierz Interference Term in Beta-Decay Spectrum of UCN}, 2009,
  \url{http://neutron.physics.ncsu.edu/UCN\_Workshop\_09/Hickerson\_SantaFe\_2%
009.pdf}.
\newblock UCN Workshop, November 6--7 2009, Santa Fe, New Mexico.

\bibitem{Profumo:2006yu}
Stefano Profumo, Michael~J. Ramsey-Musolf, and Sean Tulin.
\newblock {Supersymmetric contributions to weak decay correlation
  coefficients}.
\newblock {\em Phys.Rev.}, D75:075017, 2007, \doi{10.1103/PhysRevD.75.075017},
  \eprint{arXiv}{hep-ph/0608064}.

\bibitem{Herczeg2001vk}
P.~Herczeg.
\newblock {Beta decay beyond the standard model}.
\newblock {\em Prog. Part. Nucl. Phys.}, 46:413\unskip--\ignorespaces 457,
  2001, \doi{10.1016/S0146-6410(01)00149-1}.

\bibitem{Cirigliano:2009wk}
Vincenzo Cirigliano, James Jenkins, and Martin Gonzalez-Alonso.
\newblock {Semileptonic decays of light quarks beyond the Standard Model}.
\newblock {\em Nucl.Phys.}, B830:95\unskip--\ignorespaces 115, 2010,
  \doi{10.1016/j.nuclphysb.2009.12.020}, \eprint{arXiv}{0908.1754}.

\bibitem{Buchmuller:1985jz}
W.~Buchmuller and D.~Wyler.
\newblock {Effective Lagrangian Analysis of New Interactions and Flavor
  Conservation}.
\newblock {\em Nucl. Phys.}, B268:621, 1986,
  \doi{10.1016/0550-3213(86)90262-2}.

\bibitem{Carpentier:2010ue}
Michael Carpentier and Sacha Davidson.
\newblock {Constraints on two-lepton, two quark operators}.
\newblock {\em Eur.Phys.J.}, C70:1071\unskip--\ignorespaces 1090, 2010,
  \doi{10.1140/epjc/s10052-010-1482-4}, \eprint{arXiv}{1008.0280}.

\bibitem{Voloshin:1992sn}
M.B. Voloshin.
\newblock {Upper bound on tensor interaction in the decay $\pi \to e \nu
  \gamma$}.
\newblock {\em Phys.Lett.}, B283:120\unskip--\ignorespaces 122, 1992,
  \doi{10.1016/0370-2693(92)91439-G}.

\bibitem{Herczeg:1994ur}
P.~Herczeg.
\newblock {On the question of a tensor interaction in $\pi \to e \nu \gamma$
  decay}.
\newblock {\em Phys.Rev.}, D49:247\unskip--\ignorespaces 253, 1994,
  \doi{10.1103/PhysRevD.49.247}.

\bibitem{Campbell:2003ir}
Bruce~A. Campbell and David~W. Maybury.
\newblock {Constraints on scalar couplings from $\pi \to e \nu$}.
\newblock {\em Nucl.Phys.}, B709:419\unskip--\ignorespaces 439, 2005,
  \doi{10.1016/j.nuclphysb.2004.12.015}, \eprint{arXiv}{hep-ph/0303046}.

\bibitem{Weinberg:1958ut}
Steven Weinberg.
\newblock {Charge symmetry of weak interactions}.
\newblock {\em Phys.Rev.}, 112:1375\unskip--\ignorespaces 1379, 1958,
  \doi{10.1103/PhysRev.112.1375}.

\bibitem{Holstein:1974zf}
Barry~R. Holstein.
\newblock {Recoil Effects in Allowed beta Decay: The Elementary Particle
  Approach}.
\newblock {\em Rev. Mod. Phys.}, 46:789, 1974, \doi{10.1103/RevModPhys.46.789}.

\bibitem{Ademollo:1964sr}
M.~Ademollo and Raoul Gatto.
\newblock {Nonrenormalization Theorem for the Strangeness Violating Vector
  Currents}.
\newblock {\em Phys.Rev.Lett.}, 13:264\unskip--\ignorespaces 265, 1964,
  \doi{10.1103/PhysRevLett.13.264}.

\bibitem{Donoghue:1990ti}
John~F. Donoghue and D.~Wyler.
\newblock {Isospin breaking and the precise determination of $V_{ud}$}.
\newblock {\em Phys.Lett.}, B241:243, 1990, \doi{10.1016/0370-2693(90)91287-L}.

\bibitem{Lee:1956qn}
T.D. Lee and Chen-Ning Yang.
\newblock {Question of Parity Conservation in Weak Interactions}.
\newblock {\em Phys.Rev.}, 104:254\unskip--\ignorespaces 258, 1956,
  \doi{10.1103/PhysRev.104.254}.

\bibitem{Jackson1957zz}
J.~D. Jackson, S.~B. Treiman, and H.~W. Wyld.
\newblock {Possible tests of time reversal invariance in Beta decay}.
\newblock {\em Phys. Rev.}, 106:517\unskip--\ignorespaces 521, 1957,
  \doi{10.1103/PhysRev.106.517}.

\bibitem{VCMGMGA}
Vincenzo Cirigliano, Martin Gonzalez-Alonso, and Michael~L. Graesser, 2011.
\newblock Work in preparation.

\bibitem{Wilkinson:1982hu}
Denys~H. Wilkinson.
\newblock {Analysis of neutron beta decay}.
\newblock {\em Nucl.Phys.}, A377:474\unskip--\ignorespaces 504, 1982,
  \doi{10.1016/0375-9474(82)90051-3}.

\bibitem{Gluck:1995hs}
F.~Gluck, I.~Joo, and J.~Last.
\newblock {Measurable parameters of neutron decay}.
\newblock {\em Nucl.Phys.}, A593:125\unskip--\ignorespaces 150, 1995,
  \doi{10.1016/0375-9474(95)00354-4}.

\bibitem{Ando:2004rk}
S.~Ando et~al.
\newblock {Neutron beta decay in effective field theory}.
\newblock {\em Phys. Lett.}, B595:250\unskip--\ignorespaces 259, 2004,
  \doi{10.1016/j.physletb.2004.06.037}, \eprint{arXiv}{nucl-th/0402100}.

\bibitem{Gudkov:2005bu}
Vladimir~P. Gudkov, G.L. Greene, and J.R. Calarco.
\newblock {General classification and analysis of neutron {$\beta$}-decay
  experiments}.
\newblock {\em Phys.Rev.}, C73:035501, 2006, \doi{10.1103/PhysRevC.73.035501},
  \eprint{arXiv}{nucl-th/0510012}.

\bibitem{Czarnecki:2004cw}
Andrzej Czarnecki, William~J. Marciano, and Alberto Sirlin.
\newblock {Precision measurements and {CKM} unitarity}.
\newblock {\em Phys.Rev.}, D70:093006, 2004, \doi{10.1103/PhysRevD.70.093006},
  \eprint{arXiv}{hep-ph/0406324}.

\bibitem{Gardner:2000nk}
Susan Gardner and C.~Zhang.
\newblock {Sharpening low-energy, standard model tests via correlation
  coefficients in neutron beta decay}.
\newblock {\em Phys.Rev.Lett.}, 86:5666\unskip--\ignorespaces 5669, 2001,
  \doi{10.1103/PhysRevLett.86.5666}, \eprint{arXiv}{hep-ph/0012098}.

\bibitem{Sjue:2005ks}
S.K.L. Sjue.
\newblock {Polarized neutron beta-decay: The Proton asymmetry and recoil-order
  currents}.
\newblock {\em Phys.Rev.}, C72:045501, 2005, \doi{10.1103/PhysRevC.72.045501,
  10.1103/PhysRevC.83.019901}, \eprint{arXiv}{nucl-th/0507041}.

\bibitem{ayoung}
Albert Young, 2011.
\newblock Private communication.

\bibitem{Abele:2002wc}
H.~Abele, M.~Astruc~Hoffmann, S.~Baessler, D.~Dubbers, F.~Gluck, et~al.
\newblock {Is the unitarity of the quark mixing CKM matrix violated in neutron
  {$\beta$}-decay?}
\newblock {\em Phys.Rev.Lett.}, 88:211801, 2002,
  \doi{10.1103/PhysRevLett.88.211801}, \eprint{arXiv}{hep-ex/0206058}.

\bibitem{Liu:2010ms}
J.~Liu et~al., UCNA Collaboration.
\newblock {Determination of the Axial-Vector Weak Coupling Constant with
  Ultracold Neutrons}.
\newblock {\em Phys.Rev.Lett.}, 105:181803, 2010,
  \doi{10.1103/PhysRevLett.105.181803}, \eprint{arXiv}{1007.3790}.

\bibitem{Schumann:2007qe}
M.~Schumann, T.~Soldner, M.~Deissenroth, F.~Gluck, J.~Krempel, et~al.
\newblock {Measurement of the neutrino asymmetry parameter {$B$} in neutron
  decay}.
\newblock {\em Phys.Rev.Lett.}, 99:191803, 2007,
  \doi{10.1103/PhysRevLett.99.191803}, \eprint{arXiv}{0706.3788}.

\bibitem{Byrne:2002tx}
J.~Byrne, P.G. Dawber, M.G.D. van~der Grinten, C.G. Habeck, F.~Shaikh, et~al.
\newblock {Determination of the electron anti-neutrino angular correlation
  coefficient {$a_0$} and the parameter {$|\lambda|=|G_A/G_V|$} in free neutron
  {$\beta$-decay} from measurements of the integrated energy spectrum of recoil
  protons stored in an ion trap}.
\newblock {\em J.Phys.}, G28:1325\unskip--\ignorespaces 1349, 2002,
  \doi{10.1088/0954-3899/28/6/314}.

\bibitem{Konrad:2010wz}
G.~Konrad, W.~Heil, S.~Baessler, D.~Pocanic, and F.~Gluck.
\newblock {Impact of Neutron Decay Experiments on non-Standard Model Physics}.
\newblock 2010, \eprint{arXiv}{1007.3027}.

\bibitem{Serebrov:2004zf}
A.~Serebrov, V.~Varlamov, A.~Kharitonov, A.~Fomin, Yu. Pokotilovski, et~al.
\newblock {Measurement of the neutron lifetime using a gravitational trap and a
  low-temperature Fomblin coating}.
\newblock {\em Phys.Lett.}, B605:72\unskip--\ignorespaces 78, 2005,
  \doi{10.1016/j.physletb.2004.11.013}, \eprint{arXiv}{nucl-ex/0408009}.

\bibitem{Pichlmaier:2010zz}
A.~Pichlmaier, V.~Varlamov, K.~Schreckenbach, and P.~Geltenbort.
\newblock {Neutron lifetime measurement with the UCN trap-in-trap MAMBO II}.
\newblock {\em Phys.Lett.}, B693:221\unskip--\ignorespaces 226, 2010,
  \doi{10.1016/j.physletb.2010.08.032}.

\bibitem{Marciano:2005ec}
William~J. Marciano and Alberto Sirlin.
\newblock {Improved calculation of electroweak radiative corrections and the
  value of {$V_{ud}$}}.
\newblock {\em Phys.Rev.Lett.}, 96:032002, 2006,
  \doi{10.1103/PhysRevLett.96.032002}, \eprint{arXiv}{hep-ph/0510099}.

\bibitem{Hardy:2008gy}
J.C. Hardy and I.S. Towner.
\newblock {Superallowed $0^+ \rightarrow 0^+$ nuclear {$\beta$} decays: A New
  survey with precision tests of the conserved vector current hypothesis and
  the standard model}.
\newblock {\em Phys.Rev.}, C79:055502, 2009, \doi{10.1103/PhysRevC.79.055502},
  \eprint{arXiv}{0812.1202}.

\bibitem{Bychkov:2008ws}
M.~Bychkov, D.~Pocanic, B.A. VanDevender, V.A. Baranov, Wilhelm~H. Bertl,
  et~al.
\newblock {New Precise Measurement of the Pion Weak Form Factors in $\pi \to e
  \nu \gamma$ Decay}.
\newblock {\em Phys.Rev.Lett.}, 103:051802, 2009,
  \doi{10.1103/PhysRevLett.103.051802}, \eprint{arXiv}{0804.1815}.

\bibitem{Mateu:2007tr}
V.~Mateu and J.~Portoles.
\newblock {Form-factors in radiative pion decay}.
\newblock {\em Eur.Phys.J.}, C52:325\unskip--\ignorespaces 338, 2007,
  \doi{10.1140/epjc/s10052-007-0393-5}, \eprint{arXiv}{0706.1039}.

\bibitem{Wauters:2010gh}
F.~Wauters, I.S. Kraev, D.~Zakoucky, M.~Beck, M.~Breitenfeldt, et~al.
\newblock {Precision measurements of the $^{60}$Co $\beta$-asymmetry parameter
  in search for tensor currents in weak interactions}.
\newblock {\em Phys.Rev.}, C82:055502, 2010, \doi{10.1103/PhysRevC.82.055502},
  \eprint{arXiv}{1005.5034}.

\bibitem{Wauters:2009jw}
F.~Wauters, I.~Kraev, M.~Tandecki, E.~Traykov, S.~Van~Gorp, et~al.
\newblock {{$\beta$} asymmetry parameter in the decay of {${}^{114}$In}}.
\newblock {\em Phys.Rev.}, C80:062501, 2009, \doi{10.1103/PhysRevC.80.062501},
  \eprint{arXiv}{0901.0081}.

\bibitem{Carnoy:1991jd}
A.S. Carnoy, J.~Deutsch, T.A. Girard, and R.~Prieels.
\newblock {Limits on nonstandard weak currents from the polarization of
  {${}^{14}$O} and {${}^{10}$C} decay positrons}.
\newblock {\em Phys.Rev.}, C43:2825\unskip--\ignorespaces 2834, 1991,
  \doi{10.1103/PhysRevC.43.2825}.

\bibitem{Wichers:1986es}
V.A. Wichers, T.R. Hageman, J.~Van~Klinken, H.W. Wilschut, and D.~Atkinson.
\newblock {Bounds on Right-handed Currents from Nuclear Beta Decay}.
\newblock {\em Phys.Rev.Lett.}, 58:1821\unskip--\ignorespaces 1824, 1987,
  \doi{10.1103/PhysRevLett.58.1821}.

\bibitem{Severijns:2000vc}
N.~Severijns et~al.
\newblock {Fundamental Weak Interaction Studies using Polarised Nuclei and Ion
  Traps}.
\newblock {\em Hyperfine Interactions}, 129:223\unskip--\ignorespaces 236,
  2000, \doi{10.1023/A:1012665917625}.

\bibitem{Vetter:2008zz}
P.A. Vetter, J.R. Abo-Shaeer, S.J. Freedman, and R.~Maruyama.
\newblock {Measurement of the {$\beta$-$\nu$} correlation of {${}^{21}$Na}
  using shakeoff electrons}.
\newblock {\em Phys.Rev.}, C77:035502, 2008, \doi{10.1103/PhysRevC.77.035502},
  \eprint{arXiv}{0805.1212}.

\bibitem{Gorelov:2004hv}
A.~Gorelov, D.~Melconian, W.P. Alford, D.~Ashery, G.~Ball, et~al.
\newblock {Scalar interaction limits from the {$\beta$-$\nu$} correlation of
  trapped radioactive atoms}.
\newblock {\em Phys.Rev.Lett.}, 94:142501, 2005,
  \doi{10.1103/PhysRevLett.94.142501}, \eprint{arXiv}{nucl-ex/0412032}.

\bibitem{Adelberger:1999ud}
E.G. Adelberger et~al., ISOLDE Collaboration.
\newblock {Positron neutrino correlation in the $0^+ \to 0^+$ decay of
  {${}^{32}$Ar}}.
\newblock {\em Phys.Rev.Lett.}, 83:1299\unskip--\ignorespaces 1302, 1999,
  \doi{10.1103/PhysRevLett.83.1299}, \eprint{arXiv}{nucl-ex/9903002}.

\bibitem{Johnson:1963zza}
C.H. Johnson, Frances Pleasonton, and T.A. Carlson.
\newblock {Precision Measurement of the Recoil Energy Spectrum from the Decay
  of {${}^6$He}}.
\newblock {\em Phys.Rev.}, 132:1149\unskip--\ignorespaces 1165, 1963,
  \doi{10.1103/PhysRev.132.1149}.

\bibitem{Britton:1992pg}
D.I. Britton, S.~Ahmad, D.A. Bryman, R.A. Burnbam, E.T.H. Clifford, et~al.
\newblock {Measurement of the $\pi^+ \to e^+ \nu$ branching ratio}.
\newblock {\em Phys.Rev.Lett.}, 68:3000\unskip--\ignorespaces 3003, 1992,
  \doi{10.1103/PhysRevLett.68.3000}.

\bibitem{Czapek:1993kc}
G.~Czapek, A.~Federspiel, A.~Fluckiger, D.~Frei, B.~Hahn, et~al.
\newblock {Branching ratio for the rare pion decay into positron and neutrino}.
\newblock {\em Phys.Rev.Lett.}, 70:17\unskip--\ignorespaces 20, 1993,
  \doi{10.1103/PhysRevLett.70.17}.

\bibitem{Cirigliano:2007ga}
Vincenzo Cirigliano and Ignasi Rosell.
\newblock {{$\pi/K \to e \bar \nu_e$} branching ratios to {$O(e^2 p^4)$} in
  Chiral Perturbation Theory}.
\newblock {\em JHEP}, 0710:005, 2007, \doi{10.1088/1126-6708/2007/10/005},
  \eprint{arXiv}{0707.4464}.

\bibitem{Cirigliano:2007xi}
Vincenzo Cirigliano and Ignasi Rosell.
\newblock {Two-loop effective theory analysis of {$\pi(K) \to e \bar \nu_e
  [\gamma]$} branching ratios}.
\newblock {\em Phys.Rev.Lett.}, 99:231801, 2007,
  \doi{10.1103/PhysRevLett.99.231801}, \eprint{arXiv}{0707.3439}.

\bibitem{Laiho:2009eu}
Jack Laiho, Enrico Lunghi, and Ruth~S. Van~de Water.
\newblock {Lattice QCD inputs to the CKM unitarity triangle analysis}.
\newblock {\em Phys. Rev.}, D81:034503, 2010, \doi{10.1103/PhysRevD.81.034503},
  \eprint{arXiv}{0910.2928}.

\bibitem{Sachrajda:2011tg}
Christopher Sachrajda.
\newblock {Phenomenology from the Lattice}.
\newblock {\em PoS}, LATTICE2010:018, 2010, \eprint{arXiv}{1103.5959}.

\bibitem{Colangelo:2010et}
Gilberto Colangelo, Stephan Durr, Andreas Juttner, Laurent Lellouch, Heinrich
  Leutwyler, et~al.
\newblock {Review of lattice results concerning low energy particle physics}.
\newblock {\em Eur.Phys.J.}, C71:1695, 2011,
  \doi{10.1140/epjc/s10052-011-1695-1}, \eprint{arXiv}{1011.4408}.

\bibitem{Durr:2008zz}
S.~Durr, Z.~Fodor, J.~Frison, C.~Hoelbling, R.~Hoffmann, et~al.
\newblock {Ab Initio Determination of Light Hadron Masses}.
\newblock {\em Science}, 322:1224\unskip--\ignorespaces 1227, 2008,
  \doi{10.1126/science.1163233}, \eprint{arXiv}{0906.3599}.

\bibitem{Durr:2011ap}
S.~Durr et~al.
\newblock {Precision computation of the kaon bag parameter}.
\newblock 2011, \eprint{arXiv}{1106.3230}.

\bibitem{Kaplan:1992bt}
David~B. Kaplan.
\newblock {A Method for simulating chiral fermions on the lattice}.
\newblock {\em Phys.Lett.}, B288:342\unskip--\ignorespaces 347, 1992,
  \doi{10.1016/0370-2693(92)91112-M}, \eprint{arXiv}{hep-lat/9206013}.

\bibitem{Kaplan:1992sg}
D.B. Kaplan.
\newblock {Chiral fermions on the lattice}.
\newblock {\em Nucl.Phys.Proc.Suppl.}, 30:597\unskip--\ignorespaces 600, 1993,
  \doi{10.1016/0920-5632(93)90282-B}.

\bibitem{Shamir:1993zy}
Yigal Shamir.
\newblock {Chiral fermions from lattice boundaries}.
\newblock {\em Nucl.Phys.}, B406:90\unskip--\ignorespaces 106, 1993,
  \doi{10.1016/0550-3213(93)90162-I}, \eprint{arXiv}{hep-lat/9303005}.

\bibitem{Furman:1994ky}
Vadim Furman and Yigal Shamir.
\newblock {Axial symmetries in lattice QCD with Kaplan fermions}.
\newblock {\em Nucl.Phys.}, B439:54\unskip--\ignorespaces 78, 1995,
  \doi{10.1016/0550-3213(95)00031-M}, \eprint{arXiv}{hep-lat/9405004}.

\bibitem{Neuberger:1997fp}
Herbert Neuberger.
\newblock {Exactly massless quarks on the lattice}.
\newblock {\em Phys.Lett.}, B417:141\unskip--\ignorespaces 144, 1998,
  \doi{10.1016/S0370-2693(97)01368-3}, \eprint{arXiv}{hep-lat/9707022}.

\bibitem{Sheikholeslami:1985ij}
B.~Sheikholeslami and R.~Wohlert.
\newblock {Improved Continuum Limit Lattice Action for QCD with Wilson
  Fermions}.
\newblock {\em Nucl.Phys.}, B259:572, 1985, \doi{10.1016/0550-3213(85)90002-1}.

\bibitem{Frezzotti:2000nk}
Roberto Frezzotti, Pietro~Antonio Grassi, Stefan Sint, and Peter Weisz, Alpha
  collaboration.
\newblock {Lattice QCD with a chirally twisted mass term}.
\newblock {\em JHEP}, 0108:058, 2001, \doi{10.1088/1126-6708/2001/08/058},
  \eprint{arXiv}{hep-lat/0101001}.

\bibitem{Naik:1986bn}
Satchidananda Naik.
\newblock {On-shell Improved Lattice Action for {QCD} with {S}usskind Fermions
  and Asymptotic Freedom Scale}.
\newblock {\em Nucl.Phys.}, B316:238, 1989, \doi{10.1016/0550-3213(89)90394-5}.

\bibitem{Orginos:1998ue}
Kostas Orginos and Doug Toussaint, MILC collaboration.
\newblock {Testing improved actions for dynamical Kogut-Susskind quarks}.
\newblock {\em Phys.Rev.}, D59:014501, 1999, \doi{10.1103/PhysRevD.59.014501},
  \eprint{arXiv}{hep-lat/9805009}.

\bibitem{Follana:2006rc}
E.~Follana et~al., HPQCD.
\newblock {Highly Improved Staggered Quarks on the Lattice, with Applications
  to Charm Physics}.
\newblock {\em Phys. Rev.}, D75:054502, 2007, \doi{10.1103/PhysRevD.75.054502},
  \eprint{arXiv}{hep-lat/0610092}.

\bibitem{Khan:2006de}
A.Ali Khan, M.~Gockeler, Ph. Hagler, T.R. Hemmert, R.~Horsley, et~al.
\newblock {Axial coupling constant of the nucleon for two flavours of dynamical
  quarks in finite and infinite volume}.
\newblock {\em Phys.Rev.}, D74:094508, 2006, \doi{10.1103/PhysRevD.74.094508},
  \eprint{arXiv}{hep-lat/0603028}.

\bibitem{Pleiter:2011gw}
D.~Pleiter et~al., QCDSF/UKQCD Collaboration.
\newblock {Nucleon form factors and structure functions from {\(N_f=2\)} Clover
  fermions}.
\newblock {\em PoS}, LATTICE2010:153, 2010, \eprint{arXiv}{1101.2326}.

\bibitem{Brandt:2011sj}
B.B. Brandt, S.~Capitani, M.~Della~Morte, D.~Djukanovic, J.~Gegelia, et~al.
\newblock {Form factors in lattice QCD}.
\newblock 2011, \eprint{arXiv}{1106.1554}.

\bibitem{Alexandrou:2010hf}
C.~Alexandrou et~al., ETM Collaboration.
\newblock {Axial Nucleon form factors from lattice QCD}.
\newblock {\em Phys.Rev.}, D83:045010, 2011, \doi{10.1103/PhysRevD.83.045010},
  \eprint{arXiv}{1012.0857}.

\bibitem{Lin:2008uz}
Huey-Wen Lin, Tom Blum, Shigemi Ohta, Shoichi Sasaki, and Takeshi Yamazaki.
\newblock {Nucleon structure with two flavors of dynamical domain-wall
  fermions}.
\newblock {\em Phys.Rev.}, D78:014505, 2008, \doi{10.1103/PhysRevD.78.014505},
  \eprint{arXiv}{0802.0863}.

\bibitem{Yamazaki:2008py}
T.~Yamazaki et~al., RBC+UKQCD Collaboration.
\newblock {Nucleon axial charge in 2+1 flavor dynamical lattice QCD with domain
  wall fermions}.
\newblock {\em Phys.Rev.Lett.}, 100:171602, 2008,
  \doi{10.1103/PhysRevLett.100.171602}, \eprint{arXiv}{0801.4016}.

\bibitem{Aoki:2010xg}
Yasumichi Aoki, Tom Blum, Huey-Wen Lin, Shigemi Ohta, Shoichi Sasaki, et~al.
\newblock {Nucleon isovector structure functions in (2+1)-flavor QCD with
  domain wall fermions}.
\newblock {\em Phys.Rev.}, D82:014501, 2010, \doi{10.1103/PhysRevD.82.014501},
  \eprint{arXiv}{1003.3387}.

\bibitem{Edwards:2005ym}
R.G. Edwards et~al., LHPC Collaboration.
\newblock {The Nucleon axial charge in full lattice QCD}.
\newblock {\em Phys.Rev.Lett.}, 96:052001, 2006,
  \doi{10.1103/PhysRevLett.96.052001}, \eprint{arXiv}{hep-lat/0510062}.

\bibitem{Edwards:2006qx}
R.G. Edwards, G.~Fleming, Ph. Hagler, John~W. Negele, K.~Orginos, et~al.
\newblock {Nucleon structure in the chiral regime with domain wall fermions on
  an improved staggered sea}.
\newblock {\em PoS}, LAT2006:121, 2006, \eprint{arXiv}{hep-lat/0610007}.

\bibitem{Bratt:2010jn}
J.D. Bratt et~al., LHPC Collaboration.
\newblock {Nucleon structure from mixed action calculations using 2+1 flavors
  of asqtad sea and domain wall valence fermions}.
\newblock {\em Phys.Rev.}, D82:094502, 2010, \doi{10.1103/PhysRevD.82.094502},
  \eprint{arXiv}{1001.3620}.

\bibitem{Gockeler:2011ze}
M.~Gockeler et~al., QCDSF/UKQCD Collaboration.
\newblock {Baryon Axial Charges and Momentum Fractions with $N_f=2+1$ Dynamical
  Fermions}.
\newblock {\em PoS}, LATTICE2010:163, 2010, \eprint{arXiv}{1102.3407}.

\bibitem{Lin:2011sa}
Huey-Wen Lin and Saul~D. Cohen.
\newblock {Nucleon and Pion Form Factors from $N_f=2+1$ Anisotropic Lattices}.
\newblock 2011, \eprint{arXiv}{1104.4319}.

\bibitem{Martinelli:1994ty}
G.~Martinelli, C.~Pittori, Christopher~T. Sachrajda, M.~Testa, and A.~Vladikas.
\newblock {A General method for nonperturbative renormalization of lattice
  operators}.
\newblock {\em Nucl. Phys.}, B445:81\unskip--\ignorespaces 108, 1995,
  \doi{10.1016/0550-3213(95)00126-D}, \eprint{arXiv}{hep-lat/9411010}.

\bibitem{Gockeler:2010yr}
M.~{G\"ockeler} et~al.
\newblock {Perturbative and Nonperturbative Renormalization in Lattice QCD}.
\newblock {\em Phys. Rev.}, D82:114511, 2010, \doi{10.1103/PhysRevD.82.114511},
  \eprint{arXiv}{1003.5756}.

\bibitem{Aoki:2010yq}
Yasumichi Aoki.
\newblock {Non-perturbative renormalization in lattice QCD}.
\newblock {\em PoS}, LAT2009:012, 2009, \eprint{arXiv}{1005.2339}.

\bibitem{Nakamura:2010zzi}
K.~Nakamura et~al., Particle Data Group.
\newblock {Review of particle physics}.
\newblock {\em J.Phys.}, G37:075021, 2010,
  \doi{10.1088/0954-3899/37/7A/075021}.

\bibitem{Lin:2007ap}
Huey-Wen Lin and Konstantinos Orginos.
\newblock {First Calculation of Hyperon Axial Couplings from Lattice QCD}.
\newblock {\em Phys.Rev.}, D79:034507, 2009, \doi{10.1103/PhysRevD.79.034507},
  \eprint{arXiv}{0712.1214}.

\bibitem{DOEReport}
G.R. Young et~al.
\newblock Scientific grand challenges: forefront questions in nuclear science
  and the role of computing at the extreme scale,
  \url{http://extremecomputing.labworks.org/nuclearphysics/PNNL_18739_onlineve%
rsion_opt.pdf}.
\newblock Workshop held January 26--28, 2009, sponsored by the U.S. Department
  of Energy, Office of Nuclear Physics and the Office of Advanced Scientific
  Computing.

\bibitem{Andreev:2007wg}
V.A. Andreev et~al., MuCap Collaboration.
\newblock {Measurement of the rate of muon capture in hydrogen gas and
  determination of the proton's pseudoscalar coupling {\(g_P\)}}.
\newblock {\em Phys.Rev.Lett.}, 99:032002, 2007,
  \doi{10.1103/PhysRevLett.99.032002}, \eprint{arXiv}{0704.2072}.

\bibitem{Czarnecki:2007th}
Andrzej Czarnecki, William~J. Marciano, and Alberto Sirlin.
\newblock {Electroweak radiative corrections to muon capture}.
\newblock {\em Phys.Rev.Lett.}, 99:032003, 2007,
  \doi{10.1103/PhysRevLett.99.032003}, \eprint{arXiv}{0704.3968}.

\bibitem{Bernard:2001rs}
Veronique Bernard, Latifa Elouadrhiri, and Ulf.G. Meissner.
\newblock {Axial structure of the nucleon: Topical Review}.
\newblock {\em J.Phys.}, G28:R1\unskip--\ignorespaces R35, 2002,
  \doi{10.1088/0954-3899/28/1/201}, \eprint{arXiv}{hep-ph/0107088}.

\bibitem{Gorringe:2002xx}
Tim Gorringe and Harold~W. Fearing.
\newblock {Induced pseudoscalar coupling of the proton weak interaction}.
\newblock {\em Rev.Mod.Phys.}, 76:31\unskip--\ignorespaces 91, 2004,
  \doi{10.1103/RevModPhys.76.31}, \eprint{arXiv}{nucl-th/0206039}.

\bibitem{Jonkmans:1996my}
G.~Jonkmans, S.~Ahmad, D.S. Armstrong, G.~Azuelos, Wilhelm~H. Bertl, et~al.
\newblock {Radiative muon capture on hydrogen and the induced pseudoscalar
  coupling}.
\newblock {\em Phys.Rev.Lett.}, 77:4512\unskip--\ignorespaces 4515, 1996,
  \doi{10.1103/PhysRevLett.77.4512}, \eprint{arXiv}{nucl-ex/9608005}.

\bibitem{Clark:2005as}
J.H.D. Clark, D.S. Armstrong, T.P. Gorringe, M.D. Hasinoff, P.M. King, et~al.
\newblock {Ortho-para transition rate in mu-molecular hydrogen and the proton's
  induced pseudoscalar coupling {$g_P$}}.
\newblock {\em Phys.Rev.Lett.}, 96:073401, 2006,
  \doi{10.1103/PhysRevLett.96.073401}, \eprint{arXiv}{nucl-ex/0509025}.

\bibitem{Sasaki:2003jh}
Shoichi Sasaki, Kostas Orginos, Shigemi Ohta, and Tom Blum, the
  RIKEN-BNL-Columbia-KEK Collaboration.
\newblock {Nucleon axial charge from quenched lattice QCD with domain wall
  fermions}.
\newblock {\em Phys.Rev.}, D68:054509, 2003, \doi{10.1103/PhysRevD.68.054509},
  \eprint{arXiv}{hep-lat/0306007}.

\bibitem{Yamazaki:2009zq}
Takeshi Yamazaki, Yasumichi Aoki, Tom Blum, Huey-Wen Lin, Shigemi Ohta, et~al.
\newblock {Nucleon form factors with 2+1 flavor dynamical domain-wall
  fermions}.
\newblock {\em Phys.Rev.}, D79:114505, 2009, \doi{10.1103/PhysRevD.79.114505},
  \eprint{arXiv}{0904.2039}.

\bibitem{Anselmino:2008jk}
M.~Anselmino, M.~Boglione, U.~D'Alesio, A.~Kotzinian, F.~Murgia, et~al.
\newblock {Update on transversity and Collins functions from SIDIS and
  {\(e^+e^-\)} data}.
\newblock {\em Nucl.Phys.Proc.Suppl.}, 191:98\unskip--\ignorespaces 107, 2009,
  \doi{10.1016/j.nuclphysbps.2009.03.117}, \eprint{arXiv}{0812.4366}.

\bibitem{:2008dn}
M.~Alekseev et~al., COMPASS Collaboration.
\newblock {Collins and Sivers asymmetries for pions and kaons in muon-deuteron
  DIS}.
\newblock {\em Phys.Lett.}, B673:127\unskip--\ignorespaces 135, 2009,
  \doi{10.1016/j.physletb.2009.01.060}, \eprint{arXiv}{0802.2160}.

\bibitem{Diefenthaler:2007rj}
Markus Diefenthaler, HERMES Collaboration.
\newblock {HERMES measurements of Collins and Sivers asymmetries from a
  transversely polarised hydrogen target}.
\newblock pages 579\unskip--\ignorespaces 582, 2007, \eprint{arXiv}{0706.2242}.

\bibitem{Seidl:2008xc}
R.~Seidl et~al., Belle Collaboration.
\newblock {Measurement of Azimuthal Asymmetries in Inclusive Production of
  Hadron Pairs in {\(e^+e^-\)} Annihilation at {\(\sqrt s = 10.58\)~GeV}}.
\newblock {\em Phys.Rev.}, D78:032011, 2008, \doi{10.1103/PhysRevD.78.032011},
  \eprint{arXiv}{0805.2975}.

\bibitem{Cloet:2007em}
I.C. Cloet, Wolfgang Bentz, and Anthony~William Thomas.
\newblock {Transversity quark distributions in a covariant quark-diquark
  model}.
\newblock {\em Phys.Lett.}, B659:214\unskip--\ignorespaces 220, 2008,
  \doi{10.1016/j.physletb.2007.09.071}, \eprint{arXiv}{0708.3246}.

\bibitem{Wakamatsu:2007nc}
M.~Wakamatsu.
\newblock {Comparative analysis of the transversities and the longitudinally
  polarized distribution functions of the nucleon}.
\newblock {\em Phys.Lett.}, B653:398\unskip--\ignorespaces 403, 2007,
  \doi{10.1016/j.physletb.2007.08.013}, \eprint{arXiv}{0705.2917}.

\bibitem{He:1994gz}
Han-xin He and Xiang-Dong Ji.
\newblock {The Nucleon's tensor charge}.
\newblock {\em Phys.Rev.}, D52:2960\unskip--\ignorespaces 2963, 1995,
  \doi{10.1103/PhysRevD.52.2960}, \eprint{arXiv}{hep-ph/9412235}.

\bibitem{Gockeler:2005cj}
M.~Gockeler et~al., QCDSF Collaboration, UKQCD Collaboration.
\newblock {Quark helicity flip generalized parton distributions from two-flavor
  lattice QCD}.
\newblock {\em Phys.Lett.}, B627:113\unskip--\ignorespaces 123, 2005,
  \doi{10.1016/j.physletb.2005.09.002}, \eprint{arXiv}{hep-lat/0507001}.

\bibitem{Detmold:2001jb}
William Detmold, W.~Melnitchouk, John~W. Negele, Dru~Bryant Renner, and
  Anthony~William Thomas.
\newblock {Chiral extrapolation of lattice moments of proton quark
  distributions}.
\newblock {\em Phys.Rev.Lett.}, 87:172001, 2001,
  \doi{10.1103/PhysRevLett.87.172001}, \eprint{arXiv}{hep-lat/0103006}.

\bibitem{Detmold:2002nf}
William Detmold, W.~Melnitchouk, and Anthony~William Thomas.
\newblock {Moments of isovector quark distributions from lattice QCD}.
\newblock {\em Phys.Rev.}, D66:054501, 2002, \doi{10.1103/PhysRevD.66.054501},
  \eprint{arXiv}{hep-lat/0206001}.

\bibitem{Adler:1975he}
Stephen~L. Adler, E.W. Colglazier~Jr., J.B. Healy, Inga Karliner, Judy
  Lieberman, et~al.
\newblock {Renormalization Constants for Scalar, Pseudoscalar, and Tensor
  Currents}.
\newblock {\em Phys.Rev.}, D11:3309, 1975, \doi{10.1103/PhysRevD.11.3309}.

\bibitem{Edwards:2008ja}
Robert~G. Edwards, Balint Joo, and Huey-Wen Lin.
\newblock {Tuning for Three-flavors of Anisotropic Clover Fermions with
  Stout-link Smearing}.
\newblock {\em Phys.Rev.}, D78:054501, 2008, \doi{10.1103/PhysRevD.78.054501},
  \eprint{arXiv}{0803.3960}.

\bibitem{Lin:2008pr}
Huey-Wen Lin et~al., Hadron Spectrum Collaboration.
\newblock {First results from 2+1 dynamical quark flavors on an anisotropic
  lattice: Light-hadron spectroscopy and setting the strange-quark mass}.
\newblock {\em Phys.Rev.}, D79:034502, 2009, \doi{10.1103/PhysRevD.79.034502},
  \eprint{arXiv}{0810.3588}.

\bibitem{Aoki:2010dy}
Y.~Aoki et~al., RBC.
\newblock {Continuum Limit Physics from 2+1 Flavor Domain Wall QCD}.
\newblock {\em Phys. Rev.}, D83:074508, 2011, \doi{10.1103/PhysRevD.83.074508},
  \eprint{arXiv}{1011.0892}.

\bibitem{Charles:2004jd}
J.~Charles et~al., CKMfitter Group.
\newblock {{$CP$} violation and the {CKM} matrix: Assessing the impact of the
  asymmetric {$B$} factories}.
\newblock {\em Eur. Phys. J.}, C41:1\unskip--\ignorespaces 131, 2005,
  \doi{10.1140/epjc/s2005-02169-1}, \eprint{arXiv}{hep-ph/0406184}.

\bibitem{Abe:1997gt}
F.~Abe et~al., CDF Collaboration.
\newblock {Limits on quark-lepton compositeness scales from dileptons produced
  in 1.8 TeV $p\bar{p}$ collisions}.
\newblock {\em Phys.Rev.Lett.}, 79:2198\unskip--\ignorespaces 2203, 1997,
  \doi{10.1103/PhysRevLett.79.2198}.

\bibitem{Khachatryan:2010te}
Vardan Khachatryan et~al., CMS.
\newblock {Search for Quark Compositeness with the Dijet Centrality Ratio in
  {$pp$} Collisions at {$\sqrt s = 7\ \rm TeV$}}.
\newblock {\em Phys. Rev. Lett.}, 105:262001, 2010,
  \doi{10.1103/PhysRevLett.105.262001}, \eprint{arXiv}{1010.4439}.

\bibitem{Aad:2011aj}
Georges Aad et~al., ATLAS.
\newblock {Search for New Physics in Dijet Mass and Angular Distributions in
  {$pp$} Collisions at {$\sqrt s = 7\ \rm TeV$} Measured with the ATLAS
  Detector}.
\newblock {\em New J. Phys.}, 13:053044, 2011,
  \doi{10.1088/1367-2630/13/5/053044}, \eprint{arXiv}{1103.3864}.

\bibitem{Khachatryan:2011as}
Vardan Khachatryan et~al., CMS.
\newblock {Measurement of Dijet Angular Distributions and Search for Quark
  Compositeness in {$pp$} Collisions at 7 TeV}.
\newblock {\em Phys. Rev. Lett.}, 106:201804, 2011,
  \doi{10.1103/PhysRevLett.106.201804}, \eprint{arXiv}{1102.2020}.

\bibitem{Collaboration:2011tt}
Georges Aad et~al., ATLAS.
\newblock {Search for Contact Interactions in Dimuon Events from {$pp$}
  Collisions at {$\sqrt s = 7\ \rm TeV$} with the ATLAS Detector}.
\newblock 2011, \eprint{arXiv}{1104.4398}.

\bibitem{Broadhurst:1994se}
David~J. Broadhurst and A.G. Grozin.
\newblock {Matching QCD and HQET heavy - light currents at two loops and
  beyond}.
\newblock {\em Phys.Rev.}, D52:4082\unskip--\ignorespaces 4098, 1995,
  \doi{10.1103/PhysRevD.52.4082}, \eprint{arXiv}{hep-ph/9410240}.

\bibitem{EllisSterlingWebber}
R.K. Ellis, W.J. Stirling, and B.R. Webber.
\newblock {\em QCD and Collider Physics}.
\newblock Cambridge Monographs on Particle Physics, Nuclear Physics and
  Cosmology. Cambridge University Press, 2003.

\bibitem{Aad:2011yg}
Georges Aad et~al., ATLAS.
\newblock {Search for a heavy gauge boson decaying to a charged lepton and a
  neutrino in {$1\ {\rm fb}^{-1}$} of {$pp$} collisions at {$\sqrt s = 7\ \rm
  TeV$} using the ATLAS detector}.
\newblock 2011, \eprint{arXiv}{1108.1316}.

\bibitem{CMSWprime}
Vardan Khachatryan et~al., CMS Collaboration.
\newblock {Search for W{$^\prime$} in the leptonic channels in {$pp$}
  Collisions at {$\sqrt{s}=7$ TeV}},
  \url{http://cdsweb.cern.ch/record/1369201}.
\newblock CERN Report number CMS-PAS-EXO-11-024, 2011.

\bibitem{Khachatryan:2010fa}
Vardan Khachatryan et~al., CMS Collaboration.
\newblock {Search for a heavy gauge boson W' in the final state with an
  electron and large missing transverse energy in {$pp$} collisions at {$\sqrt
  s = 7$ TeV}}.
\newblock {\em Phys.Lett.}, B698:21\unskip--\ignorespaces 39, 2011,
  \doi{10.1016/j.physletb.2011.02.048}, \eprint{arXiv}{1012.5945}.

\bibitem{Martin:2009iq}
A.~D. Martin, W.~J. Stirling, R.~S. Thorne, and G.~Watt.
\newblock {Parton distributions for the LHC}.
\newblock {\em Eur. Phys. J.}, C63:189\unskip--\ignorespaces 285, 2009,
  \doi{10.1140/epjc/s10052-009-1072-5}, \eprint{arXiv}{0901.0002}.

\bibitem{Pumplin:2002vw}
J.~Pumplin, D.R. Stump, J.~Huston, H.L. Lai, Pavel~M. Nadolsky, et~al.
\newblock {New generation of parton distributions with uncertainties from
  global QCD analysis}.
\newblock {\em JHEP}, 0207:012, 2002, \doi{10.1088/1126-6708/2002/07/012},
  \eprint{arXiv}{hep-ph/0201195}.

\end{thebibliography}

\end{document}